\documentclass[a4paper]{JHEP3}

\JHEPspecialurl{}

\usepackage{epsfig,multicol,bbm}

\newcommand\fverb{\setbox\fverbbox=\hbox\bgroup\verb}
\newcommand\fverbdo{\egroup\medskip\noindent%
            \fbox{\unhbox\fverbbox}\ }
\newcommand\fverbit{\egroup\item[\fbox{\unhbox\fverbbox}]}
\newbox\fverbbox

\title{Dynamics of false vacuum bubbles\\ : beyond the thin shell approximation}

\author{Jakob Hansen\footnote{jakobidetsortehul@gmail.com}\\ Department of Physics, Waseda University, Tokyo 169-8555, Japan \\ {\rm and} \\ Advanced Research Team, KISTI, Daejeon 305-806, Republic of Korea}
\author{Dong-il Hwang\footnote{eastone83@gmail.com} , Dong-han Yeom\footnote{innocent@muon.kaist.ac.kr} \\ Department of Physics, KAIST, Daejeon 305-701, Republic of Korea}

\received{\today}
\revised{}
\accepted{}

\abstract{We numerically study the dynamics of false vacuum bubbles which are inside an almost flat background; we assumed spherical symmetry and the size of the bubble is smaller than the size of the background horizon.
According to the thin shell approximation and the null energy condition, if the bubble is outside of a Schwarzschild black hole, unless we assume Farhi-Guth-Guven tunneling, expanding and inflating solutions are impossible.
In this paper, we extend our method to beyond the thin shell approximation: we include the dynamics of fields and assume that the transition layer between a true vacuum and a false vacuum has non-zero thickness.
If a shell has sufficiently low energy, as expected from the thin shell approximation, it collapses (Type~1).
However, if the shell has sufficiently large energy, it tends to expand.
Here, via the field dynamics, field values of inside of the shell slowly roll down to the true vacuum and hence the shell does not inflate (Type~2).
If we add sufficient exotic matters to regularize the curvature near the shell, inflation may be possible without assuming Farhi-Guth-Guven tunneling.
In this case, a wormhole is dynamically generated around the shell (Type~3).
By tuning our simulation parameters, we could find transitions between Type~1 and Type~2, as well as between Type~2 and Type~3. Between Type~2 and Type~3, we could find another class of solutions (Type~4).
Finally, we discuss the generation of a bubble universe and the violation of unitarity.
We conclude that the existence of a certain combination of exotic matter fields violates unitarity.}


\dedicated{}

\begin{document}

\section{\label{sec:intro}Introduction}

Our Universe experiences an accelerated expansion; also, our Universe seems to have had a period of exponential expansion, so called \textit{inflation} \cite{Guth:1980zm}. The simplest explanation of these phenomena are to assume that our Universe is or was in a vacuum with non-zero vacuum energy \cite{Linde:1981mu}, as the vacuum energy effectively gives a cosmological constant. After the discovery of \textit{landscape} \cite{Susskind:2003kw}, it becomes natural to assume that there are a lot of different vacua.

Once a complex potential structure is allowed, it is inevitable to include some kind of quantum tunneling from one vacuum to other vacuum. Some authors have studied these phenomena \cite{Coleman:1980aw}\cite{Lee:1987qc}. For simplicity, we assume that the background is a positive false vacuum with a small cosmological constant or an almost flat background. This de Sitter space violates the energy conservation; thus any kinds of tunneling can be allowed in principle, i.e. tunneling from low to high vacuum is possible and of course, vice versa \cite{Lee:1987qc}.

If a true vacuum bubble is generated in the de Sitter background, then the true vacuum bubble will expand and dominate all over the background \cite{Coleman:1980aw}. Here, we need to observe the causal structure of de Sitter(inside)-de Sitter(outside) combination \cite{Ansoldi:2007qu}. However, what will happen if a false vacuum bubble is generated? If the false vacuum bubble collapses, we need to observe the de Sitter-Schwarzschild-de Sitter combination \cite{Blau:1986cw}\cite{Aguirre:2005xs}\cite{Freivogel:2005qh}, where the Schwarzschild implies a Schwarzschild black hole due to the collapse of the transition region. On the other hand, if the size of a false vacuum bubble becomes on the order of the horizon size of the background \cite{Lee:1987qc} again we need to observe the de Sitter-de Sitter combination.

The most difficult situation is when the size of a false vacuum bubble becomes less than the horizon size of the background but greater than the horizon size of the false vacuum \cite{Sato:1981bf}. In this case, the false vacuum bubble has to inflate in physical coordinates but the outside observer will only see the Schwarzschild structure. Then, if we assume the null energy condition, this structure seems to be a kind of de Sitter-Schwarzschild-de Sitter space, where the Schwarzschild means the Schwarzschild wormhole \cite{Sato:1981bf}\cite{Blau:1986cw}\cite{Aguirre:2005xs}\cite{Freivogel:2005qh}; of course, it is a mathematically allowed solution. However, when the null energy condition is violated \cite{Lee:2006vka}, its exact causal structure is not well-known, but maybe a dynamical generation of a wormhole will be accompanied.

Next natural question is the meaning of the inside de Sitter space. It implies the generation of a bubble universe which is separated from our Universe, while an outside observer sees a black hole \cite{Blau:1986cw}. Is it possible to happen in a laboratory? If we assume the null energy condition and global hyperbolicity, in a general relativistic sense, it seems to be impossible since the initial condition needs a kind of singularity from a singularity theorem \cite{Farhi:1986ty}. This kind of bubble is known as an unbuildable bubble \cite{Freivogel:2005qh}. In a semi-classical sense, it seems to be possible via tunneling from a buildable bubble to an unbuildable bubble, i.e. tunneling from the outside to the inside of a Schwarzschild wormhole \cite{Farhi:1989yr}\cite{Aguirre:2005xs}\cite{Freivogel:2005qh}.

However, if the background is the anti de Sitter space and if one assumes AdS/CFT \cite{Maldacena:1997re}, false vacuum bubbles are expected to evolve by a unitary way. Therefore, one may guess that tunneling from a buildable state to an unbuildable state should be excluded in the anti de Sitter background \cite{Banks:2002nm}\cite{Freivogel:2005qh}. Also, it is reasonable to apply this principle to a background de Sitter space.

Up to now, these results were based on the thin shell approximation. According to the thin shell approximation, the transition region needs a kind of energy shell to satisfy the Einstein equations. However, in a real situation, the transition layer will have a non-zero thickness and the field values of inside of the shell will have non-trivial dynamics. Of course, it is very difficult to solve such dynamics by hand, and hence \textit{one needs a numerical approach} \cite{Piran}\cite{HHSY}\cite{Hansen:2005am}. If we extend our calculations beyond the thin shell approximation, then we can describe not only the geometry (metric and shell), but also the field dynamics which was ignored in the thin shell approximation.

In this paper, we prepare a false vacuum bubble inside of an almost flat background. According to the thin shell approximation, as one tunes the initial parameters, basically two behaviors are expected: namely the \textit{collapse} or \textit{expansion} of a shell \cite{Blau:1986cw}. We could reproduce the collapsing solution easily. However, one interesting issue is whether the expansion of a false vacuum bubble is possible or not. According to the thin shell approximation with the null energy condition, this seemed to be impossible unless assuming tunneling from a buildable state to an unbuildable state \cite{Farhi:1989yr}\cite{Freivogel:2005qh}.

Here, in this paper, we remark two important points:
\begin{itemize}
\item First, an expanding bubble solution which contains the inflating region is difficult to obtain not by the reason of geometry, but by reasons of field dynamics. Field values of a false vacuum bubble is unstable as the shell expands.
\item Second, general relativity does not exclude a generation of an inflating bubble if one assumes exotic matter fields. We show that a wormhole is dynamically generated to induce a bubble universe along the shell. It does not necessarily require tunneling from the outside to the inside of a Schwarzschild wormhole.
\end{itemize}
Since a generation of an inflating bubble implies the violation of unitarity, the authors suspect that some holographic arguments on the unitary evolution seem to have potential dangerous; or some holographic arguments restricts our assumptions on the initial state of a bubble.

%Of course, one may think that such initial states cannot be prepared for unknown physical reasons. However, though our setup requires a kind of tunneling for the initial state, it does not imply a tunneling from the outside to the inside of a Schwarzschild wormhole, but it just implies a setting for field values in a flat background. Therefore, the authors suspect that some holographic arguments on the unitary evolution seem to have potential dangerous; or some holographic arguments restricts our assumptions on the initial state of a bubble.

This paper is structured as follows. In Section~\ref{sec:basic}, we study previous results of the thin shell approximation. In Section~\ref{sec:beyond}, we introduce numerical setup to extend beyond the thin shell approximation and introduce simulation parameters. In Section~\ref{sec:causal}, we observe and classify solutions of false vacuum bubbles and discuss interesting physical issues. Finally, in Section~\ref{sec:discussion}, we comment our contributions from studies beyond the thin shell approximation and discuss the unitarity issue with the generation of a bubble universe.

\section{\label{sec:basic}Basic results of the thin shell approximation}

\subsection{\label{sec:thin}Thin shell approximation}

We assume spherical symmetry and observe the dynamics of a false vacuum bubble inside of the true vacuum background. Traditionally, some authors studied this problem by using the thin shell approximation \cite{Israel:1966rt}\cite{Sato:1981bf}\cite{Blau:1986cw}\cite{Aguirre:2005xs}\cite{Alberghi:1999kd}\cite{Freivogel:2005qh}. The thin shell approximation is based on the following two assumptions:
\begin{enumerate}
\item The inside false vacuum region is governed by the de Sitter metric and the outside true vacuum region is governed by the Schwarzschild metric.
\item Between the two regions, there is a thin mass shell which has a proper surface tension.
\end{enumerate}

The metric ansatz for the inside and the outside regions, respectively, are
\begin{eqnarray}
ds_{\mathrm{i}}^{2} = - f_{\mathrm{i}}(r)dt_{\mathrm{i}}^{2} + \frac{1}{f_{\mathrm{i}}(r)}dr^{2} + r^{2}d\Omega^{2},
\end{eqnarray}
and
\begin{eqnarray}
ds_{\mathrm{o}}^{2} = - f_{\mathrm{o}}(r)dt_{\mathrm{o}}^{2} + \frac{1}{f_{\mathrm{o}}(r)}dr^{2} + r^{2}d\Omega^{2},
\end{eqnarray}
where $f_{\mathrm{i}}(r) = 1 - r^{2}/l^{2}$ and $f_{\mathrm{o}}(r) = 1 - 2m / r$.

The metric ansatz for transition region is
\begin{eqnarray}
ds_{\mathrm{shell}}^{2} = - d\tau^{2} + R(\tau)^{2}d\Omega^{2},
\end{eqnarray}
where $r=R(\tau)$ holds. We get the equation of motion for the shell
\begin{eqnarray}
\sqrt{\dot{R}^{2} + f_{\mathrm{i}}(R)} - \sqrt{\dot{R}^{2} + f_{\mathrm{o}}(R)} = \kappa R,
\end{eqnarray}
where $\kappa = 4 \pi \sigma$ and $\sigma$ is the surface tension of the transition region. This can be reduced to the following form:
\begin{eqnarray}
\dot{R}^{2} + V_{\mathrm{eff}}(R) = 0,
\end{eqnarray}
where
\begin{eqnarray}
V_{\mathrm{eff}}(r) = f_{\mathrm{o}}(r) - \frac{(f_{\mathrm{i}}(r) - f_{\mathrm{o}}(r) - \kappa^{2} r^{2})^{2}}{4 \kappa^{2} r^{2}}.
\end{eqnarray}
However, to maintain the information of the sign of each roots, we need to compare the extrinsic curvature for the outside and the inside of the shell. The extrinsic curvatures are defined as follows:
\begin{eqnarray}
\beta_{\mathrm{i}} = \frac{f_{\mathrm{i}}(R) - f_{\mathrm{o}}(R) + \kappa^{2} R^{2}}{2 \kappa R} = \pm \sqrt{\dot{R}^{2} + f_{\mathrm{i}}(R)},
\end{eqnarray}
and
\begin{eqnarray}
\beta_{\mathrm{o}} = \frac{f_{\mathrm{i}}(R) - f_{\mathrm{o}}(R) - \kappa^{2} R^{2}}{2 \kappa R} = \pm \sqrt{\dot{R}^{2} + f_{\mathrm{o}}(R)}.
\end{eqnarray}
Now, to satisfy the Einstein equations,
\begin{eqnarray}
\beta_{\mathrm{i}} - \beta_{\mathrm{o}} = \kappa R
\end{eqnarray}
should hold. The sign of the extrinsic curvatures in $R \rightarrow 0$ or $R \rightarrow \infty$ limit intuitively tell us the asymptotic direction of the (expanding or collapsing) shell \cite{Freivogel:2005qh}.

In general, the effective potential $V_{\mathrm{eff}}$ is a convex function for a time-like shell \cite{Blau:1986cw}\cite{Freivogel:2005qh}; therefore, it allows a collapsing solution or an expanding solution. Then there are basically $5$ possibilities: (a) from expanding to collapsing, (b) from collapsing to expanding, (c) from collapsing to collapsing, (d) from expanding to expanding, and (e) a static solution in an unstable equilibrium. (a) and (b) are symmetric solutions, whereas (c) and (d) are asymmetric solutions. For simplicity, we omit the unstable equilibrium case (e).

\begin{figure}
\begin{center}
\includegraphics[scale=0.5]{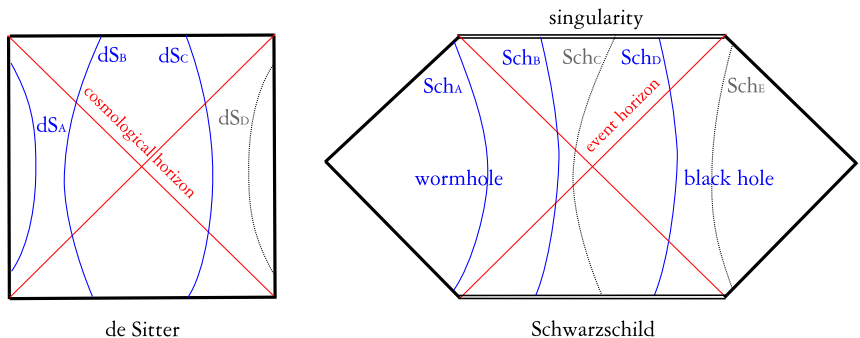}
\caption{\label{fig:thinshell}Solutions of the thin shell approximation (symmetric cases). Since $\beta_{\mathrm{i}}$ is always positive in $R \rightarrow 0$ limit, $\mathrm{dS}_{\mathrm{D}}$ is disallowed; $\beta_{\mathrm{o}}$ is always positive in $R \rightarrow 0$ limit, $\mathrm{Sch}_{\mathrm{C}}$ is disallowed; $\beta_{\mathrm{o}}$ is always negative in $R \rightarrow \infty$ limit, $\mathrm{Sch}_{\mathrm{E}}$ is disallowed.}
\end{center}
\end{figure}
\begin{figure}
\begin{center}
\includegraphics[scale=0.5]{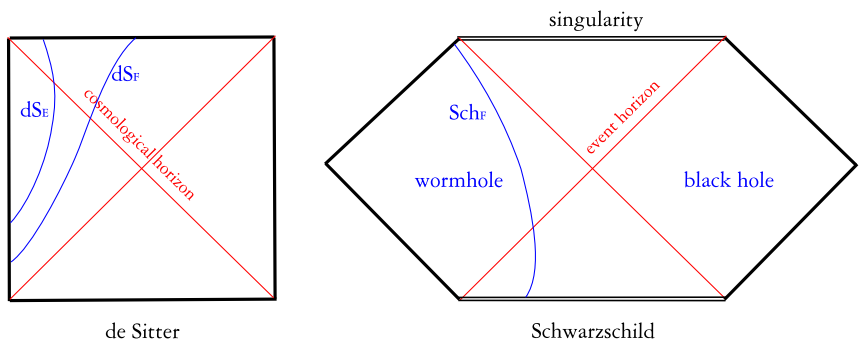}
\caption{\label{fig:thinshell2}Solutions of the thin shell approximation (asymmetric cases).}
\end{center}
\end{figure}

Firstly, let us classify symmetric solutions. The left diagram of Figure~\ref{fig:thinshell} is for the de Sitter space, and the right diagram is for the Schwarzschild space. For a collapsing case, $\mathrm{dS}_{\mathrm{A}}$ or $\mathrm{dS}_{\mathrm{D}}$ are possible; and $\mathrm{Sch}_{\mathrm{B}}$, $\mathrm{Sch}_{\mathrm{C}}$, or $\mathrm{Sch}_{\mathrm{D}}$ are possible. Also, for an expanding case, $\mathrm{dS}_{\mathrm{B}}$ or $\mathrm{dS}_{\mathrm{C}}$ are possible; and $\mathrm{Sch}_{\mathrm{A}}$ or $\mathrm{Sch}_{\mathrm{E}}$ are possible. However, according to the behavior of the extrinsic curvatures in $R\rightarrow 0$ or $R\rightarrow \infty$ limit, we can remove the solutions of $\mathrm{dS}_{\mathrm{D}}$, $\mathrm{Sch}_{\mathrm{C}}$, and $\mathrm{Sch}_{\mathrm{E}}$. Therefore, there are $4$ possible solutions: $\mathrm{dS}_{\mathrm{A}}-\mathrm{Sch}_{\mathrm{B}}$, $\mathrm{dS}_{\mathrm{A}}-\mathrm{Sch}_{\mathrm{D}}$, $\mathrm{dS}_{\mathrm{B}}-\mathrm{Sch}_{\mathrm{A}}$, $\mathrm{dS}_{\mathrm{C}}-\mathrm{Sch}_{\mathrm{A}}$. The case $\mathrm{dS}_{\mathrm{A}}-\mathrm{Sch}_{\mathrm{B}}$ is a collapsing bubble solution, where the collapsing shell is inside of a Schwarzschild wormhole.
The case $\mathrm{dS}_{\mathrm{A}}-\mathrm{Sch}_{\mathrm{D}}$ is a collapsing bubble solution, where the collapsing shell induces a Schwarzschild black hole. The case $\mathrm{dS}_{\mathrm{B}}-\mathrm{Sch}_{\mathrm{A}}$ is an expanding bubble solution, where the shell expands inside of a Schwarzschild wormhole, and the shell becomes greater than the horizon size of the inside de Sitter space. The case $\mathrm{dS}_{\mathrm{C}}-\mathrm{Sch}_{\mathrm{A}}$ is an expanding bubble solution, where the shell expands inside of a Schwarzschild wormhole, and the shell expands outside of the cosmological horizon for the $r=0$ observer.

Secondly, let us classify asymmetric solutions (Figure~\ref{fig:thinshell2}). The most interesting case is the creation of a bubble universe. In this case, we need to consider from expanding to expanding solution. Here, $\mathrm{dS}_{\mathrm{E}}$, $\mathrm{dS}_{\mathrm{F}}$, and $\mathrm{Sch}_{\mathrm{F}}$ are allowed; thus giving us the case of $\mathrm{dS}_{\mathrm{E}}-\mathrm{Sch}_{\mathrm{F}}$ and $\mathrm{dS}_{\mathrm{F}}-\mathrm{Sch}_{\mathrm{F}}$ as allowed transition solutions. We can interpret these as expanding solutions which begin from a singularity \cite{Blau:1986cw}\cite{Farhi:1986ty}.

\subsection{\label{sec:buildability}Buildability of initial states and Farhi-Guth-Guven tunneling}

\begin{figure}
\begin{center}
\includegraphics[scale=0.5]{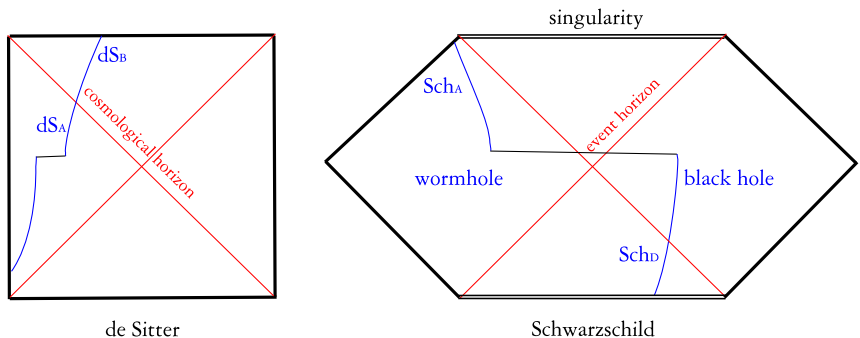}
\caption{\label{fig:FGG}Farhi-Guth-Guven tunneling.}
\end{center}
\end{figure}

If we want to build a bubble universe in a laboratory, the initial state should begin from the inside of the horizon of the de Sitter space, and the outside of the horizon of the Schwarzschild black hole or the Schwarzschild wormhole. And the final state should end at the outside of the horizon of the de Sitter space and the inside of the Schwarzschild wormhole. Then the only reasonable solution is the unbound solution $\mathrm{dS}_{\mathrm{E}}-\mathrm{Sch}_{\mathrm{F}}$ or $\mathrm{dS}_{\mathrm{F}}-\mathrm{Sch}_{\mathrm{F}}$.

However, this is not the final answer to the generation of a bubble universe. According to Farhi and Guth \cite{Farhi:1986ty}, whenever the null energy condition and global hyperbolicity hold, if a shell becomes greater than the horizon size of the inside de Sitter space, the horizon becomes a kind of anti-trapped surface, and the bubble should begin from an initial singularity. Therefore, if a solution ends with a bubble universe, its initial state should be a singular state. We call this \textit{an unbuildable state} and the opposite case is called \textit{a buildable state} \cite{Freivogel:2005qh}.

To overcome this problem and generate a bubble universe by using a constructible way, we need to include tunneling \cite{Farhi:1989yr}. The tunneling is to paste a buildable solution $\mathrm{dS}_{\mathrm{A}}-\mathrm{Sch}_{\mathrm{D}}$ and an unbuildable solution $\mathrm{dS}_{\mathrm{B}}-\mathrm{Sch}_{\mathrm{A}}$ \cite{Aguirre:2005xs}. The former does not hold the conditions of the singularity theorem, and thus, one may assume that its initial state is buildable. Some authors have calculated the probability using some approximations of quantum gravity, and found consistent results \cite{Farhi:1989yr}. This is known as Farhi-Guth-Guven tunneling (Figure~\ref{fig:FGG}).

\subsection{\label{sec:problems}Conclusions of the thin shell approximation}

The followings are conclusions of the thin shell approximation with the null energy condition:
\begin{enumerate}
\item A shell will either collapse or expand; if it has sufficient energy, it can expand to an unbounded size.
\item A shell cannot expand except if it is in a Schwarzschild wormhole.
\item When a false vacuum bubble is generated in an almost flat background, if one wants to make a bubble universe, it should tunnel into the Schwarzschild wormhole.
\end{enumerate}

One interesting question is this: is it possible to prepare any field configurations, thin or thick shell, so that the bubble is able to expand forever? Let us assume an initial bubble which is inside of the de Sitter horizon and outside of the Schwarzschild black hole. Then, the only allowed solution from the thin shell approximation is $\mathrm{dS}_{\mathrm{A}}-\mathrm{Sch}_{\mathrm{D}}$, i.e. a collapsing bubble solution \cite{Blau:1986cw}, and hence it cannot expand forever according to the thin shell approximation \cite{Aguirre:2005xs}\cite{Freivogel:2005qh}. Now the next natural step is to move to beyond the thin shell approximation.

\section{\label{sec:beyond}Beyond the thin shell approximation}

In this paper, by going beyond the thin shell approximation, we observe some field configurations where the bubble which is outside of the Schwarzschild radius can expand forever. Then, what is the difference between the thin shell approximation and \textit{beyond} the thin shell approximation?

We specify the following key assumptions for beyond the thin shell approximation:
\begin{enumerate}
\item The field of the inside region is in a false vacuum and the field of outside is in a true vacuum; we do not assume special metric structures and we consider whole dynamics of metric and fields.
\item The transition region has a non-zero thickness.
\end{enumerate}

One possible interpretation is that, from the first assumption, the field value of the inside region is not static and not stable; then, the inside is no longer an exact de Sitter space. The other possibility is that, from the second assumption, the thick transition region induces a wormhole in a dynamical way. In this paper, we will demonstrate that (1) if we do not violate the null energy condition, the first possibility happens and (2) if we do violate the null energy condition, both possibilities happen.

%By tuning the field amplitude or the thickness of the transition region, one may tune the energy of the shell. If it has sufficient energy, it should expand to an unbounded size. However, it is not inside of the Schwarzschild wormhole. One possible interpretation of this problem is that, from the first assumption, the field value of the inside region is not static and not stable; then, the inside is no longer an exact de Sitter space. The other possibility is that, from the second assumption, the thick transition region induces a wormhole in a dynamical way. In this paper, we will demonstrate that (1) if we do not violate the null energy condition, the first possibility happens and (2) if we do violate the null energy condition, the second possibility also happens.

%Of course, if the null energy condition is violated, it is known that the expanding and inflating bubble solution is allowed even in the thin shell approximation. However, if there is no Farhi-Guth-Guven tunneling, it was unclear how to connect the outside Schwarzschild structure and the inside inflating region; we need a wormhole, but the thin shell approximation does not tell how to build a wormhole. We show that the non-zero thickness transition region induces a throat of a wormhole by a dynamical way.

Now we discuss our model and setup.

\subsection{\label{sec:setup}Setup}

We describe a Lagrangian with a scalar field $\Phi$ with a potential $V(\Phi)$ \cite{Hawking:1973uf}\cite{Wald:1984rg}:
\begin{eqnarray} \label{Lagrangian}
\mathcal{L} = - \Phi_{;a}\Phi_{;b}g^{ab}-2V(\Phi).
\end{eqnarray}
From this Lagrangian we can derive the equations of motion for the scalar field:
\begin{eqnarray} \label{scalar}
\Phi_{;ab}g^{ab}-V^{'}(\Phi) = 0.
\end{eqnarray}
Also, the energy-momentum tensor becomes
\begin{eqnarray} \label{energy_momentum}
T_{ab}=\Phi_{;a}\Phi_{;b}-\frac{1}{2}g_{ab}(\Phi_{;c}\Phi_{;d}g^{cd}+2V(\Phi)).
\end{eqnarray}

Now, we will describe our numerical setup.
We start from the double-null coordinates (our convention is $[u,v,\theta,\varphi]$),
\begin{eqnarray} \label{double_null}
ds^{2} = -\alpha^{2}(u,v) du dv + r^{2}(u,v) d\Omega^{2},
\end{eqnarray}
assuming spherical symmetry.
Here, $u$ is the in-going null direction and $v$ is the out-going null direction.

We define main functions as follows (we follow the numerical approach of previous authors \cite{Hamade:1995ce}\cite{Piran}\cite{HHSY}\cite{Hansen:2005am}.): the metric function $\alpha$, the area function $r$, and the massless scalar field $S \equiv \sqrt{4\pi} \Phi$. Also, we use some conventions: $d \equiv \alpha_{,v}/\alpha$, $h \equiv \alpha_{,u}/\alpha$, $f \equiv r_{,u}$, $g \equiv r_{,v}$, $W \equiv S_{,u}$, $Z \equiv S_{,v}$.

From this setup, the following components can be calculated:
\begin{eqnarray} \label{G_and_T}
G_{uu}&=&-\frac{2}{r} (f_{,u}-2fh), \nonumber \\
G_{uv}&=&\frac{1}{2r^{2}} \left( 4 rf_{,v} + \alpha^{2} + 4fg \right), \nonumber \\
G_{vv}&=&-\frac{2}{r} (g_{,v}-2gd), \nonumber \\
G_{\theta\theta}&=&-4\frac{r^{2}}{\alpha^{2}} \left(d_{,u}+\frac{f_{,v}}{r}\right), \nonumber \\
T_{uu}&=&\frac{1}{4\pi} W^{2}, \nonumber \\
T_{uv}&=&\frac{\alpha^{2}}{2} V(S), \nonumber \\
T_{vv}&=&\frac{1}{4\pi} Z^{2}, \nonumber \\
T_{\theta\theta} &=& \frac{r^{2}}{2\pi\alpha^{2}} WZ - r^{2} V(S),
\end{eqnarray}
where
\begin{eqnarray}
V(S) = V(\Phi) |_{\Phi = S/\sqrt{4\pi}}.
\end{eqnarray}

From the equation of the scalar field, we get the following equation:
\begin{eqnarray} \label{scalar_2}
rZ_{,u}+fZ+gW+ \pi \alpha^{2}rV^{'}(S)=0.
\end{eqnarray}
Note that, $V^{'}(S) = dV(S)/dS$.

Finally, we use the Einstein equation,
\begin{eqnarray} \label{Einstein}
G_{\mu\nu}=8\pi T_{\mu\nu}.
\end{eqnarray}

Including this, we can list all equations for our numerical simulations.
\begin{enumerate}
\item \emph{Einstein equations:}
\begin{eqnarray}
d_{,u} = h_{,v} &=& \frac{fg}{r^{2}} + \frac{\alpha^2}{4r^{2}} -WZ, \nonumber \\
g_{,v} &=& 2dg - rZ^{2}, \nonumber \\
g_{,u} = f_{,v} &=& -\frac{fg}{r} - \frac{\alpha^{2}}{4r} + 2\pi\alpha^2 r V(S), \nonumber \\
f_{,u} &=& 2fh - rW^{2}.
\end{eqnarray}

\item \emph{Scalar field equations:}
\begin{eqnarray}
Z_{,u} = W_{,v} = - \frac{fZ}{r} - \frac{gW}{r} - \pi \alpha^{2}V^{'}(S).
\end{eqnarray}
\end{enumerate}

\subsection{\label{sec:initial}The initial value problem and integration schemes}

We prepare a false vacuum bubble along the initial ingoing surface, where the outside is flat background. One can interpret that a combination of fields is generated in the almost flat background via quantum tunneling. We need initial conditions for each function on initial $u=u_{\mathrm{i}}=0$ and $v=v_{\mathrm{i}}=0$ surfaces.
There are two kinds of information: geometry ($\alpha, r, g, f, h, d$) and matter ($S, W, Z$).

On the geometry side, we have gauge freedom to choose the initial $r$ function; although all constant $u$ and $v$ lines are null, there remains freedom to choose the distances between null lines.
We choose $r(u,v_{\mathrm{i}})=ur_{u0}+r_{0}$ and $r(u_{\mathrm{i}},v)=vr_{v0}+r_{0}$. Here, we fix $r_{0}=10$.
Then, $g(u_{\mathrm{i}},v)=r_{v0}$ and $f(u,v_{\mathrm{i}})=r_{u0}$ are naturally obtained.
We assume that the asymptotic outside is flat: $\alpha(u_{\mathrm{i}},v_{\mathrm{i}})=1$.
Since the mass function($M(u,v) = (r/2) (1+4r_{,u}r_{,v}/\alpha^{2})$) \cite{Waugh:1986jh} should vanish at the initial surface $(u_{\mathrm{i}},v)$, we can choose $r_{u0}=-1/2$, $r_{v0}=1/2$.

On the matter side, we fix $S(u_{\mathrm{i}},v)=0$ and $S(u,v_{\mathrm{i}})$ will be defined in the next subsection. Then, one can calculate $S(u,v_{\mathrm{i}})$, $W(u,v_{\mathrm{i}})$, $S(u_{\mathrm{i}},v)$, and $Z(u_{\mathrm{i}},v)$ for initial states. Then, as one fixes $S(u,v_{\mathrm{i}})$, from the Einstein equations, one can obtain $\alpha(u,v_{\mathrm{i}})$ from $2fh = rW^{2}$ (since $r_{,uu}=0$ at the initial surface). And then, the other functions can be evolved using equations on $\alpha_{,uv}$, $r_{,uu}$ or $r_{,vv}$, and $S_{,uv}$.

We can choose two integration schemes. First, we can get $\alpha$ from $d$, $r$ from the equation for $r_{,vv}$, and $S$ from $Z$. Second, we can get $\alpha$ from $h$, $r$ from the equation for $r_{,uu}$, and $S$ from $W$. We call the former $v$-scheme, whereas the latter $u$-scheme. We mainly used the $v$-scheme. However, these results should be same. We compared them to check the consistency of simulations in Appendix \ref{sec:convergence}. Here, we used the 2nd order Runge-Kutta method \cite{nr}.

\subsection{\label{sec:free}Free parameters}

\DOUBLEFIGURE[t]{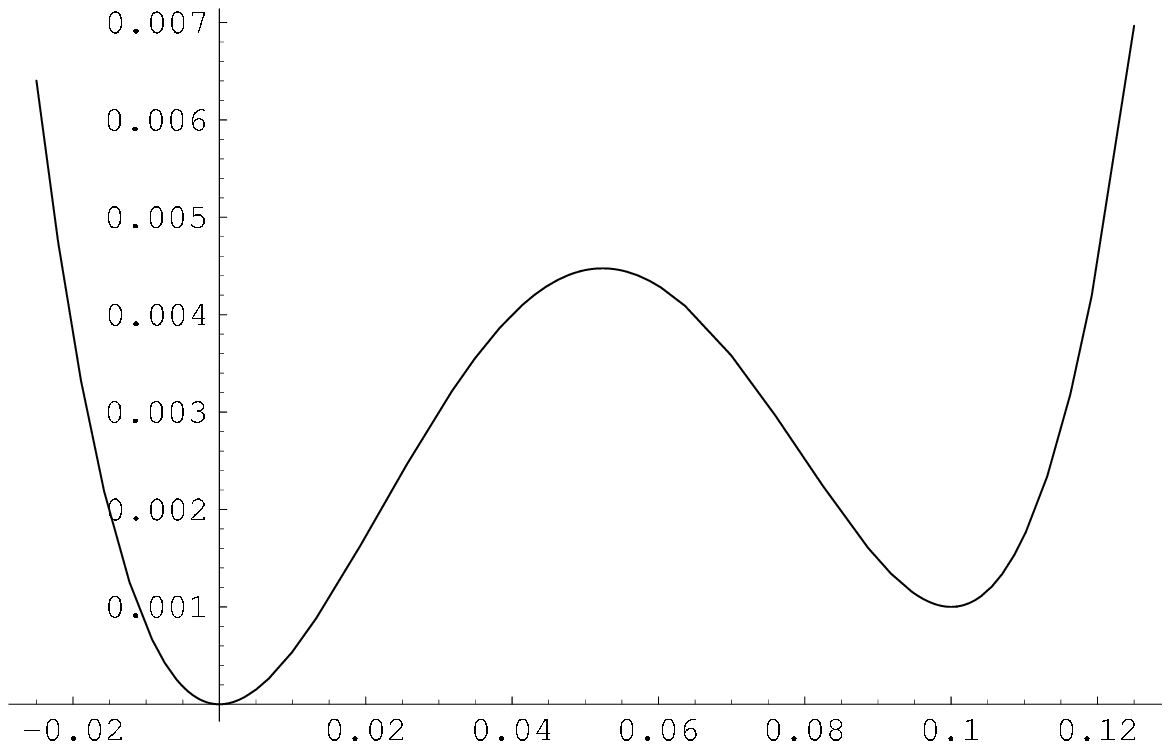,scale=0.6}{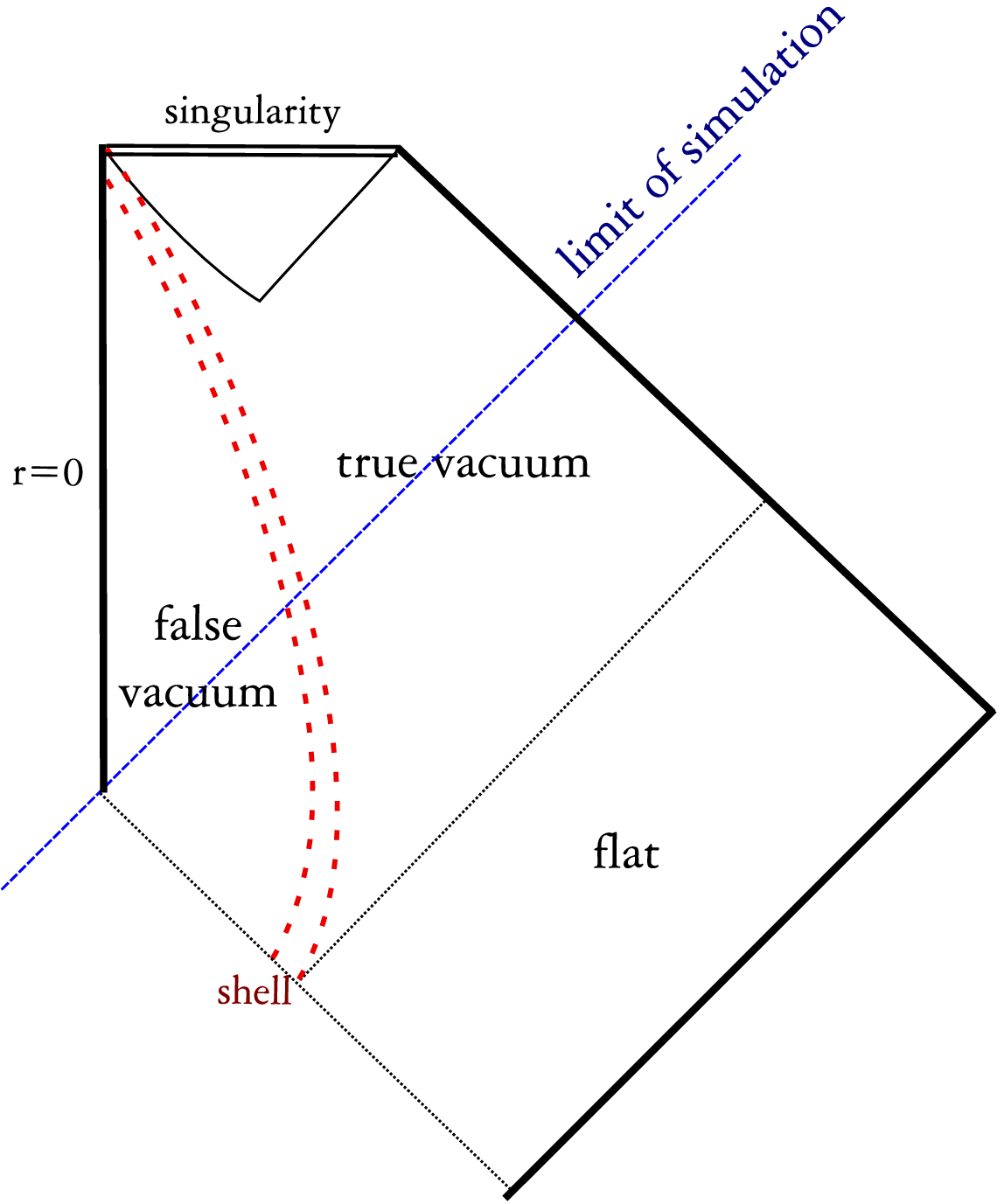,scale=0.5}{\label{fig:potential_poly}$V_{\mathrm{poly}}(S)$ with $(A=10^{5}, S_{0}=0.1, \Lambda=0.001)$.}{\label{fig:sol1}A collapsing shell solution (Type~1).}

Now we specify initial parameters.

First, we specify the potential function. We want a potential that has two stable minima, where the true vacuum of a field value $\Phi=0$ has $0$ vacuum energy and the false vacuum of a field value $\Phi=\Phi_{0}$ has vacuum energy $\Lambda$.

Second, we specify the initial field configuration $\Phi(u,v_{\mathrm{i}})$. For simplicity, we assume that the inner part has a field value $\Phi_{0}$ and the outer part has a field value $0$, i.e. the true vacuum. However, to go beyond the thin shell approximation, we need a transition region from the true vacuum to the false vacuum. So, we define the function $\Phi(u,v_{\mathrm{i}})$ as follows:
\begin{eqnarray}
\Phi(u,v_{\mathrm{i}}) = \left\{ \begin{array}{ll}
0 & u < u_{\mathrm{shell}},\\
\Phi_{0} G(u) & u_{\mathrm{shell}} \leq u < u_{\mathrm{shell}}+\Delta u,\\
\Phi_{0} & u_{\mathrm{shell}}+\Delta u \leq u,
\end{array} \right.
\end{eqnarray}
where $G(u)$ is a pasting function which goes from $0$ to $1$ by a smooth way. We choose $G(u)$ by
\begin{eqnarray}
G(u) = \sin^{2} \left[\frac{\pi(u-u_{\mathrm{shell}})}{2\Delta u}\right],
\end{eqnarray}
and we choose $u_{\mathrm{shell}} = 5$ for all simulations in this paper.\footnote{Here, the initial energy of the shell is proportional to $(\Phi_{0}/\Delta u)^{2}$ since the energy of shell depends on the gradient of the field, as we can see in the $T_{uu}$ component.
Note that if the radial function of the transition region increases, we interpret that the shell expands; if the radial function decreases, we interpret that the shell collapses. Since $r_{,u}<0$ and $r_{,v}>0$ at the outside of the shell, if the transition region approaches the out-going null direction, the shell expands; if the transition region approaches the in-going null direction, the shell collapses.}

\begin{figure}
\begin{center}
\includegraphics[scale=0.37]{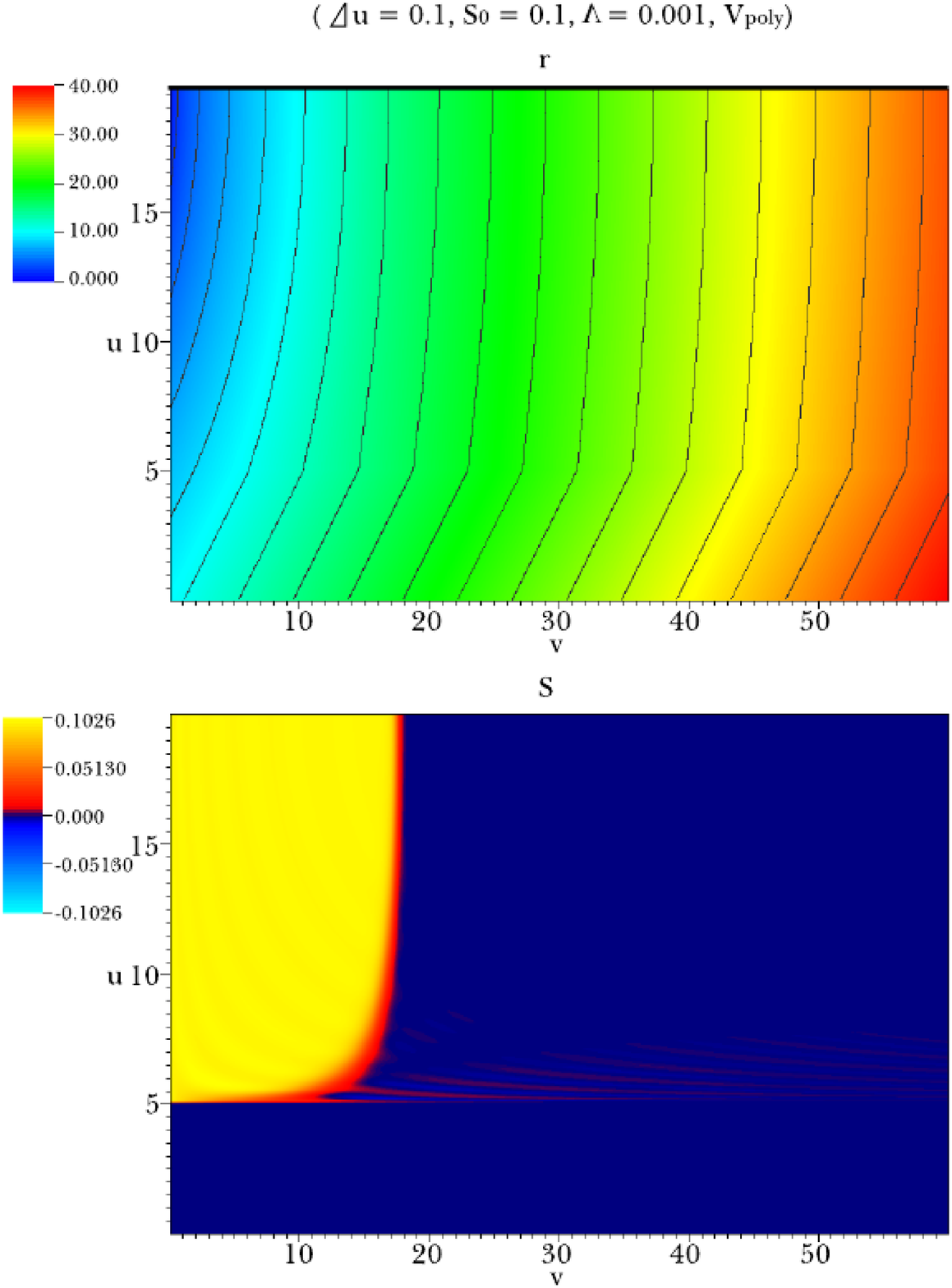}
\caption{\label{fig:sol1_sim}Simulations of $r$ and $S$ for $(\Delta u=0.1, S_{0}=0.1, \Lambda=0.001)$ with potential $V_{\mathrm{poly}}(S)$. Black curves of upper diagram are contours of $r$. The lower diagram is for $S$. Here, the yellow region is in the false vacuum, and the blue region is in the true vacuum. This shows the collapsing shell solution.}
\end{center}
\end{figure}

Now, we can specify all free parameters of our simulation.
\begin{enumerate}
  \item A potential $V(\Phi)$
  \item Thickness of the transition region $\Delta u$
  \item A field value of the false vacuum $\Phi_{0}$
  \item Vacuum energy of inside $\Lambda$
\end{enumerate}
From now, for convenience, we use $S=\sqrt{4\pi}\Phi$ rather than $\Phi$ for numerical simulations; but, of course, we can change both conventions easily. Therefore, if one fixes a potential $V(S)$, then three parameters $(\Delta u, S_{0}=\sqrt{4\pi}\Phi_{0}, \Lambda)$ fully defines a simulation.

\section{\label{sec:causal}Causal structures and physical issues}

%\FIGURE{\epsfig{file=sol1.eps,scale=0.5} \caption{A collapsing shell solution (Type 1).} \label{fig:sol1}}

In this section, we will classify $4$ types of solutions. First, as we discussed in the thin shell approximation in Section~\ref{sec:basic}, we observe a collapsing shell solution. We call this Type~1. Second, by giving sufficient energy to a shell, we will get an expanding shell solution. We call this Type~2. However, our simulations do not contradict with the expectations of the thin shell approximation, since field values of the inside false vacuum become unstable. To induce inflation, we will prepare $N$ matter shells and $N$ \textit{exotic} matter shells. In this setup, an inflating shell and creation of a bubble universe are possible. We call this Type~3. Finally, by tuning the initial parameters, we observe a transition from Type~2 to Type~3; between these two types, we could find another solution which we call Type~4.

\begin{figure}
\begin{center}
\includegraphics[scale=0.27]{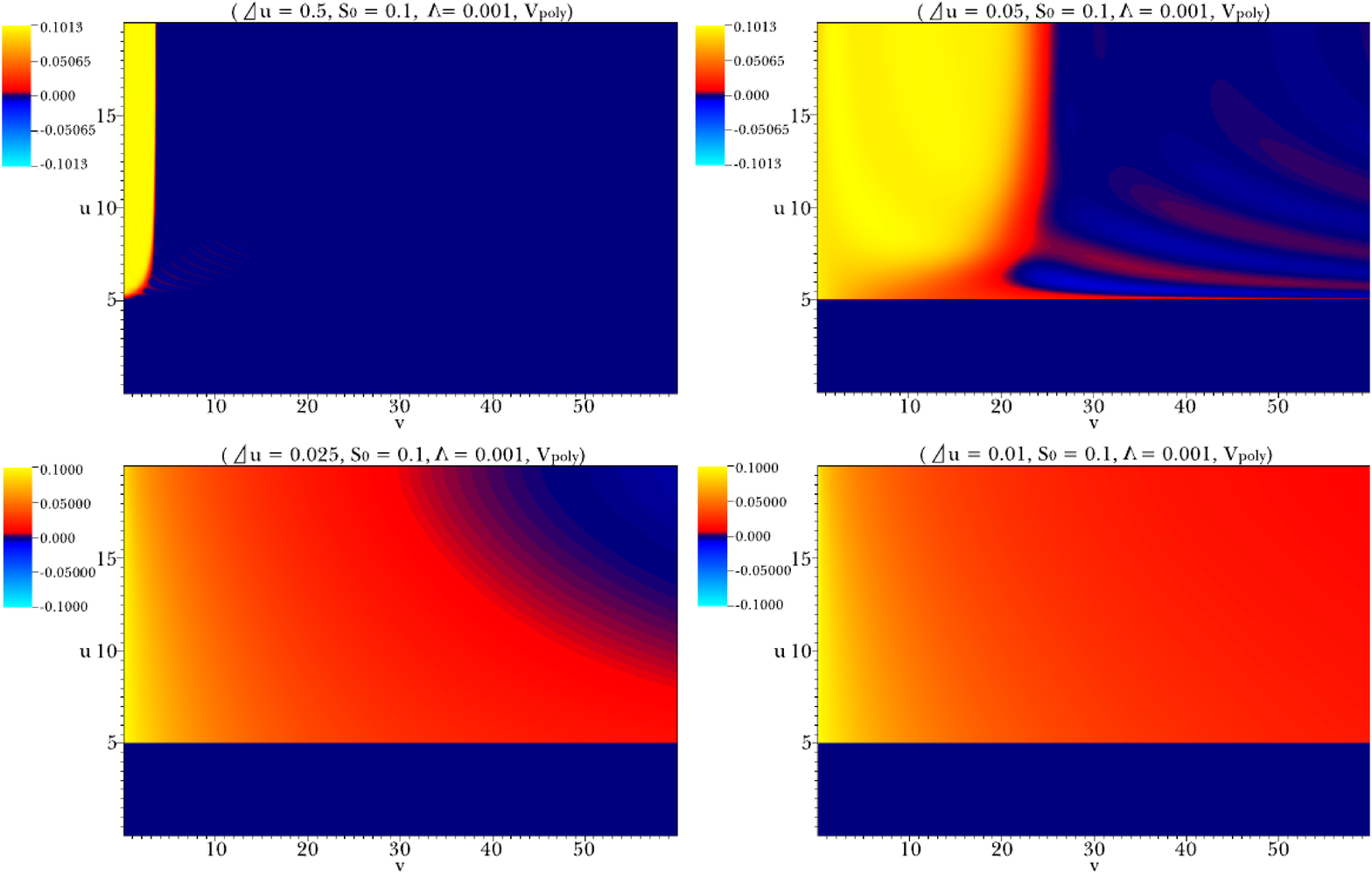}
\caption{\label{fig:change_Delta_u}Variation of the thickness of the shell $\Delta u$. As $\Delta u$ decreases, the shell expands, and field values of the inside false vacuum become unstable.}
\end{center}
\end{figure}

\subsection{\label{sec:type1}Type~1: a collapsing shell}

\subsubsection{\label{sec:collapsing}Collapsing shell solutions}

We begin with the following potential $V_{\mathrm{poly}}(\Phi)$ of polynomial form:
\begin{eqnarray} \label{poly}
V_{\mathrm{poly}}(\Phi) = A\Phi^{2} \left[ \Phi^{2} - 2 \left( \frac{\Lambda}{A\Phi_{0}^{3}}+\Phi_{0} \right) \Phi - 2\Phi_{0}^{2} + 3 \left( \frac{\Lambda}{A\Phi_{0}^{2}}+\Phi_{0}^{2} \right) \right].
\end{eqnarray}
This potential has a stable minimum around $\Phi_{0}$ with vacuum energy $\Lambda$. The three parameters $A, \Phi_{0},$ and $\Lambda$ define one specific potential. For example, if one chooses $A=10^{5}, S_{0}=0.1$, and $\Lambda=0.001$, then the following potential $V_{\mathrm{poly}}(S)$ is obtained (Figure~\ref{fig:potential_poly}). As we change $S_{0}$ and $\Lambda$, by tuning $A$, we could tune the ratio between the height of the unstable equilibrium and the height of the false vacuum to be $\sim 4.5$.

First, we calculate $(\Delta u=0.1, S_{0}=0.1, \Lambda=0.001)$. Figure~\ref{fig:sol1_sim} shows the results.

The upper diagram of Figure~\ref{fig:sol1_sim} is for the function $r$. Gradients of each contour lines are changed around $u=5$, since there is the transition region from the true vacuum to the false vacuum. Though the gradients are changed from lower to upper region, each contour line in the upper region is almost straight and parallel. This implies that the upper region of $u=5$ does not inflate.\footnote{Here, we could not simulate beyond $r=0$, since it gives a mathematical singularity of our equations; and hence, from $r=0$ point, there is a cutoff line where its beyond is impossible to calculate. This problem comes from the 4-dimensional spherical symmetry. However, this should be resolved as we restore full 4-dimensional gravity and hence it may not be a fundamental limitation.} %\footnote{However, this may not be the fundamental limitation (Figure \ref{fig:sol0}). Let us assume that all information of the blue line $(u=0,v)$ is copied along an another line (the red line) $(u,v=0)$ with $u=-v$. Then, in principle, we can extend our calculation beyond the region. Therefore, we strongly suspect that our setup is globally hyperbolic in principle. However, to implement by a numerical way, there remain some technical problems and we need more studies on this issue.}

\begin{figure}
\begin{center}
\includegraphics[scale=0.27]{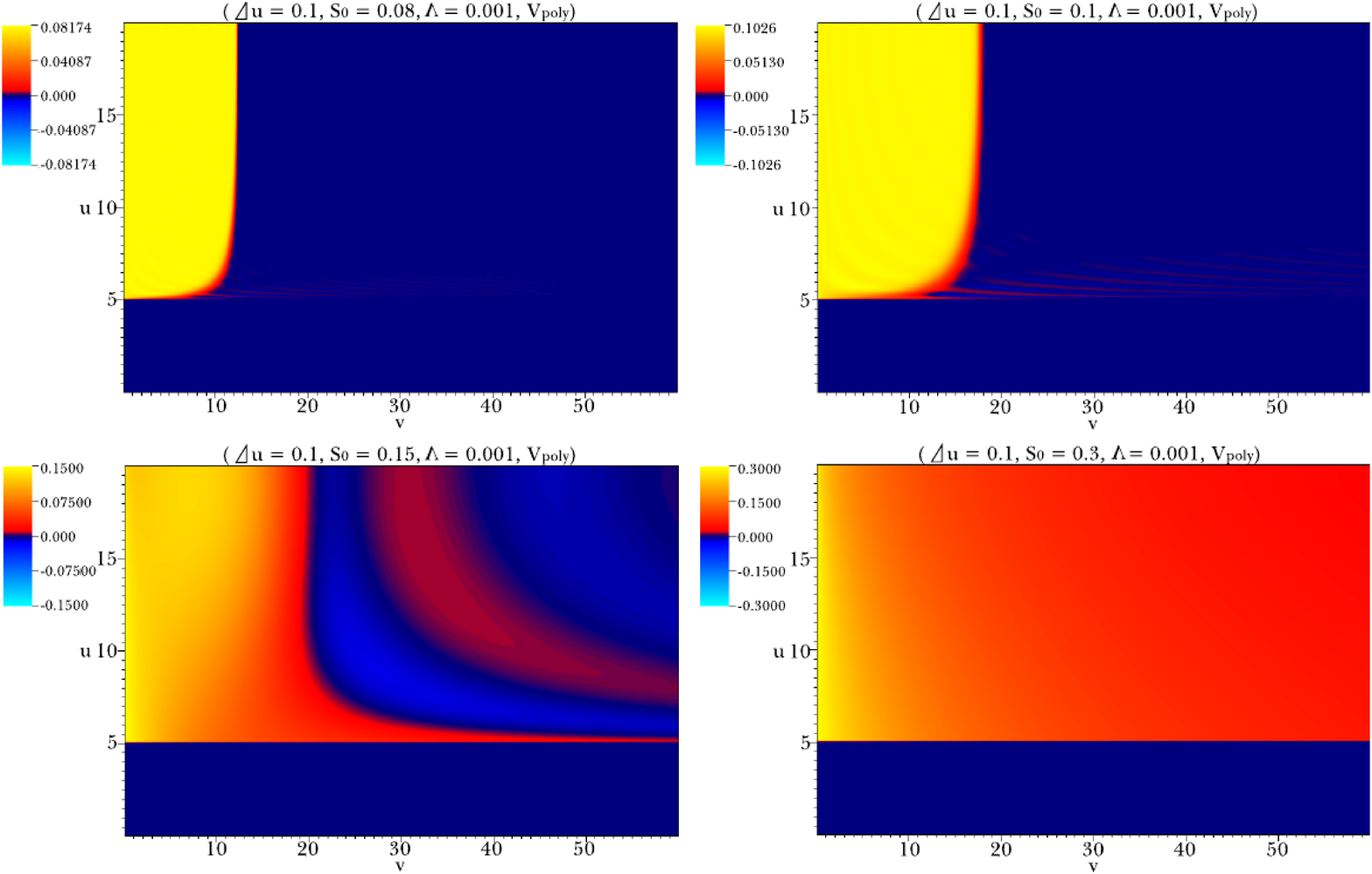}
\caption{\label{fig:change_S_0}Variation of the field value of the inside false vacuum $S_{0}$. As $S_{0}$ increases, the shell expands, and field values of the inside false vacuum become unstable.}
\end{center}
\end{figure}

The lower diagram of Figure~\ref{fig:sol1_sim} is for the function $S$. The yellow region is the false vacuum, and the black region is the true vacuum. One may notice that there are small fluctuations in the yellow region and the black region. However, even though there are fluctuations, the field value of the inside false vacuum region is almost constant. Thus, the false vacuum is quite stable.

One can notice that the false vacuum bubble collapses along a time-like direction. Then, eventually it will form a black hole, though we cannot see the black hole in our simulation. This result is demonstrated in Figure~\ref{fig:sol1}. We call this solution Type~1. This result is consistent with $\mathrm{dS}_{\mathrm{A}}-\mathrm{Sch}_{\mathrm{D}}$ solution of the thin shell approximation.

\subsubsection{\label{sec:transition}Transition from stable to unstable field values}

As we tune the three parameters $(\Delta u, S_{0}=\sqrt{4\pi}\Phi_{0}, \Lambda)$, we can observe the change of the collapsing shell solution. Note that all $r$ diagrams or causal structures are similar, and hence we omit $r$ functions and present only diagrams of the field $S$.

\begin{figure}
\begin{center}
\includegraphics[scale=0.27]{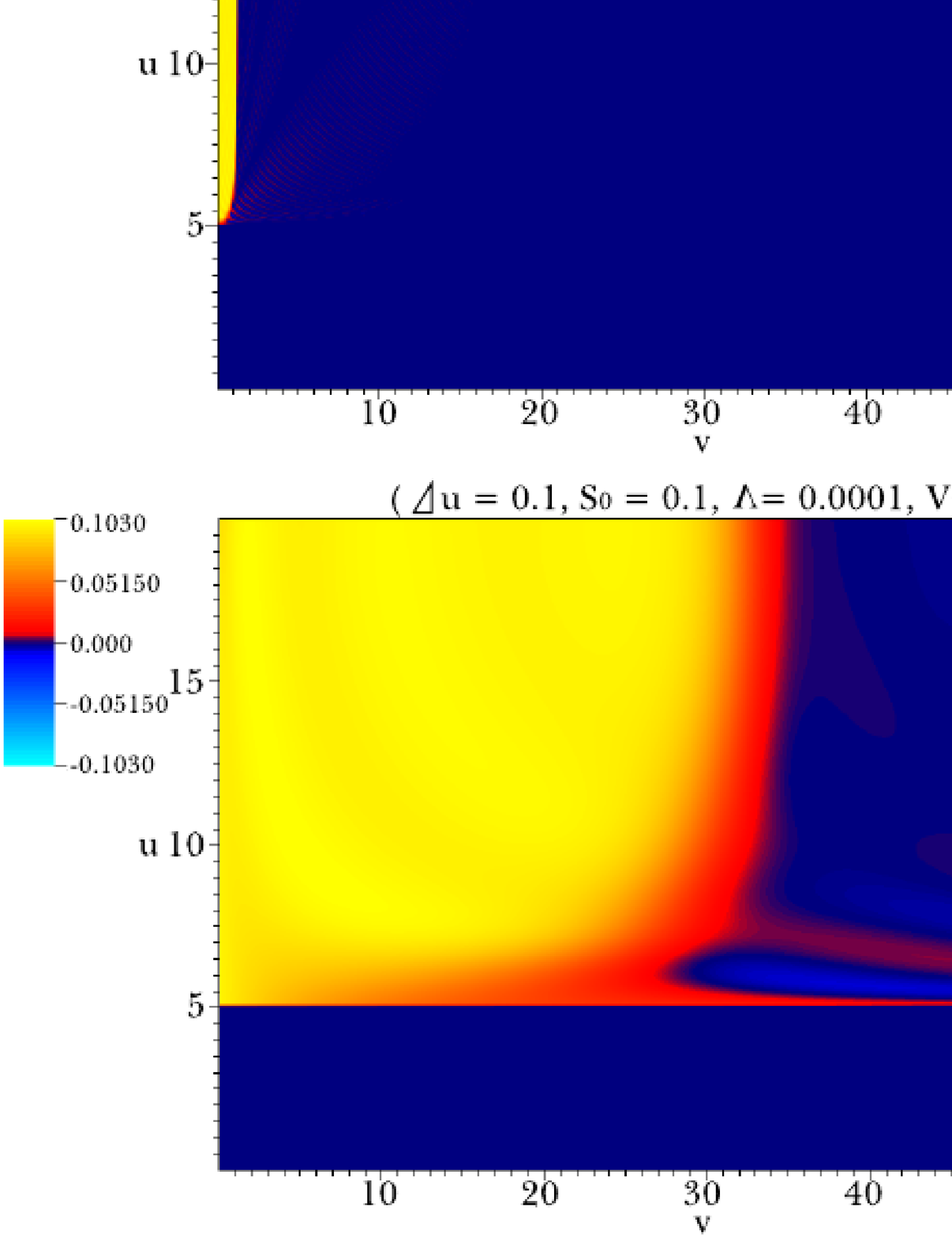}
\caption{\label{fig:change_Lambda}Variation of the vacuum energy of the inside false vacuum $\Lambda$. As $\Lambda$ decreases, the shell expands, and field values of the inside false vacuum become unstable.}
\end{center}
\end{figure}

First, we have varied the thickness of the shell $\Delta u$ (results shown in Figure~\ref{fig:change_Delta_u} and Figure~\ref{fig:sol1_sim}):
\begin{itemize}
\item $(\Delta u=0.5, S_{0}=0.1, \Lambda=0.001)$,
\item $(\Delta u=0.1, S_{0}=0.1, \Lambda=0.001)$,
\item $(\Delta u=0.05, S_{0}=0.1, \Lambda=0.001)$,
\item $(\Delta u=0.025, S_{0}=0.1, \Lambda=0.001)$,
\item $(\Delta u=0.01, S_{0}=0.1, \Lambda=0.001)$.
\end{itemize}
Note that the energy of a shell is approximately proportional to $S_{0}^{2} /\Delta u^{2}$ since the energy-momentum tensor contributes order $W^{2}$. Therefore, small $\Delta u$ and large $S_{0}$ implies more energetic shells. As the thickness of the shell becomes smaller and smaller, the shell tends to collapse more slowly. In $(\Delta u=0.05, S_{0}=0.1, \Lambda=0.001)$, fluctuations outside of the shell (i.e. near the true vacuum) becomes larger than fluctuations of $(\Delta u=0.1, S_{0}=0.1, \Lambda=0.001)$. In $(\Delta u=0.025, S_{0}=0.1, \Lambda=0.001)$ and $(\Delta u=0.01, S_{0}=0.1, \Lambda=0.001)$, the shell seems to be expanded and slowly rolls down to the true vacuum. Note that the inside region (high $u$) rolls down faster than the region near the shell (especially visible for the case $\Delta u = 0.025$).

\begin{figure}
\begin{center}
\includegraphics[scale=0.5]{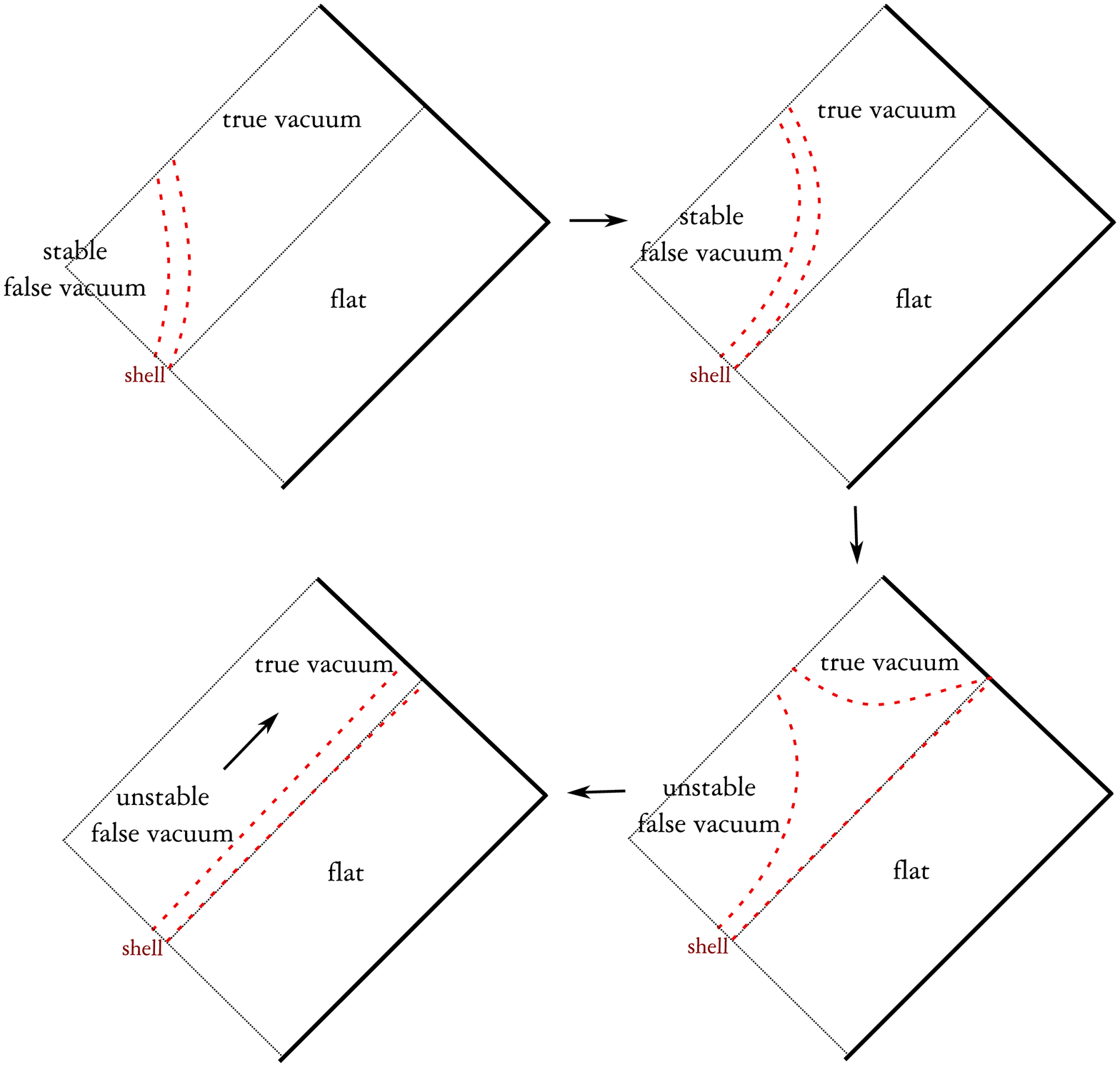}
\caption{\label{fig:sol1_transition}Transition from Type~1 to Type~2, as the energy of the shell decreases.}
\end{center}
\end{figure}

Second, we have varied the field value $S_{0}$ of the inside false vacuum region (Figure~\ref{fig:change_S_0}):
\begin{itemize}
\item $(\Delta u=0.1, S_{0}=0.08, \Lambda=0.001)$,
\item $(\Delta u=0.1, S_{0}=0.1, \Lambda=0.001)$,
\item $(\Delta u=0.1, S_{0}=0.15, \Lambda=0.001)$,
\item $(\Delta u=0.1, S_{0}=0.3, \Lambda=0.001)$.
\end{itemize}
As the field value becomes larger and larger, the shell slowly collapses. In $(\Delta u=0.1, S_{0}=0.15, \Lambda=0.001)$, the shell expands, but since the inside field values are unstable, the field quickly rolls down to the true vacuum and oscillate around $S=0$. The case $(\Delta u=0.1, S_{0}=0.3, \Lambda=0.001)$ is again an expanding solution with slow-rolling field values.

\begin{figure}
\begin{center}
\includegraphics[scale=0.37]{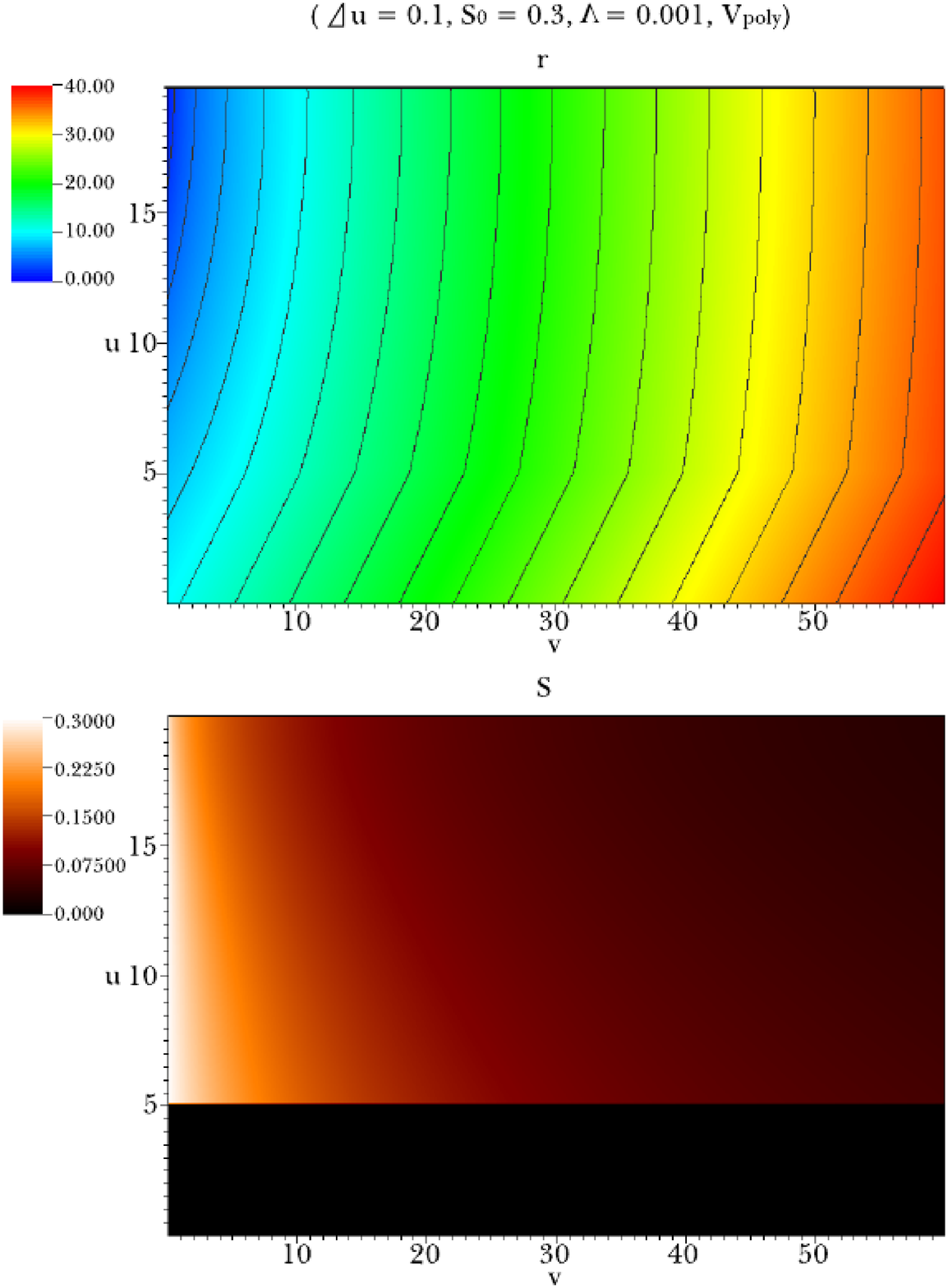}
\caption{\label{fig:sol2_sim}Simulations of $r$ and $S$ for $(\Delta u=0.1, S_{0}=0.3, \Lambda=0.001)$ with potential $V_{\mathrm{poly}}(S)$.}
\end{center}
\end{figure}

Third, we have varied the vacuum energy $\Lambda$ of the inside false vacuum region (Figure~\ref{fig:change_Lambda} and Figure~\ref{fig:sol1_sim}):
\begin{itemize}
\item $(\Delta u=0.1, S_{0}=0.1, \Lambda=0.05)$,
\item $(\Delta u=0.1, S_{0}=0.1, \Lambda=0.01)$,
\item $(\Delta u=0.1, S_{0}=0.1, \Lambda=0.001)$,
\item $(\Delta u=0.1, S_{0}=0.1, \Lambda=0.0001)$,
\item $(\Delta u=0.1, S_{0}=0.1, \Lambda=0.00001)$.
\end{itemize}
When there is large vacuum energy, the shell tends to collapse quickly. As one has small $\Lambda$, the shell tends to expand $(\Delta u=0.1, S_{0}=0.1, \Lambda=0.0001)$ and the inside field value eventually becomes unstable $(\Delta u=0.1, S_{0}=0.1, \Lambda=0.00001)$.
According to the thin shell approximation, as the vacuum energy decreases, the critical energy which determines collapse or expansion becomes decreases \cite{Blau:1986cw}. Therefore, as $\Lambda$ decreases, the expanding solution will be easily obtained and hence our observation is consistent in terms of the thin shell approximation.

Therefore, one can conclude that as the energy of a shell increases, the shell tends to expand. This result is consistent with the arguments of the thin shell approximation. We observed basic tendencies as follows. First, if the shell has sufficiently low energy than a critical value, i.e. large $\Delta u$, small $S_{0}$, and large $\Lambda$ in a certain limit, then the shell tends to collapse. In other cases, the shell will tend to expand. Second, if the shell collapses, the inside false vacuum region is stable; while if the shell expands, the inside false vacuum region is unstable and the field values slowly roll down to the true vacuum. In the latter case, the rolling begins from the inside to the outside. We include a transition diagram between two extreme cases (Figure~\ref{fig:sol1_transition}).

\DOUBLEFIGURE[t]{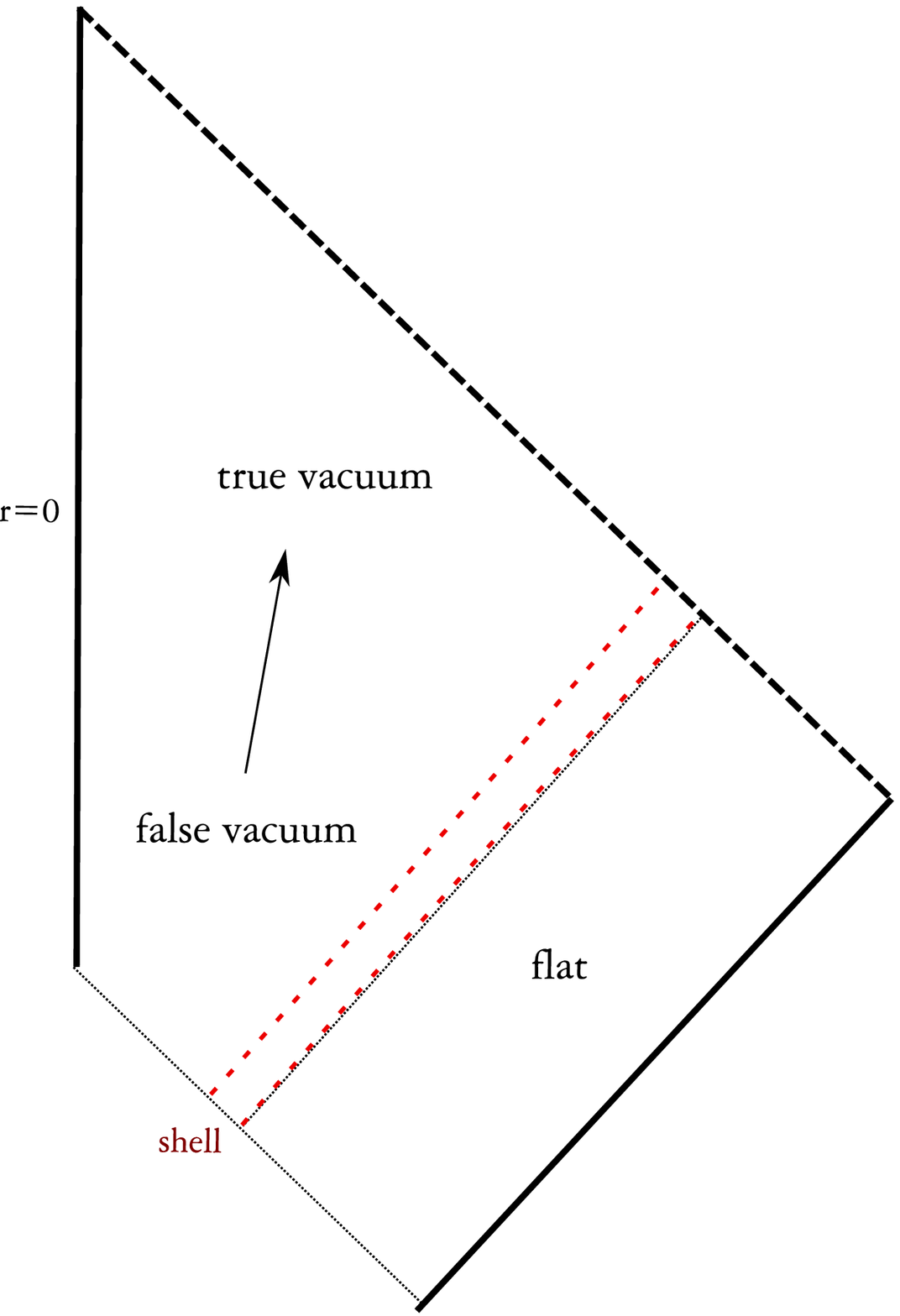,scale=0.5}{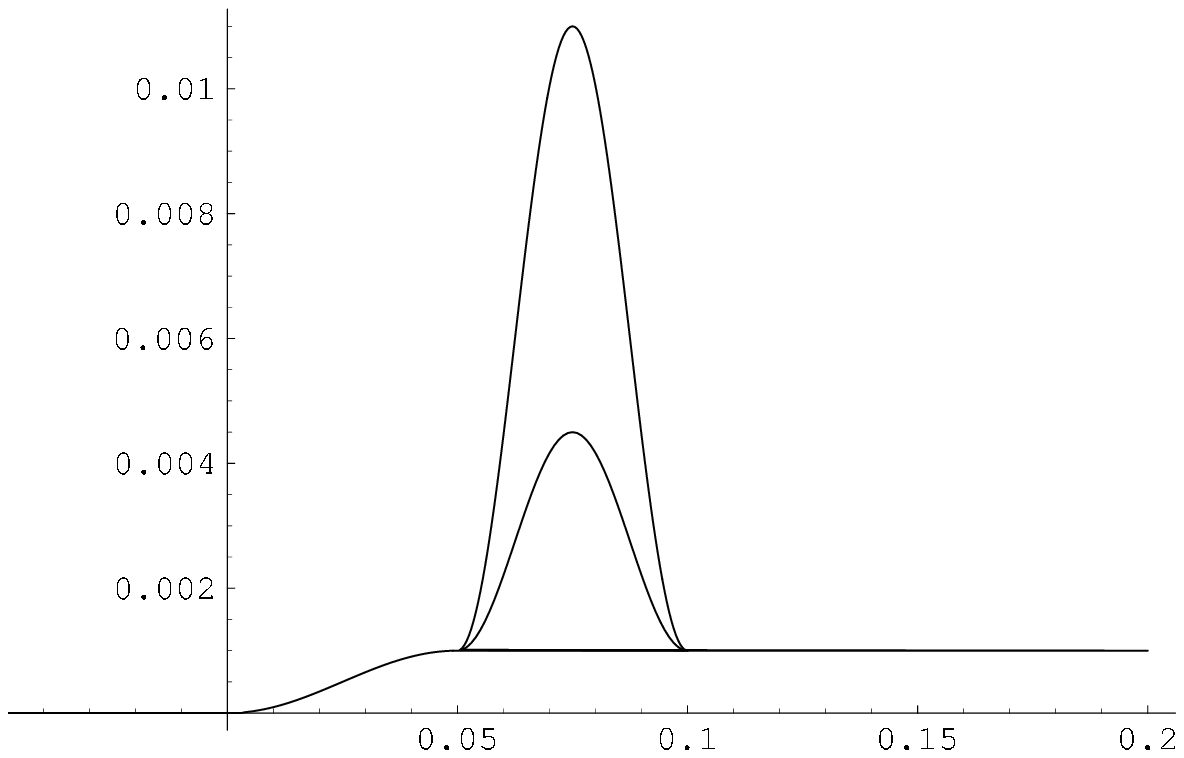,scale=0.6}{\label{fig:sol2}An expanding shell with unstable field values (Type~2).}{\label{fig:potential_wall}$V_{\mathrm{wall}}(S)$ with $(w_{1}=0.05, w_{2}=0.025, \Lambda_{\mathrm{bump}}=0, 0.0035, 0.01)$.}

\subsection{\label{sec:type2}Type~2: an expanding shell with unstable field values}

\subsubsection{\label{sec:unstable}Expanding shells with unstable field values}

From the discussions of the previous section, we observed that as the energy of the shell increases, the shell tends to expand. In this section, we observe details of the expanding unstable field value solution. We will call this Type~2.

We simulate the following parameters: $(\Delta u=0.1, S_{0}=0.3, \Lambda=0.001)$ (Figure~\ref{fig:sol2_sim}).

The upper diagram is a contour diagram of the function $r$. The causal structure is similar to Type~1. The lower diagram is of the function $S$. The field value slowly rolls down to the true vacuum. The inside region (high $u$) rolls down faster than the outside region. This result is schematically shown in Figure~\ref{fig:sol2}.

\begin{figure}
\begin{center}
\includegraphics[scale=0.27]{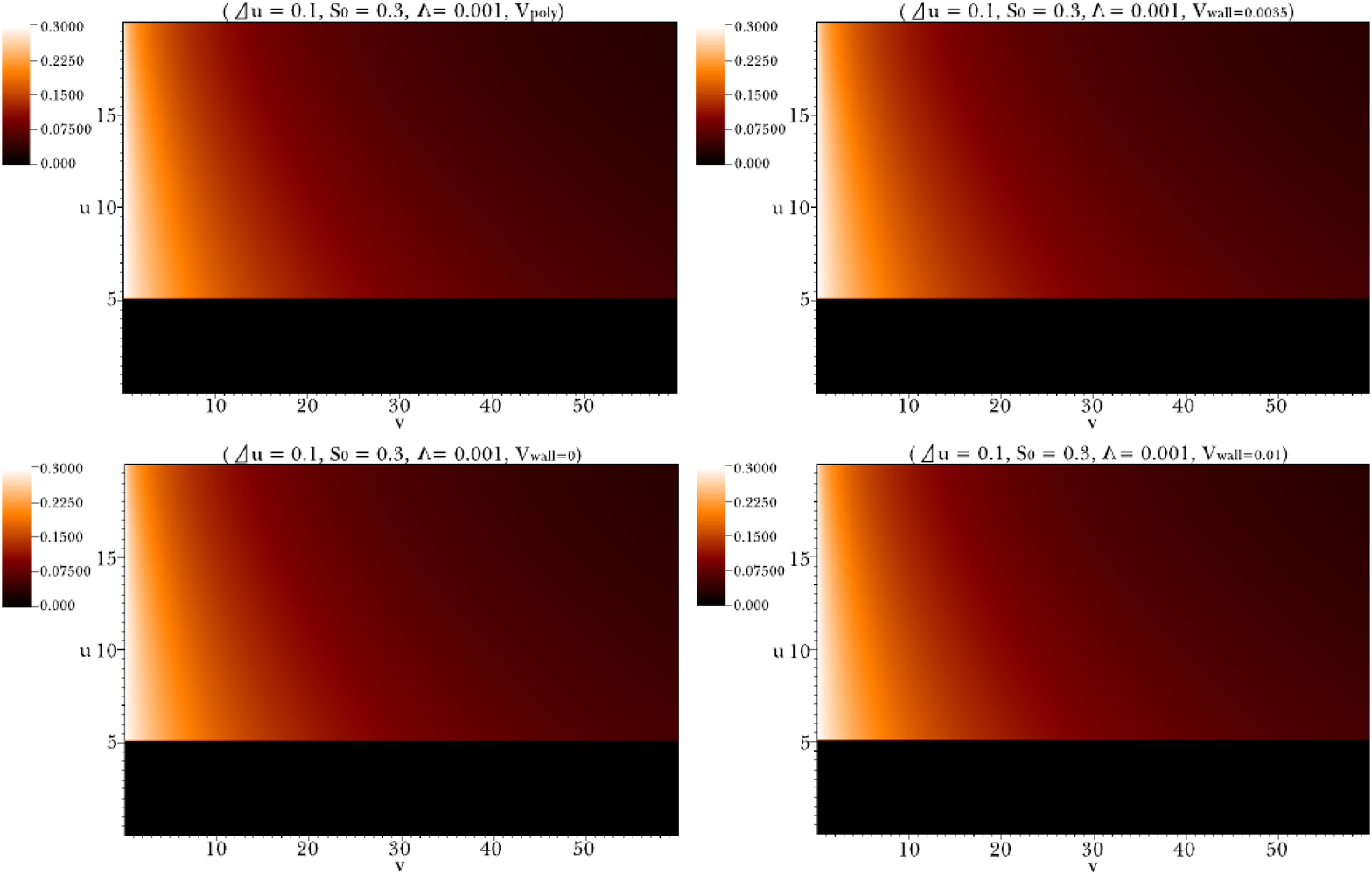}
\caption{\label{fig:change_potential}Variation of potentials.}
\end{center}
\end{figure}

\subsubsection{\label{sec:dependence}Dependence on potentials}

One question is whether the unstable behavior of inside field values depends on the potential structure or not. For this purpose, we introduce the following form of potential (Figure~\ref{fig:potential_wall}):
\begin{eqnarray}
V_{\mathrm{wall}}(S) = \left\{ \begin{array}{ll}
0 & S \leq 0, \\
\frac{\Lambda}{2} \left[ 1 - \cos \left(\frac{S \pi}{w_{1}}\right) \right] & 0 < S \leq w_{1}, \\
\Lambda + \frac{\Lambda_{\mathrm{bump}}}{2} \left[ 1 - \cos \left( \frac{(S - w_{1}) \pi}{w_{2}} \right) \right] & w_{1} < S \leq w_{1} + 2w_{2}, \\
\Lambda & w_{1}+2w_{2} < S.
\end{array} \right.
\end{eqnarray}
Here, we have used $w_{1}=0.05$ and $w_{2}=0.025$. Therefore, as we increase $\Lambda_{\mathrm{bump}}$, the barrier of potential becomes higher and higher. If the field dynamics is not changed via these different heights of barriers, it will imply that the dynamics of unstable field values do not sensitively depend on the specific potential structure.

We calculated the following cases (Figure~\ref{fig:change_potential}):
\begin{itemize}
\item $\Lambda_{\mathrm{bump}}=0$,
\item $\Lambda_{\mathrm{bump}}=0.0035$,
\item $\Lambda_{\mathrm{bump}}=0.01$,
\end{itemize}
where the other conditions are fixed by $(\Delta u=0.1, S_{0}=0.3, \Lambda=0.001)$. We observed that their field dynamics are almost the same. Therefore, the unstable behavior seems to depend on just the properties of the shell and does not sensitively depend on potential structures.

\subsubsection{\label{sec:stability}Stability analysis}

One interesting question for an expanding bubble is whether there is inflation or not. In Type~2, there is no inflation. Then, what is the reason of this?

In this paper, we will focus on the case when the field value moves slowly. Note that it does not mean the same as conventional slow-roll inflation, it just means that the change of the field amplitude via dynamics of the field is sufficiently small.\footnote{Via the instability of fast-rolling fields (see Appendix \ref{sec:app_type1}), calculations of fast-rolling fields require too large computing power, and the authors could not obtain any evidence whether there exist fast-rolling inflation in our setup.} To induce inflation, we need to maintain the field values of the inside false vacuum, and as we observed, it does not sensitively depend on the form of the potential. Therefore, for this analysis, we use a simplified potential as shown in Figure~\ref{fig:potential_analytic}.
Here, $\Phi_{,u}|_{\mathrm{shell}} \cong S_{0}/\Delta u$ and $V^{'}_{\mathrm{shell}} \cong \Lambda/S_{0}$.

\FIGURE[t]{\epsfig{file=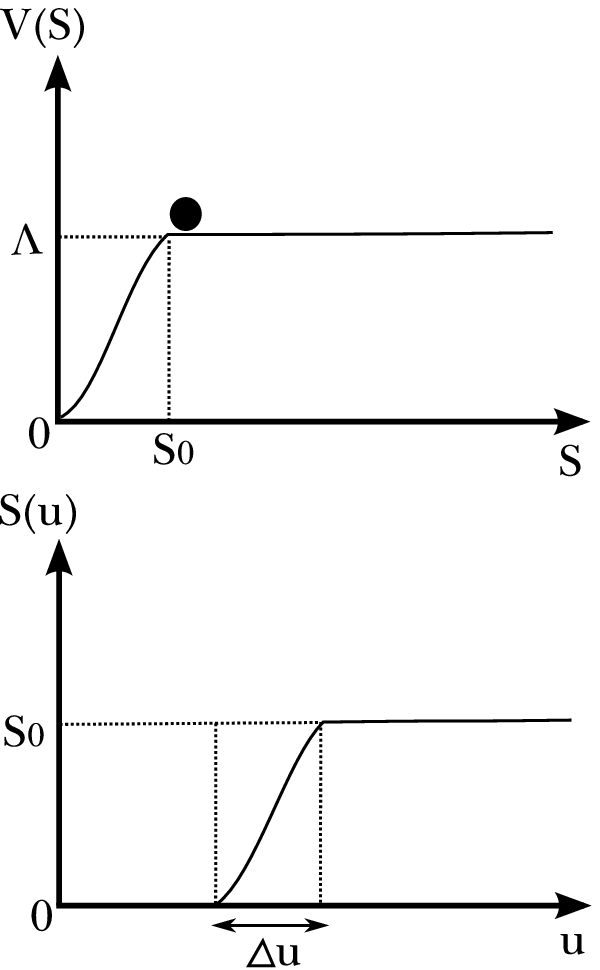,scale=0.75} \caption{A simplified potential and initial conditions.} \label{fig:potential_analytic}}

We want to see whether the near of the shell induces inflation or not. We will discuss brief outline of this section. First, to induce the stable field values, the back-reaction of the field should be sufficiently small, i.e.
\begin{eqnarray}
\frac{|\Phi_{,uv}\Delta u \Delta v|}{|\Phi_{0}|} \ll 1
\end{eqnarray}
should hold, where $\Delta u$ is the thickness of the shell and $\Delta v$ is the required time to observe inflation. Second, to observe the effect of the vacuum energy, we require $\alpha \sim 1$; if it does not hold, then all contributions of the vacuum energy will be suppressed by small $\alpha$. These two constraints will give sufficient forms of the initial conditions $\Delta u, \Phi_{0},$ and $\Lambda$. Finally, we will see that, even though we require the conditions, it is difficult to obtain inflation, since we have to require sufficiently long $\Delta v$.

Let us observe details. If one integrate the scalar field equation along $du$, it contributes only $\Delta u$. Then the initial velocity of field $\Phi_{,v}$ becomes, approximately,
\begin{eqnarray}
\Phi_{,v} \sim - \frac{r_{,v}}{r}S_{0} - \pi \alpha^{2} \frac{\Lambda}{S_{0}} \Delta u
\end{eqnarray}
around the shell. Therefore, if the field value of the false vacuum or the vacuum energy of the inside false vacuum is too large, the inside field values are unstable. To make the field values stable, we need small $S_{0}$ and small $\Delta u$ limit with $\Lambda \Delta u \ll S_{0} \ll 1$; we need to choose a proper relation between $S_{0}, \Delta u,$ and $\Lambda$.

Since $r_{,uu} = 0$ and $r_{,u} = -1/2$ along the initial constant $v$ line,
\begin{eqnarray}
- (\ln \alpha)_{,u} - r \left( \frac{S_{0}}{\Delta u} \right)^{2} \cong 0
\end{eqnarray}
holds. Then approximately,
\begin{eqnarray}
\alpha \sim \exp \left(-r \frac{S_{0}^{2}}{\Delta u} \right).
\end{eqnarray}
Therefore, one may need the condition
\begin{eqnarray}
r_{\mathrm{shell}} \frac{S_{0}^{2}}{\Delta u} \lesssim 1.
\end{eqnarray}
If it does not hold, then all terms which depend on potential $V$ are decoupled from all equations. If the field amplitude $S_{0}$ is sufficiently small, one can assume the condition. Now let us assume $\alpha \sim 1$. Note that, this condition implies that the curvature around the shell is sufficiently regular; if $\alpha \ll 1$, then the curvature $R \sim 1/\alpha^{2}$ cannot be regular around the shell.

If one compares two values
\begin{eqnarray}
\frac{|\frac{r_{,v}}{r}S_{0} \Delta v|}{|S_{0}|} \sim \frac{|\Delta v|}{|r_{\mathrm{shell}}|},
\end{eqnarray}
this ratio will be sufficiently small as one assumes $\Delta v / r_{\mathrm{shell}} \ll 1$ and $r_{,v} \sim 1$. Here, $\Delta v$ is the required time for an observation of inflation. Also, we need to compare two values
\begin{eqnarray}
\frac{|\alpha^{2} \frac{\Lambda}{S_{0}}\Delta u \Delta v|}{|\Phi|} \sim \frac{|\Lambda \Delta u \Delta v|}{|S_{0}^{2}|} \sim |\Lambda r_{\mathrm{shell}} \Delta v|.
\end{eqnarray}
Here, $\Lambda \sim 1/r_{\mathrm{shell}}^{2}$ is a natural guess, and $\Delta v / r_{\mathrm{shell}} \ll 1$ is needed.
Then, $|\Phi_{,uv}\Delta u \Delta v|/|\Phi|$ becomes sufficiently small and we can justify a stable field values.

Therefore, we can suggest some necessary conditions for a stable vacuum:
\begin{enumerate}
  \item $\alpha \sim 1$, or $r_{\mathrm{shell}} S_{0}^{2}/\Delta u \lesssim 1$, i.e. the curvature around the shell is sufficiently regular,
  \item $r_{,v} \sim 1$,
  \item $\Lambda \sim 1/r_{\mathrm{shell}}^{2}$,
  \item $\Delta v / r_{\mathrm{shell}} \ll 1$.
\end{enumerate}
Then it is natural to assume the following initial conditions:
\begin{eqnarray}
\Delta u \sim \epsilon^{2}, \quad S_{0} \sim \frac{\epsilon}{\sqrt{r_{\mathrm{shell}}}}, \quad \Lambda \sim \frac{1}{r_{\mathrm{shell}}^{2}}
\end{eqnarray}
with large $r_{\mathrm{shell}}$ and small $\epsilon$ limit.

However, this situation is difficult to implement in real situations. If we write the equation for $r_{,uv}$ and use the necessary conditions, one can obtain the following equation near the inside of the shell:
\begin{eqnarray}
r_{,uv} = f_{,v} &=& -\frac{r_{,u}r_{,v}}{r} - \frac{\alpha^{2}}{4r} + 2 \pi \alpha^{2} r V \nonumber \\
&\sim& -\frac{f}{r} - \frac{1}{4r} + 2 \pi r \Lambda.
\end{eqnarray}
If there is inflation, then initial $f = -0.5$ will increase to $0$. As one integrate both sides along $v$, the first and the second term are almost same order before the beginning of inflation. Therefore, more important contribution of $f$ comes from the third term. And, near the very thin shell, it is reasonable to choose $r \sim C v + r_{\mathrm{shell}}$, where $C$ is a constant. Then approximately,
\begin{eqnarray}
f(v) - f(v_{\mathrm{i}}) \sim C \left(\frac{v}{r_{\mathrm{shell}}}\right)^{2} + \left(\frac{v}{r_{\mathrm{shell}}}\right).
\end{eqnarray}
In this limit, $\Delta v$ for $f(\Delta v) = 0$ is on the order of $r_{\mathrm{shell}}$. Therefore, $\Delta v / r_{\mathrm{shell}}$ may not be sufficiently small. This is a basic intuitive reason why in our setup it is difficult to induce inflation.

\subsection{\label{sec:type3}Type~3: an inflating shell}

\subsubsection{\label{sec:physics}Physics of exotic matters}

Before we discuss a new type of solutions, we comment on the physics of exotic matters. If a matter violates the null energy condition, it is called a phantom matter or an exotic matter. This kind of matter were discussed in many contexts including general relativity \cite{Fewster:2005vh} and cosmology \cite{Caldwell:1999ew}.

We prepare the following Lagrangian for an exotic matter field $\Psi$:
\begin{eqnarray}
\mathcal{L'} = + \Psi_{;a}\Psi_{;b}g^{ab} + 2V(\Psi).
\end{eqnarray}
From this Lagrangian, we can derive the equation of motion for the scalar field:
\begin{eqnarray}
\Psi_{;ab}g^{ab}-V^{'}(\Psi) = 0.
\end{eqnarray}
Also, the energy-momentum tensor becomes
\begin{eqnarray}
T_{ab}=-\Psi_{;a}\Psi_{;b}+\frac{1}{2}g_{ab}(\Psi_{;c}\Psi_{;d}g^{cd}+2V(\Psi)).
\end{eqnarray}
Note that the potential $V(\Psi)$ contributes to the negative vacuum energy. Since our aim is to induce inflation, for convenience, we choose the potential by $V(\Psi)=0$.

According to previous researches, if there exist exotic matter, a static wormhole, a warp drive or a time machine may be possible \cite{Fewster:2005vh}. However, these analyses were based on a pre-existing metric ansatz which may not be justified from an almost flat background. In our setup, we do not assume a strange geometry from the initial condition; we begin with an almost flat background and assume a combination of fields.

By assuming the violation of the null energy condition, an expanding and inflating bubble solution may be justified without Farhi-Guth-Guven tunneling \cite{Lee:2006vka}. However, still the dynamical causal structure of these inflating solutions are not known. In the following sections, we will discuss the correct causal structure of the expanding and inflating bubble solution. Here, a wormhole is dynamically generated along the shell.

\subsubsection{\label{sec:Nshell}The $N$-shell bubble}

To induce inflation, one may need to hold the field values for sufficiently long time, i.e. we should control the rolling of the inside field values.
We need some conditions: (1) the force term $V'(S)$ should be suppressed; (2) the vacuum energy of the inside of the shell should be maintained; (3) the curvature around the shell should be sufficiently regular.

We introduce a brief outline of this section. One observation is that if the amplitude of a field is sufficiently small, then we can expect that the contribution on $V'(S)$ can be sufficiently small. However, the corresponding vacuum energy also becomes small. To solve this problem, we introduce a number $N$ of shells. Then we can dilute the force term maintaining the vacuum energy as a constant value. However, in that case, the curvature around the shells cannot be small, since the contribution on the energy-momentum tensor becomes order $\sqrt{N}$. To regularize the curvature around the shells, we will introduce a number of exotic matter shells. Then we can induce sufficient setup to induce inflation.

We discuss the details as the following. Let us assume $N$ scalar fields $\phi_{i}$ with potential
\begin{eqnarray}
V_{\mathrm{phi4}}(\phi_{i}) = \lambda \left( \sum_{i=1}^{N} \phi_{i}^{4} \right),
\end{eqnarray}
where $\phi_{i}$ are scalar fields and $\lambda$ is a constant. If all fields are coherent and have the same amplitude, then $V_{\mathrm{phi4}}(\phi_{i}) = N \lambda \phi_{i}^{4}$ holds. Here, the field equations are
\begin{eqnarray}
(\phi_{i})_{;ab}g^{ab} - 4 \lambda \phi_{i}^{3} = 0,
\end{eqnarray}
where $\phi_{i} \sim 1/N^{1/4}$ with $N$ fields.
Also, assume $N$ exotic matter fields $\zeta_{j}$.
Then, the equations of motion are
\begin{eqnarray}
(\zeta_{j})_{;ab}g^{ab} = 0.
\end{eqnarray}
Again, we assume $\zeta_{j} \sim 1/N^{1/4}$, and assume that all $\zeta_{j}$ are coherent with the same amplitude.

Let us assume initial conditions for $\phi_{i}$ and $\zeta_{j}$ by
\begin{eqnarray}
\phi_{i}(u,v_{\mathrm{i}}) = \left\{ \begin{array}{ll}
0 & u < u_{\mathrm{shell}},\\
(\phi_{0}/N^{1/4}) G(u) & u_{\mathrm{shell}} \leq u < u_{\mathrm{shell}}+\Delta u,\\
\phi_{0}/N^{1/4} & u_{\mathrm{shell}}+\Delta u \leq u,
\end{array} \right.
\end{eqnarray}
\begin{eqnarray}
\zeta_{j}(u,v_{\mathrm{i}}) = \left\{ \begin{array}{ll}
0 & u < u_{\mathrm{shell}},\\
(\beta \phi_{0}/N^{1/4}) G(u) & u_{\mathrm{shell}} \leq u < u_{\mathrm{shell}}+\Delta u,\\
\beta \phi_{0}/N^{1/4} & u_{\mathrm{shell}}+\Delta u \leq u.
\end{array} \right.
\end{eqnarray}
Here, we define $1 - \beta^{2} = 1/\sqrt{N}$.

Then approximately,
\begin{eqnarray}
\frac{|4 \lambda \phi_{i}^{3}|}{|(\phi_{i})_{;ab}g^{ab}|} \sim \frac{1}{N^{1/2}} \rightarrow 0
\end{eqnarray}
with large $N$. Thus, the force term can be sufficiently suppressed in this setup.

Contributions to Einstein equations should be carefully checked.
\begin{eqnarray}
(\ln \alpha)_{,uv} & = & \frac{r_{,u}r_{,v}}{r^{2}} + \frac{\alpha^2}{4r^{2}} - 4 \pi N (\phi_{i})_{,u}(\phi_{i})_{,v} + 4 \pi N (\zeta_{j})_{,u}(\zeta_{j})_{,v} \nonumber \\
& = & \frac{r_{,u}r_{,v}}{r^{2}} + \frac{\alpha^2}{4r^{2}} - 4 \pi \sqrt{N} \Phi_{,u}\Phi_{,v} + 4 \pi \sqrt{N} \Phi'_{,u}\Phi'_{,v}, \nonumber \\
r_{,vv} & = & 2 r_{,v} \frac{\alpha_{,v}}{\alpha} - 4 \pi r (N (\phi_{i})_{,v}^{2} - N (\zeta_{j})_{,v}^{2}) = 2 r_{,v} \frac{\alpha_{,v}}{\alpha} - 4 \pi r \sqrt{N} \Phi_{,v}^{2} + 4 \pi r \sqrt{N} {\Phi'}_{,v}^{2}, \nonumber \\
r_{,uu} & = & 2 r_{,u} \frac{\alpha_{,u}}{\alpha} - 4 \pi r (N (\phi_{i})_{,u}^{2} - N (\zeta_{j})_{,u}^{2}) = 2 r_{,u} \frac{\alpha_{,u}}{\alpha} - 4 \pi r \sqrt{N} \Phi_{,u}^{2} + 4 \pi r \sqrt{N} {\Phi'}_{,u}^{2}, \nonumber \\
r_{,uv} & = & -\frac{r_{,u}r_{,v}}{r} - \frac{\alpha^{2}}{4r} + 2\pi\alpha^2 r N \lambda \phi_{i}^{4} = -\frac{r_{,u}r_{,v}}{r} - \frac{\alpha^{2}}{4r} + 2\pi\alpha^2 r \lambda \Phi^{4}, \nonumber \\
\Phi_{,uv} & = & - \frac{r_{,u}\Phi_{,v}}{r} - \frac{r_{,v}\Phi_{,u}}{r} - \frac{1}{\sqrt{N}} \frac{\sqrt{\pi}}{2} \alpha^{2} (4 \lambda \Phi^{3}), \nonumber \\
\Phi'_{,uv} & = & - \frac{r_{,u}\Phi'_{,v}}{r} - \frac{r_{,v}\Phi'_{,u}}{r},
\end{eqnarray}
where we define effective fields $\Phi(u,v) \equiv N^{1/4} \phi_{i}(u,v)$ and $\Phi'(u,v) \equiv N^{1/4} \zeta_{j}(u,v)$. We assume the following initial conditions:
\begin{eqnarray}
\Phi(u,v_{\mathrm{i}}) = \left\{ \begin{array}{ll}
0 & u < u_{\mathrm{shell}},\\
\phi_{0} G(u) & u_{\mathrm{shell}} \leq u < u_{\mathrm{shell}}+\Delta u,\\
\phi_{0} & u_{\mathrm{shell}}+\Delta u \leq u,
\end{array} \right.
\end{eqnarray}
\begin{eqnarray}
\Phi'(u,v_{\mathrm{i}}) = \left\{ \begin{array}{ll}
0 & u < u_{\mathrm{shell}},\\
\beta \phi_{0} G(u) & u_{\mathrm{shell}} \leq u < u_{\mathrm{shell}}+\Delta u,\\
\beta \phi_{0} & u_{\mathrm{shell}}+\Delta u \leq u.
\end{array} \right.
\end{eqnarray}

\begin{figure}
\begin{center}
\includegraphics[scale=0.37]{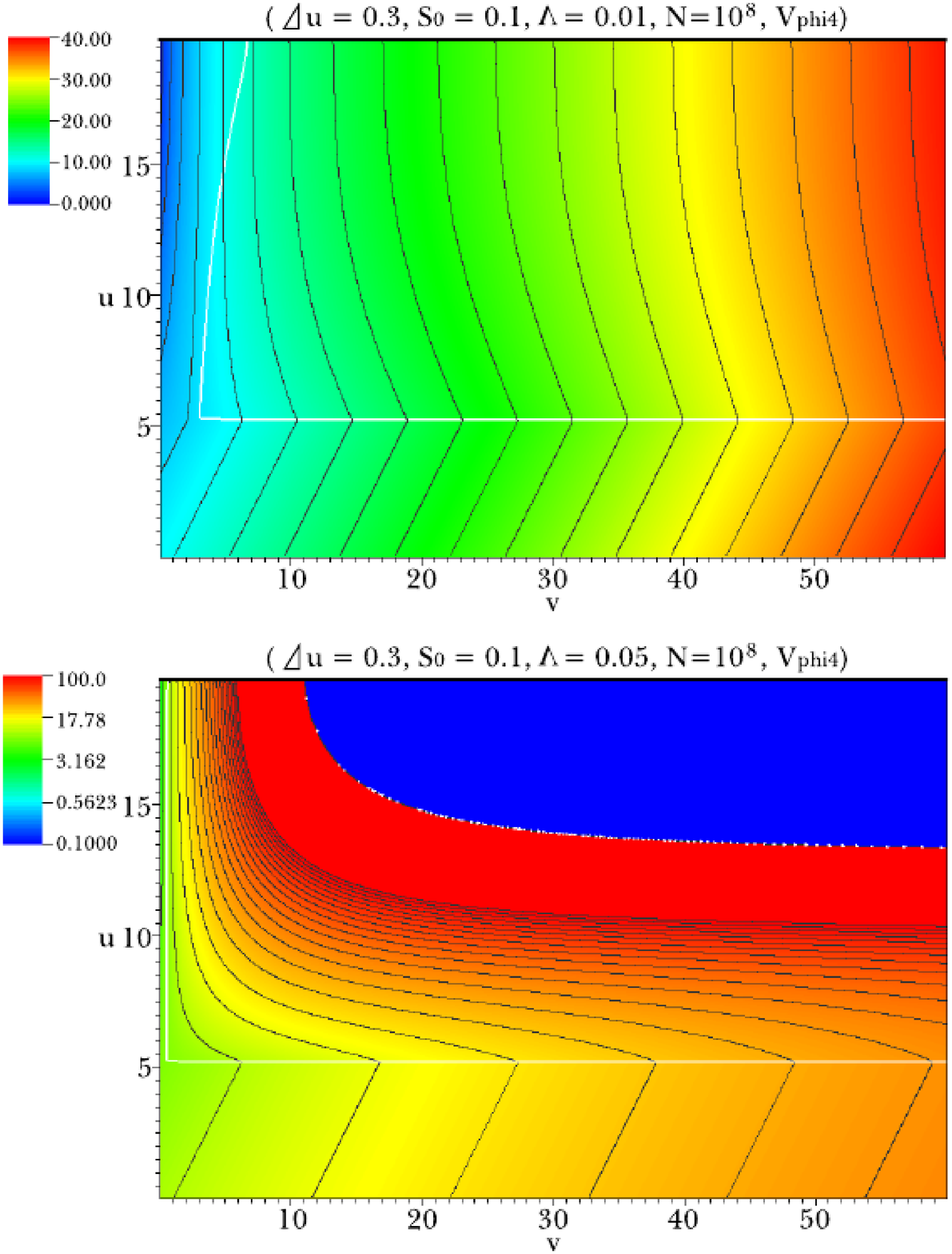}
\caption{\label{fig:sol3_sim}$r$ diagrams for $(\Delta u=0.3, S_{0}=0.1, \Lambda=0.01, N=10^{8})$ and $(\Delta u=0.3, S_{0}=0.1, \Lambda=0.05, N=10^{8})$ with $V_{\mathrm{phi4}}(S)$ potential. White curves are $r_{,u}=0$ horizons. Blue region in the lower diagram is space-like future infinity. This shows an inflating shell solution.}
\end{center}
\end{figure}

\begin{figure}
\begin{center}
\includegraphics[scale=0.5]{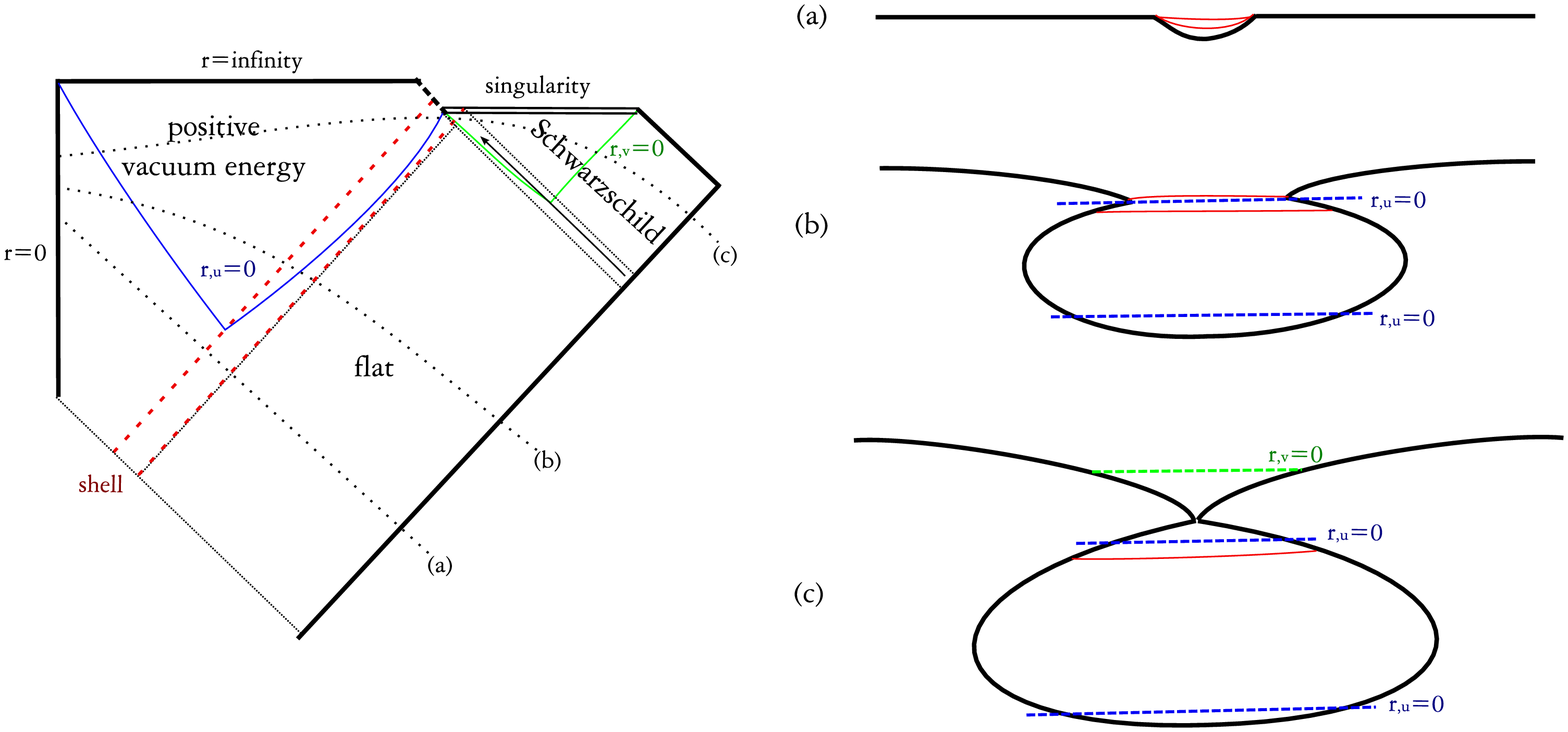}
\caption{\label{fig:sol3}An inflating shell solution (Type~3). After push a terminal matter, the evolution of the shell is ended via a black hole. (a), (b), and (c) are schematic diagrams of area for each space-like section.}
\end{center}
\end{figure}

Now they are equivalent to the following scheme:
\begin{eqnarray}
d_{,u} = h_{,v} &=& \frac{fg}{r^{2}} + \frac{\alpha^2}{4r^{2}} -\sqrt{N}WZ +\left(1-\frac{1}{\sqrt{N}}\right) \sqrt{N}W'Z', \nonumber \\
g_{,v} &=& 2dg - r\sqrt{N}Z^{2} +r\left(1-\frac{1}{\sqrt{N}}\right)\sqrt{N}Z'^{2}, \nonumber \\
g_{,u} = f_{,v} &=& -\frac{fg}{r} - \frac{\alpha^{2}}{4r} + 2\pi\alpha^2 r \frac{\Lambda}{S_{0}^{4}}S^{4}, \nonumber \\
f_{,u} &=& 2fh - r\sqrt{N}W^{2} + r\left(1-\frac{1}{\sqrt{N}}\right)\sqrt{N}W'^{2}, \nonumber \\
Z_{,u} = W_{,v} &=& - \frac{fZ}{r} - \frac{gW}{r} - 4\pi \alpha^{2} \frac{1}{\sqrt{N}} \frac{\Lambda}{S_{0}^{4}}S^{3}, \nonumber \\
Z'_{,u} = W'_{,v} &=& - \frac{fZ'}{r} - \frac{gW'}{r}.
\end{eqnarray}
\begin{eqnarray}
S, S'(u,v_{\mathrm{i}}) = \left\{ \begin{array}{ll}
0 & u < u_{\mathrm{shell}},\\
S_{0} G(u) & u_{\mathrm{shell}} \leq u < u_{\mathrm{shell}}+\Delta u,\\
S_{0} & u_{\mathrm{shell}}+\Delta u \leq u,
\end{array} \right.
\end{eqnarray}
since the equation for $\Phi'$ is linear and hence one can re-scale about factor $\beta$. Here, $S_{0} = \sqrt{4\pi}\Phi_{0}$ is the field amplitude and $\lambda = (4\pi)^{2}\Lambda / S_{0}^{4}$ is the vacuum energy of the inside of the shell.

\begin{figure}
\begin{center}
\includegraphics[scale=0.27]{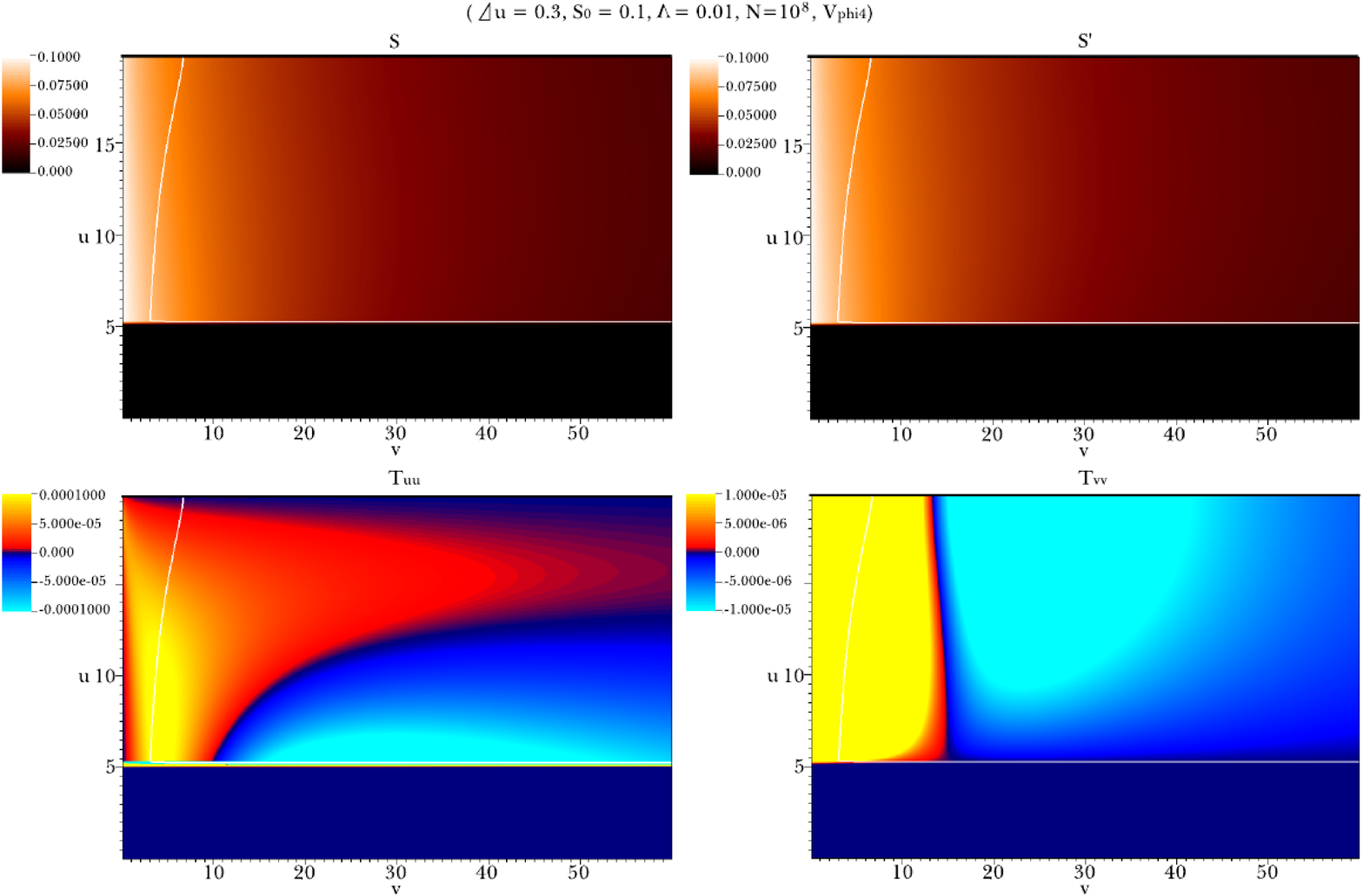}
\caption{\label{fig:sol3_field_and_energy}Function $S$, $S'$, $T_{uu}$, and $T_{vv}$ of $(\Delta u=0.3, S_{0}=0.1, \Lambda=0.01, N=10^{8})$ with $V_{\mathrm{phi4}}(S)$ potential. Yellow regions of $T_{uu}$ and $T_{vv}$ have field values more than $10^{-4}$ and $10^{-5}$; skyblue regions of $T_{uu}$ and $T_{vv}$ are less than $-10^{-4}$ and $-10^{-5}$; the sign of the energy-momentum tensor is changed around black region.}
\includegraphics[scale=0.27]{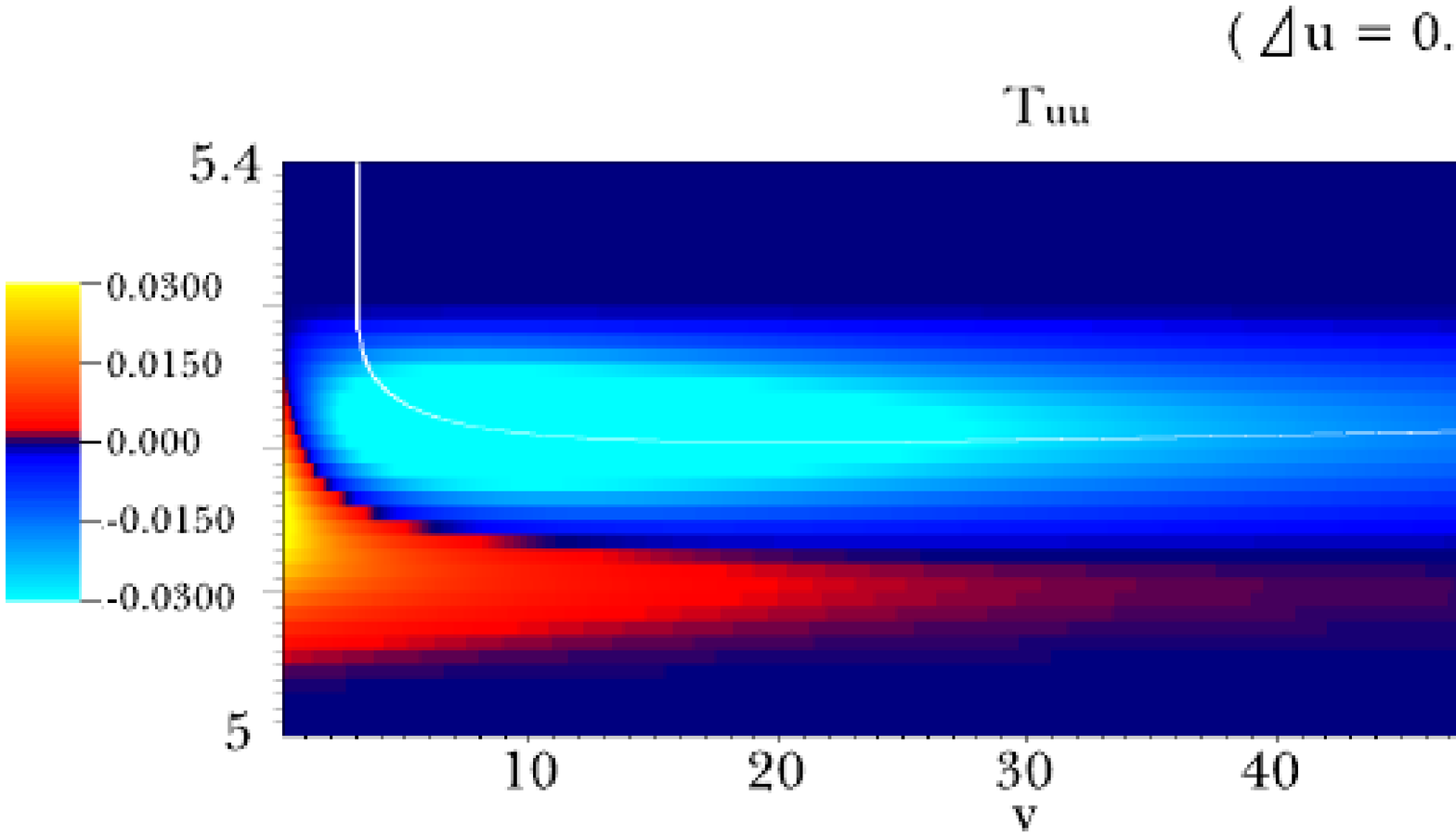}
\caption{\label{fig:sol3_shell}$T_{uu}$ and $T_{vv}$ around the shell.}
\end{center}
\end{figure}

Let us observe a naive expectation of this formulation. The field equation for $\Phi$ is effectively free via $1/\sqrt{N}$ term. Therefore, effectively, $\Phi \sim \Phi_{\mathrm{free}} + \mathcal{O}(1/\sqrt{N})$ is obtained. Contributions to the energy-momentum tensor is on the order of
\begin{eqnarray}
\sqrt{N} \left(\Phi_{\mathrm{free}} + \mathcal{O}\left(\frac{1}{\sqrt{N}} \right) \right)^{2} \sim \sqrt{N} \Phi_{\mathrm{free}}^{2} + \sqrt{N} \mathcal{O}\left(\frac{1}{\sqrt{N}}\right) + ...,
\end{eqnarray}
i.e. the leading term contributes to the energy-momentum tensor by order $\sqrt{N}$, where the second term contributes order $1$ which comes from the $V'(S)$ effect. However, the first contribution will be canceled by $N$ exotic matter shells, and hence their contributions become order $1$. Therefore, even if we assume $N \rightarrow \infty$ limit, we cannot naively assume $\Phi$ as a free field, since the second correction term is comparable to the free field contributions. If $\Phi$ is a free field, as long as the initial surface maintains the null energy condition, the null energy condition will hold in all regions. However, in our cases, \textit{even if we prepare the initial energy-momentum tensor to hold the null energy condition, the effect of $V'(S)$ is not negligible; as time goes on, $V'(S)$ term may affect to roll down the field, and hence eventually the null energy condition will be violated during inflation.}

\subsubsection{\label{sec:simulations}Simulations of $N$-shell bubbles}

\begin{figure}
\begin{center}
\includegraphics[scale=0.27]{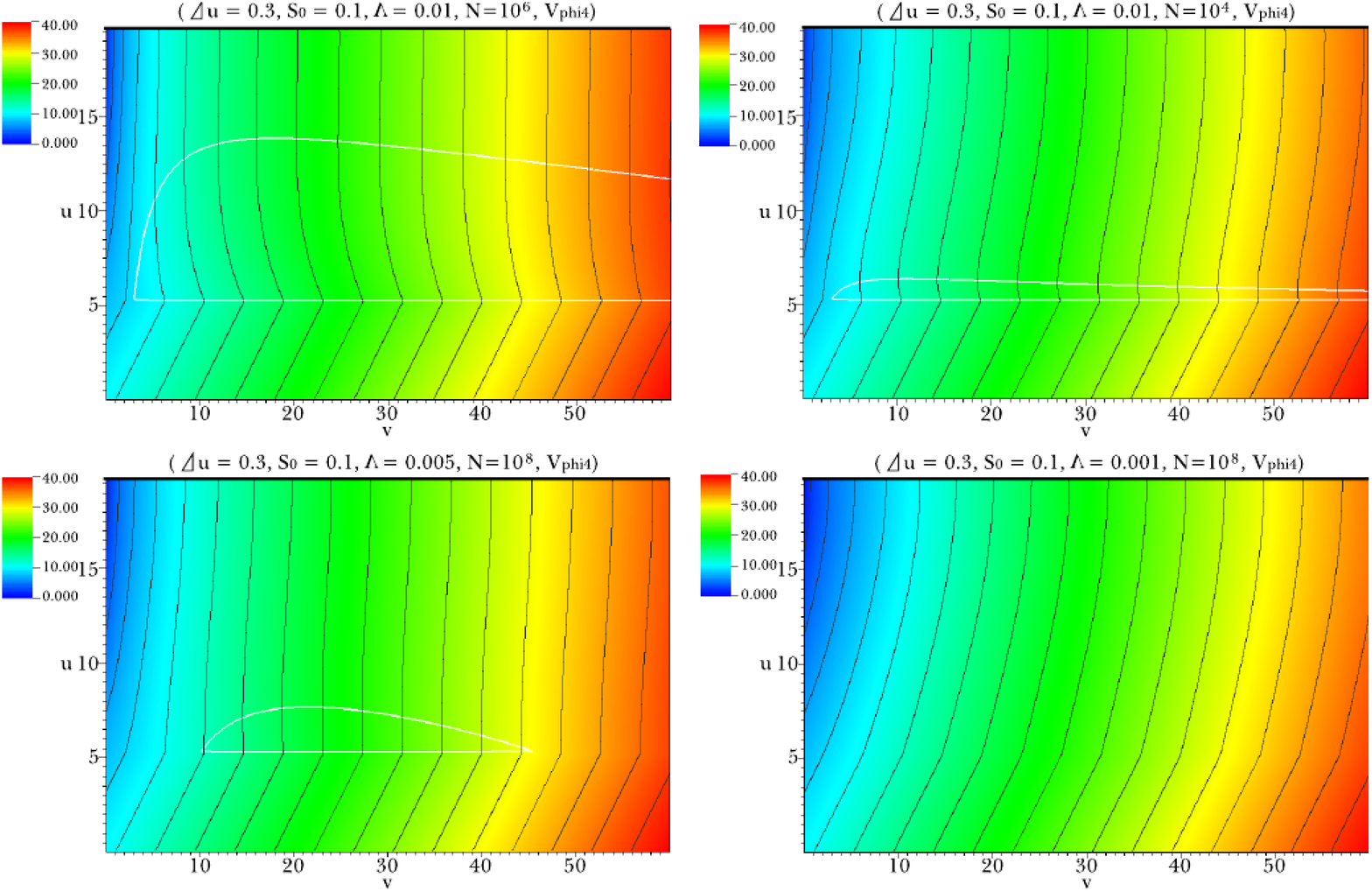}
\caption{\label{fig:transition_sol3_to_sol2}$r$ contours for each initial condition.}
\end{center}
\end{figure}

Here we observe simulations with the following conditions: $(\Delta u=0.3, S_{0}=0.1, \Lambda=0.01, N=10^{8})$ and $(\Delta u=0.3, S_{0}=0.1, \Lambda=0.05, N=10^{8})$ with $V_{\mathrm{phi4}}(S)$ potential (Figure~\ref{fig:sol3_sim}). In the lower diagram, the blue region (upper right corner) is beyond the calculation ability, since the radial function $r$ becomes exponentially large. We can interpret this as a space-like future infinity. Therefore, this is an important evidence that we induce inflation. We will call this class of solutions Type~3.

One can observe the $r_{,u}=0$ horizon: an ingoing observer sees increase of area. The inner horizon has two parts: one is almost parallel to the ingoing null direction, and the other is almost parallel to the outgoing null direction. The former corresponds to the cosmological horizon of the de Sitter space. The latter corresponds a region where area begins to increase for an ingoing observer: it is similar with the throat of a wormhole. Therefore, we can say that, during inflation, a wormhole is dynamically generated (schematically shown in Figure~\ref{fig:sol3}). One can observe that the former part is always time-like, while the latter part begins a space-like direction and finally becomes a time-like direction (in fact, the latter is almost null).

\begin{figure}
\begin{center}
\includegraphics[scale=0.5]{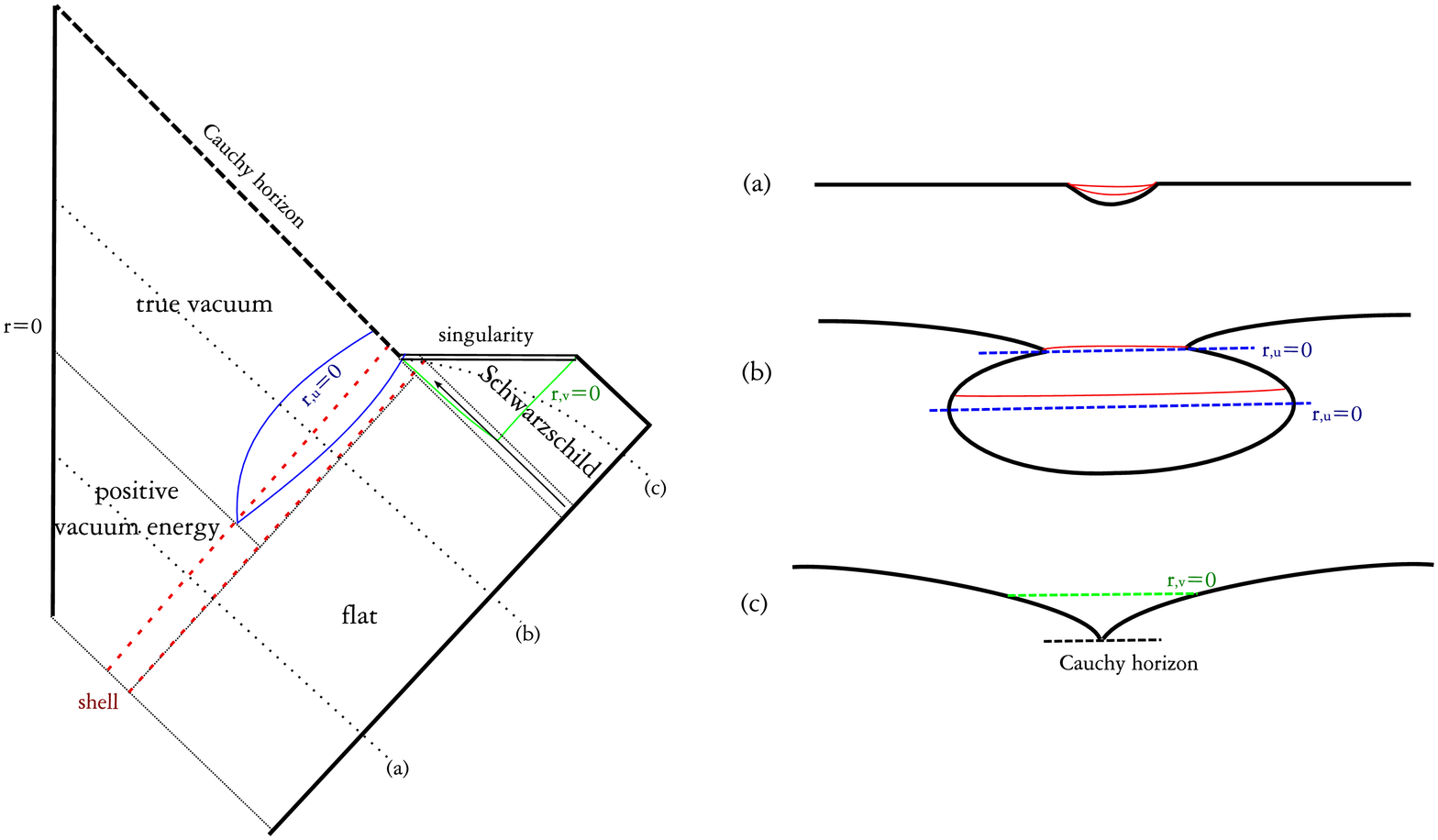}
\caption{\label{fig:sol4}A bursting shell solution (Type~4). After push a terminal matter, the evolution of the shell is ended via a black hole, and a Cauchy horizon is generated. (a), (b), and (c) are schematic diagrams of area for each section.}
\end{center}
\end{figure}

Note that, in Figure~\ref{fig:sol3}, we pushed a terminal matter to induce a black hole. We do not need that stage, but to make an end of the structure and discuss the information loss problem in the final section, we artificially inserted the process.

We observed that the wormhole throat can be generated around the mass shell. Hence, we do not need to assume tunneling from the outside to inside of a Schwarzschild wormhole. In fact, if the null energy condition is violated, this behavior may happen. However, its causal structure was not known, since the structure is related to the thick transition layer. We observed a dynamical generation of a wormhole around the transition layer, and observed the causal structure.

Also, we observed the field $S$ and $S'$, as well as the energy-momentum tensor $T_{uu}$ and $T_{vv}$ (Figure~\ref{fig:sol3_field_and_energy}). Diagrams of $S$ and $S'$ are almost similar, but $S$ rolls down more quickly than $S'$. One can see black bands from $T_{uu}$ and $T_{vv}$. Note that, if $T_{uu}$ or $T_{vv}$ is less than $0$, it implies the violation of the null energy condition. The violation of null energy condition seems to begin around the inside of the shell (Figure~\ref{fig:sol3_shell}). And the exotic matter becomes dominant as time goes on. Therefore, we highly suspect that our inflating shell solution requires a violation of the null energy condition.

\subsection{\label{sec:type4}Type~4: a bursting shell}

\subsubsection{\label{sec:transition2}Transition from Type~3 to Type~2: a bursting shell solution}

\begin{figure}
\begin{center}
\includegraphics[scale=0.27]{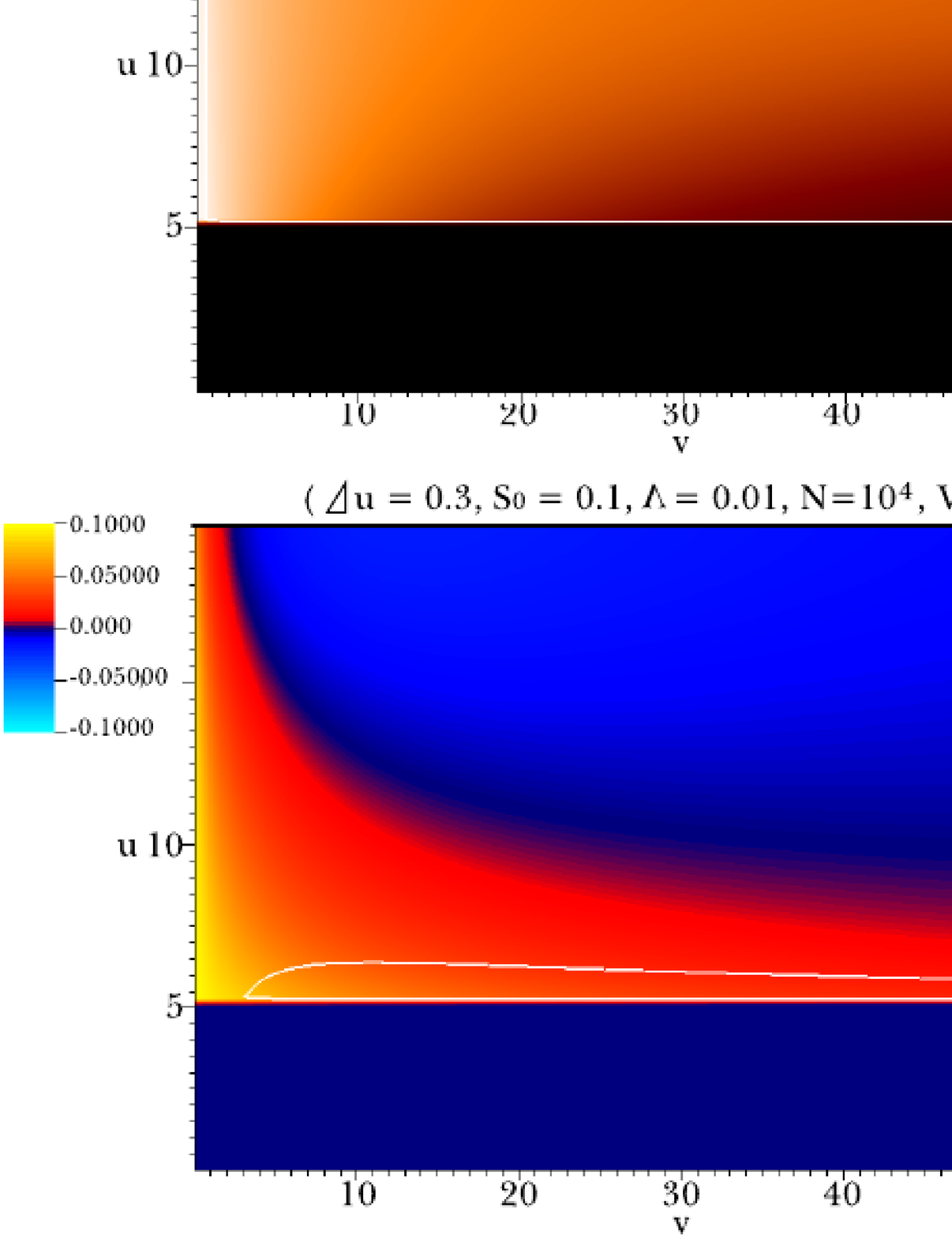}
\caption{\label{fig:sol3_to_sol4_S}Field $S$ for some conditions. As $N$ decreases, the inside field more quickly rolls down.}
\end{center}
\end{figure}

As we vary the simulation parameters, we could find another kind of solutions between Type~3 and Type~2.

First, we have changed vacuum energy of the inside $\Lambda$:
\begin{itemize}
\item $(\Delta u=0.3, S_{0}=0.1, \Lambda=0.05, N=10^{8})$,
\item $(\Delta u=0.3, S_{0}=0.1, \Lambda=0.01, N=10^{8})$,
\item $(\Delta u=0.3, S_{0}=0.1, \Lambda=0.005, N=10^{8})$,
\item $(\Delta u=0.3, S_{0}=0.1, \Lambda=0.001, N=10^{8})$.
\end{itemize}

Second, we have changed the number of shells $N$:
\begin{itemize}
\item $(\Delta u=0.3, S_{0}=0.1, \Lambda=0.01, N=10^{8})$,
\item $(\Delta u=0.3, S_{0}=0.1, \Lambda=0.01, N=10^{6})$,
\item $(\Delta u=0.3, S_{0}=0.1, \Lambda=0.01, N=10^{4})$.
\end{itemize}

Figure~\ref{fig:transition_sol3_to_sol2} shows the transition from Type~3 $(\Delta u=0.3, S_{0}=0.1, \Lambda=0.01, N=10^{8})$ to Type~2 $(\Delta u=0.3, S_{0}=0.1, \Lambda=0.001, N=10^{8})$. Between these two limits, we could find interesting structures: we will call these Type~4 (Figure~\ref{fig:sol4}).

The cases $(\Delta u=0.3, S_{0}=0.1, \Lambda=0.01, N=10^{6})$ and $(\Delta u=0.3, S_{0}=0.1, \Lambda=0.01, N=10^{4})$ show the behavior when $N$ decreases. Inflation is suppressed, but the location where inflation begins is invariant. However, in $(\Delta u=0.3, S_{0}=0.1, \Lambda=0.005, N=10^{8})$, we observe small $\Lambda$ limit, and the beginning of inflation is shifted.

If one compares stability of field $S$ (Figure~\ref{fig:sol3_to_sol4_S}), as $N$ decreases, inside field values become unstable and roll down to the true vacuum and roll around $S=0$: $(\Delta u=0.3, S_{0}=0.1, \Lambda=0.01, N=10^{4})$.

\begin{figure}
\begin{center}
\includegraphics[scale=0.27]{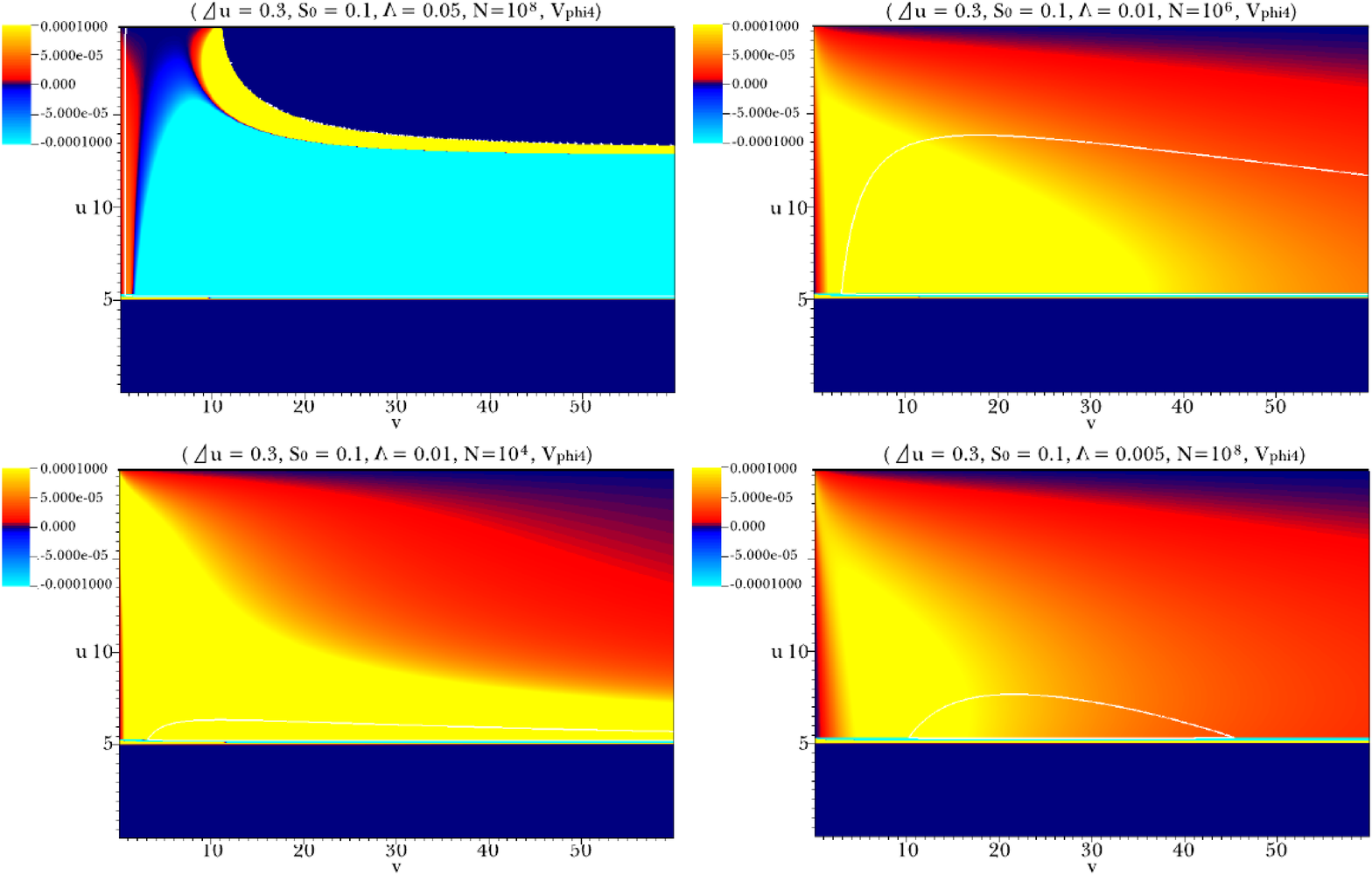}
\caption{\label{fig:sol3_to_sol4_Tuu}$T_{uu}$ plots for some conditions. Yellow regions are greater than $10^{-4}$, and skyblue regions are less than $-10^{-4}$.}
\end{center}
\end{figure}

\FIGURE[t]{\epsfig{file=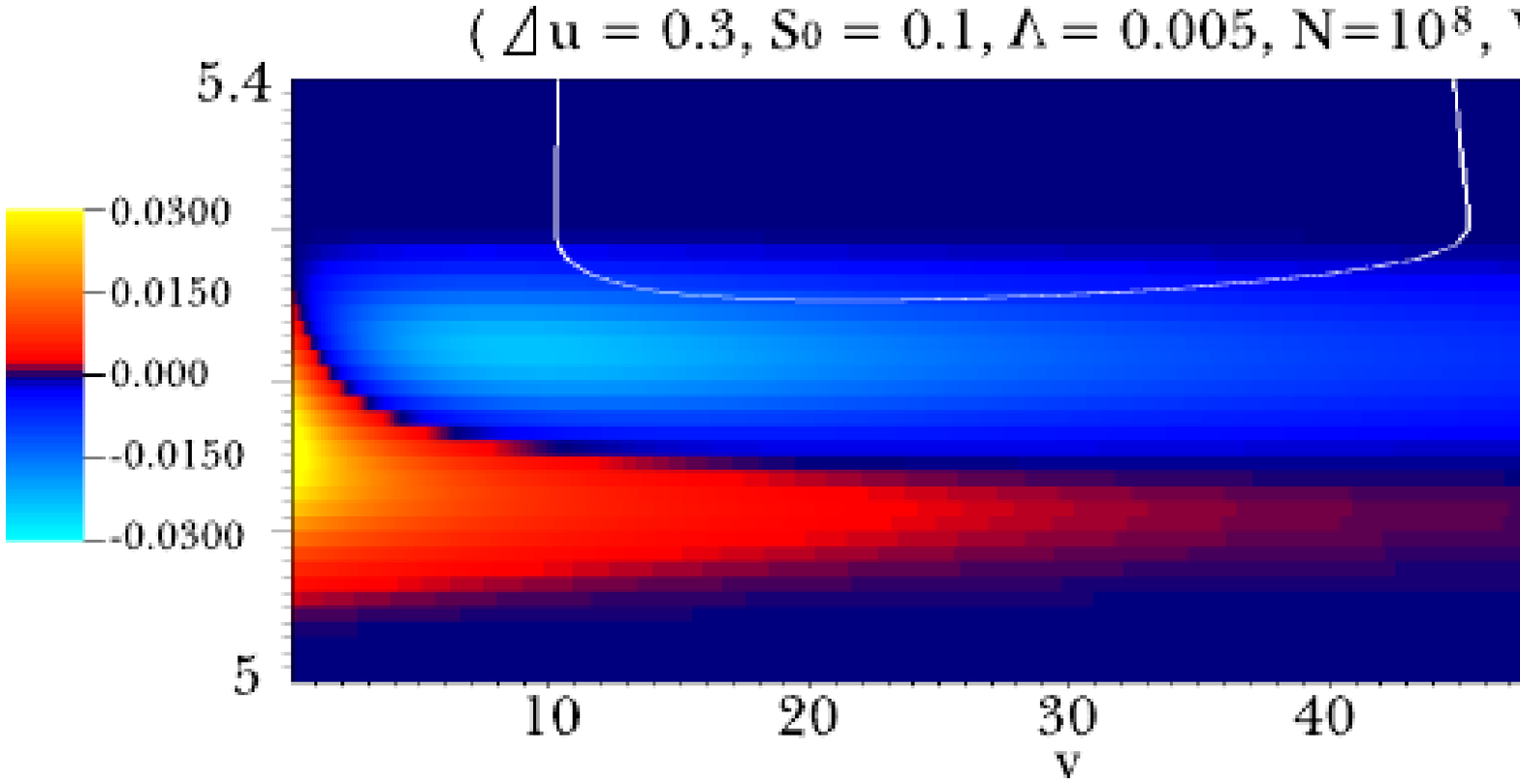,scale=0.27} \caption{$T_{uu}$ around the shell for $(\Delta u=0.3, S_{0}=0.1, \Lambda=0.005, N=10^{8})$.} \label{fig:sol4_shell}}

Figure~\ref{fig:sol3_to_sol4_Tuu} shows $T_{uu}$ components. Yellow regions are greater than $10^{-4}$ and skyblue regions are less than $-10^{-4}$. In the case of Type~3 solutions (e.g., $(\Delta u=0.3, S_{0}=0.1, \Lambda=0.05, N=10^{8})$ or $(\Delta u=0.3, S_{0}=0.1, \Lambda=0.01, N=10^{8})$), violating region of the null energy condition is quite wide. However, in Type~4 solutions (e.g., $(\Delta u=0.3, S_{0}=0.1, \Lambda=0.01, N=10^{6})$, $(\Delta u=0.3, S_{0}=0.1, \Lambda=0.01, N=10^{4})$, or $(\Delta u=0.3, S_{0}=0.1, \Lambda=0.005, N=10^{8})$), $T_{uu}$ is almost globally positive \textit{except} near the shell. Therefore, the violation of the null energy condition seems to be essential, but to see $r_{,u}=0$ horizons, the violation is needed just a narrow region around the shell (Figure~\ref{fig:sol4_shell}).

\begin{figure}
\begin{center}
\includegraphics[scale=0.27]{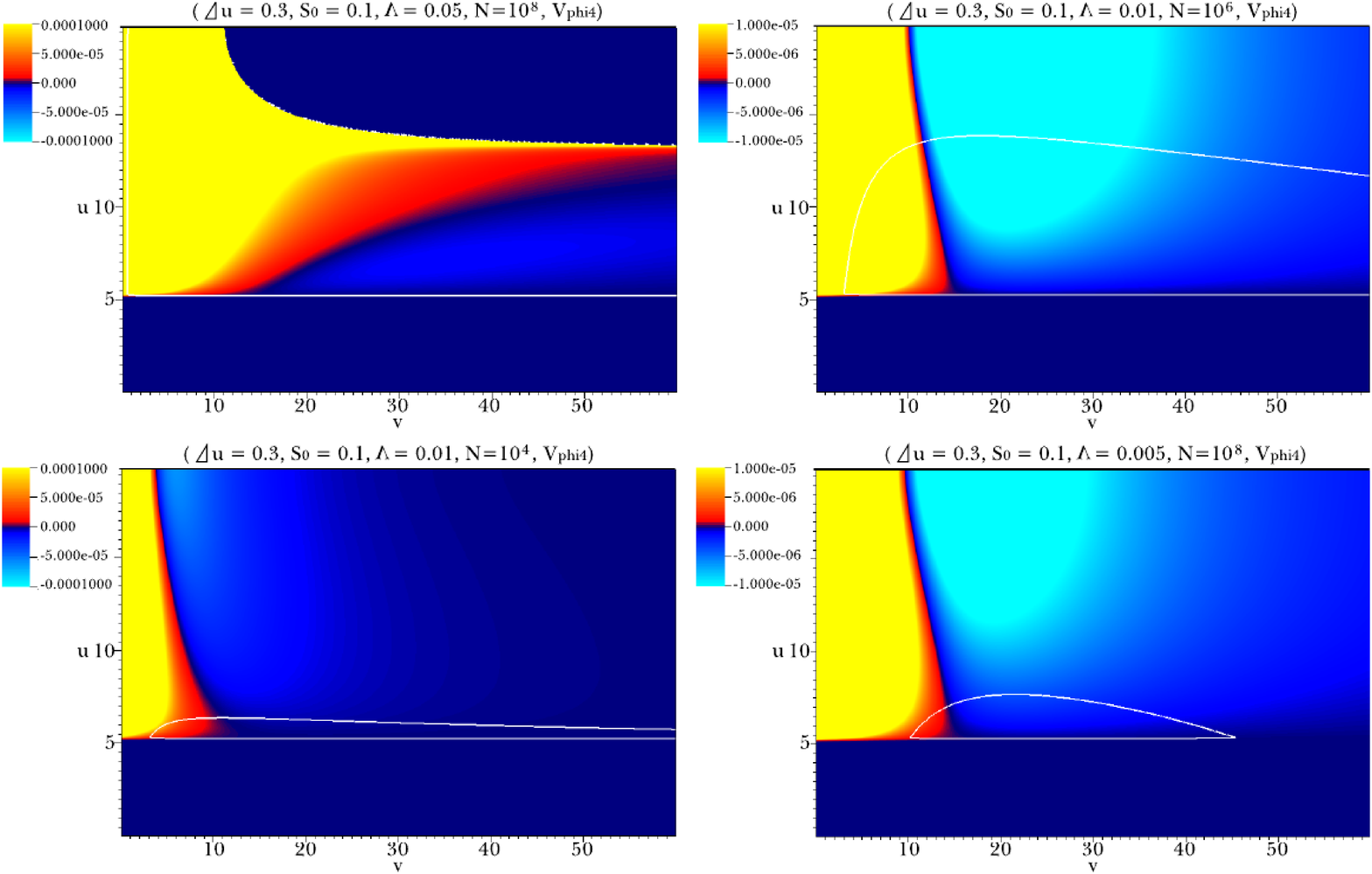}
\caption{\label{fig:sol3_to_sol4_Tvv}$T_{vv}$ plots for some conditions. Yellow regions are greater than the upper bound, and skyblue regions are less than the lower bound.}
\end{center}
\end{figure}

Figure~\ref{fig:sol3_to_sol4_Tvv} shows $T_{vv}$ components. The beginning of inflation seems not to be related to the sign of $T_{vv}$ and may not be important for inflation.

Finally, we remark on the behavior of $r_{,u}=0$ horizons. For the Type~3 case, the inner $r_{,u}=0$ horizon (the analogous horizon of the de Sitter space) is always time-like but for the Type~4 case, the inner $r_{,u}=0$ horizon bends from a time-like direction to a space-like direction. Also, the outer $r_{,u}=0$ horizon (the throat of a wormhole) bends from a space-like direction to a time-like direction (Figure~\ref{fig:sol4_shell}). These changes come from the properties of $r_{,uv}=f_{,v}$ around the horizon. Since $r_{,u}=0$, it becomes
\begin{eqnarray}
r_{,uv}|_{r_{,u}=0} = - \frac{\alpha^{2}}{4r} + 2 \pi \alpha^{2} r V(S).
\end{eqnarray}
If $r_{,uv} > 0$, then the outer $r_{,u}=0$ horizon is space-like and the inner $r_{,u}=0$ horizon is time-like; if $r_{,uv} < 0$, then the outer $r_{,u}=0$ horizon is time-like and the inner $r_{,u}=0$ horizon is space-like. The sign of $r_{,uv}$ is positive or negative if and only if
\begin{eqnarray}
8 \pi r^{2} V(S) - 1
\end{eqnarray}
is positive or negative. Therefore, if field values are sufficiently large and the vacuum energy is sufficiently large, the outer $r_{,u}=0$ horizon is space-like and the inner $r_{,u}=0$ horizon is time-like. However, as time goes on, the field values slowly roll down and then if the vacuum energy becomes sufficiently smaller than a critical value, the outer $r_{,u}=0$ horizon becomes time-like and the inner $r_{,u}=0$ horizon becomes space-like. In other words, if the field values quickly roll down to the true vacuum, the horizons tend to disappear and inflation ends. This is the basic physical difference between Type~3 and Type~4.

\begin{figure}
\begin{center}
\includegraphics[scale=0.5]{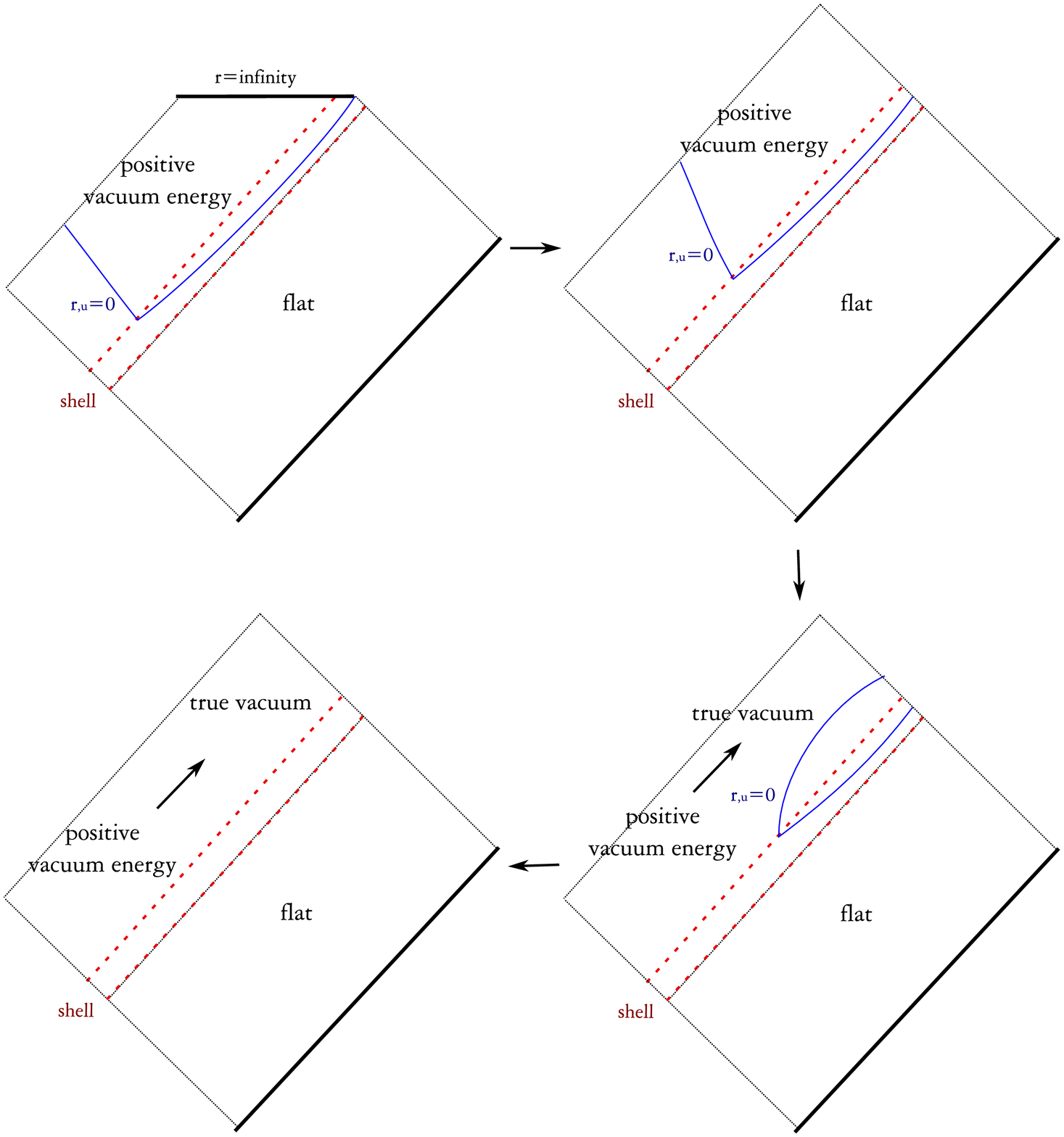}
\caption{\label{fig:sol3_transition}Transition from Type~3 to Type~2.}
\end{center}
\end{figure}

These changes are schematically shown in Figure~\ref{fig:sol3_transition}.

\section{\label{sec:discussion}Discussion}

We considered a false vacuum bubble inside of an almost flat background. Here, we assumed that the size of the bubble was smaller than the background horizon size, and of similar order as its own horizon size. A traditional approach on this problem is the thin shell approximation. In this paper, we extended our methods to beyond the thin shell approximation.

To summarize our discussion, firstly we will focus on the meaning of beyond the thin shell approximation. Secondly we will discuss about some speculations of a bubble universe and unitarity issues.

\subsection{\label{sec:beyondthinshell}Beyond the thin shell approximation}

We discussed previous results of the thin shell approximation. The thin shell approximation assumes the inside false vacuum region as the de Sitter space and the outside true vacuum region as the Schwarzschild space. According to the thin shell approximation and the null energy condition, if a shell is outside of a Schwarzschild black hole it cannot evolve to a bubble universe unless one introduces the Farhi-Guth-Guven tunneling; the only allowed solution is $\mathrm{dS}_{\mathrm{A}}-\mathrm{Sch}_{\mathrm{D}}$.

We extend the analysis beyond the thin shell approximation by using numerical methods. Two essential points of beyond thin shell approach are the added field dynamics and the thick transition layer. If a shell has sufficiently low energy, as expected from the thin shell approximation, it will collapse (Type~1). However, if the shell has sufficiently large energy, it tends to expand. However, via the field dynamics, the inside vacuum slowly rolls down to the true vacuum (Type~2). Moreover, if we add sufficient exotic matters to regularize curvatures around the shell, inflation may be possible without Farhi-Guth-Guven tunneling and a wormhole can be dynamically generated (Type~3). By tuning parameters, we could find transitions between Type~1 and Type~2, as well as between Type~2 and Type~3. Between Type~2 and Type~3, we also find another class of solutions (Type~4).

We can conclude our \textit{new} contributions on this issue. First, finding of Type~2 is a new result, thanks to going beyond the thin shell approximation. Second, we discussed the physical reason why inflation is difficult to induce without the violation of the null energy condition. Third, the exact causal structure of Type~3 is also a new feature. Some authors know that the violation of the null energy condition may allow an expanding inflating bubble, but the causal structure was not known exactly since the structure is related to the thick transition layer. We observed the causal structure of an inflating bubble as the null energy condition is violated, and it accompanies the dynamical generation of a wormhole. Fourth, finding of Type~4 is also a new result. Here, we observed the properties of $r_{,u}=0$ horizons and the energy-momentum tensor; properties of $r_{,u}=0$ horizons are related to the potential $V$, and to induce inflation, $T_{uu}<0$ seems to be the sufficient condition. Fifth, we could observe a continuous change of each types as one tunes initial parameters. In fact, the important parameters are just on shells: $\Delta u$, $\Phi_{0}$, $\Lambda$, and effects of the other parameters may not be important.

\subsection{\label{sec:generation}Generation of a bubble universe: discussion on the information loss problem}

If the false vacuum bubble is greater than the background horizon size, even though the bubble is separated from the background, it is not so meaningful since an observer of a scattering experiment is always inside of the bubble. However, in this paper, we discussed a bubble which is sufficiently smaller than the background horizon size. Therefore, in principle, it can be discussed in the context of scattering experiments, and thus the generation of a bubble universe should be discussed in the context of the information loss problem \cite{inforpara}.

The Type~3 and Type~4 solutions induce a separation between the outside true vacuum region and the inside region. The inside will have a second asymptotic region; Type~3 case is quite clear in this point. Then, information loss is inevitable unless duplication of information happens.\footnote{Black hole complementarity may be related to this issue. For further discussions, see \cite{Alberghi:1999kd}\cite{Yeom:2008qw}.} If the background is an anti de Sitter space, one may notice the AdS/CFT correspondence that implies unitarity \cite{Maldacena:1997re}. In this context, some authors said that Farhi-Guth-Guven tunneling should be excluded \cite{Freivogel:2005qh}. In the same sense, we can say that \textit{some assumptions of our setup is inconsistent with unitarity and AdS/CFT.} What are the exotic assumptions of our setup? Firstly, we assumed the existence of exotic matter fields. Secondly, we assumed some special initial conditions of scalar fields and exotic fields. Thirdly, a large number of $N$ shells was used to maintain field values.

We can comment on the second assumption. If a bubble is generated in an almost flat background, it can be described by a field combination for the inside false vacuum and the outside true vacuum. Here, if the background is a de Sitter space or an anti de Sitter space, the tunneling is not excluded via the violation of the energy conservation \cite{Lee:1987qc}. We found that initial conditions for each types are not quite different; each types transits continuously (Figure~\ref{fig:sol1_transition} and \ref{fig:sol3_transition}). Therefore, if certain initial conditions are excluded by an unknown reason, the other conditions should be excluded, too; if a combination for Type~2 is possible in principle, then it is difficult to find a reason why a similar initial condition that generates Type~3 should be excluded.

We can comment on the large number $N$. To see the $r_{,u}=0$ horizon, we do not need such large $N$. Also, there are some discussions that string theory seems not exclude a large number of fields \cite{Yeom:2009zp}.

The most strange assumption is the existence of exotic matter fields. Now, we can say that if one assumes the existence of the exotic matter fields, the creation of a bubble universe, dynamical generation of a wormhole and the violation of unitarity become possible. We show a clear contradiction between the existence of a certain combination of exotic matter fields and unitarity or AdS/CFT.

\DOUBLEFIGURE[t]{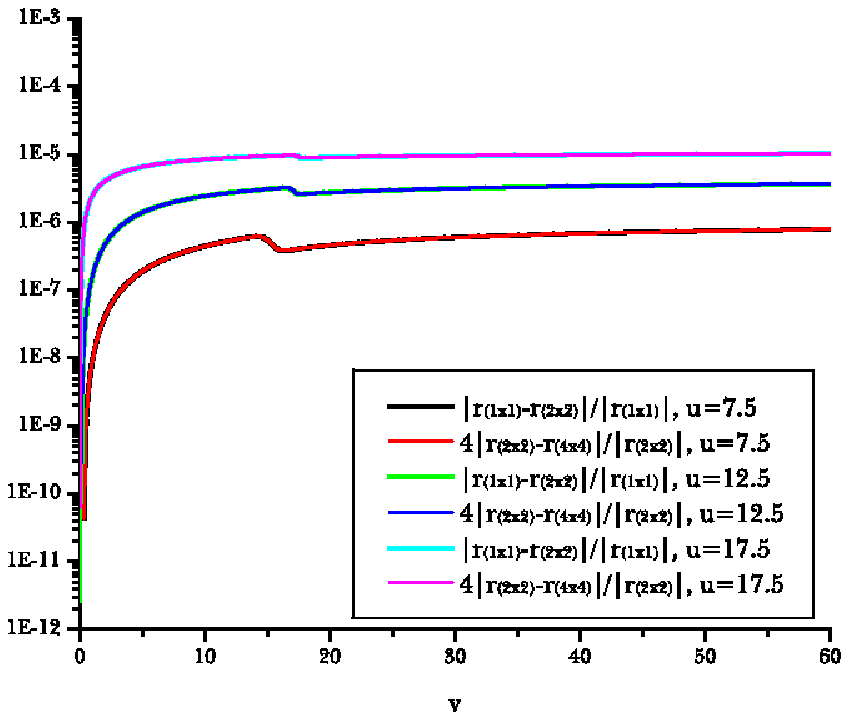,scale=0.75}{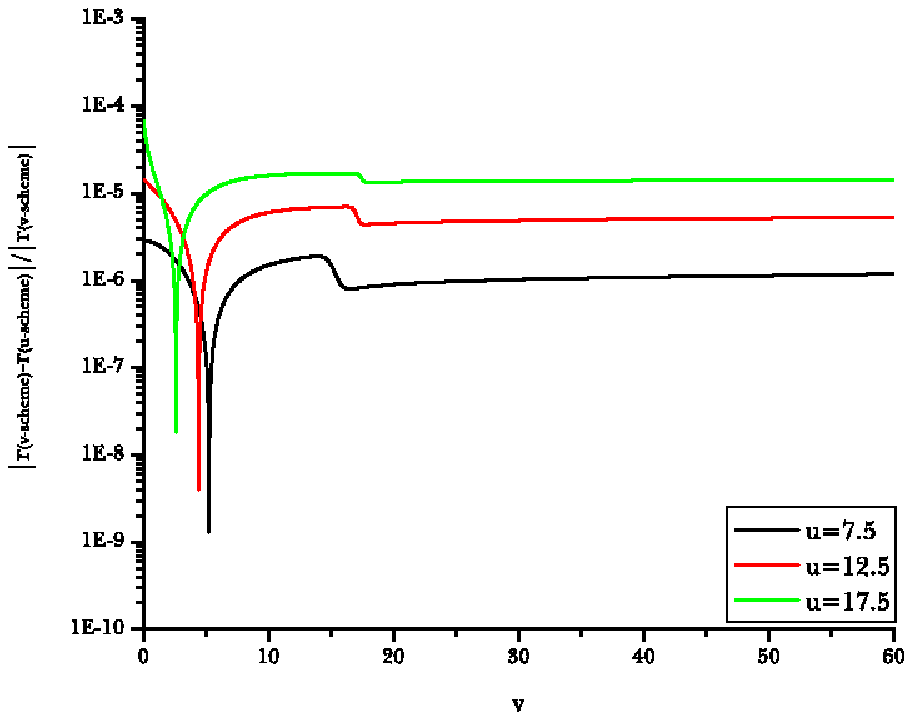,scale=0.75} {\label{fig:convergence_R_type1}Convergence tests of $r$ for the condition $(\Delta u=0.1, S_{0}=0.1, \Lambda=0.001)$ with potential $V_{\mathrm{poly}}(S)$. This shows the second order convergence.}{\label{fig:consistency_R_type1}Consistency tests of $r$ by comparing $r_{v-\mathrm{scheme}}$ and $r_{u-\mathrm{scheme}}$.}
\DOUBLEFIGURE[t]{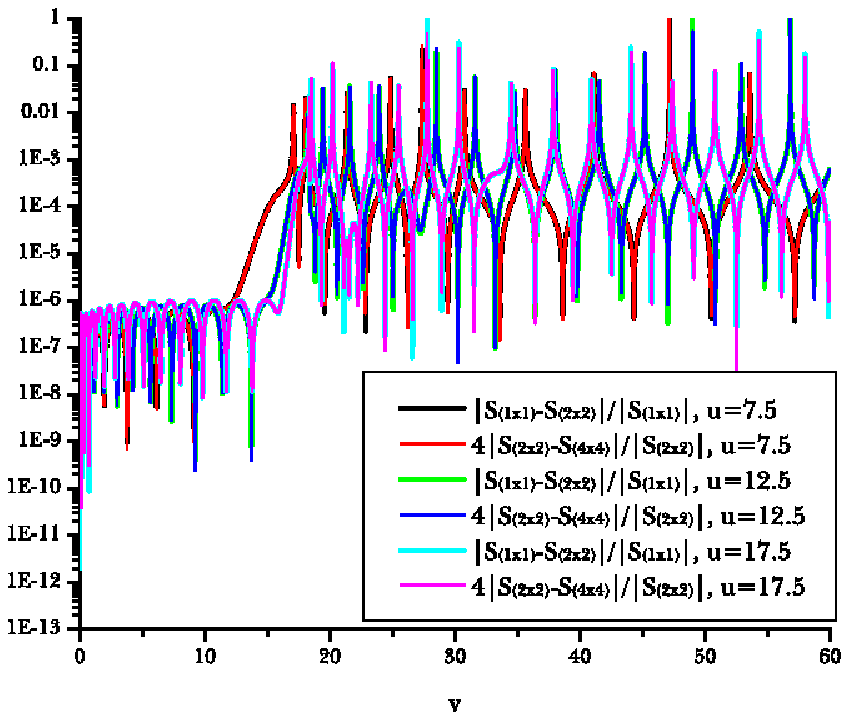,scale=0.75}{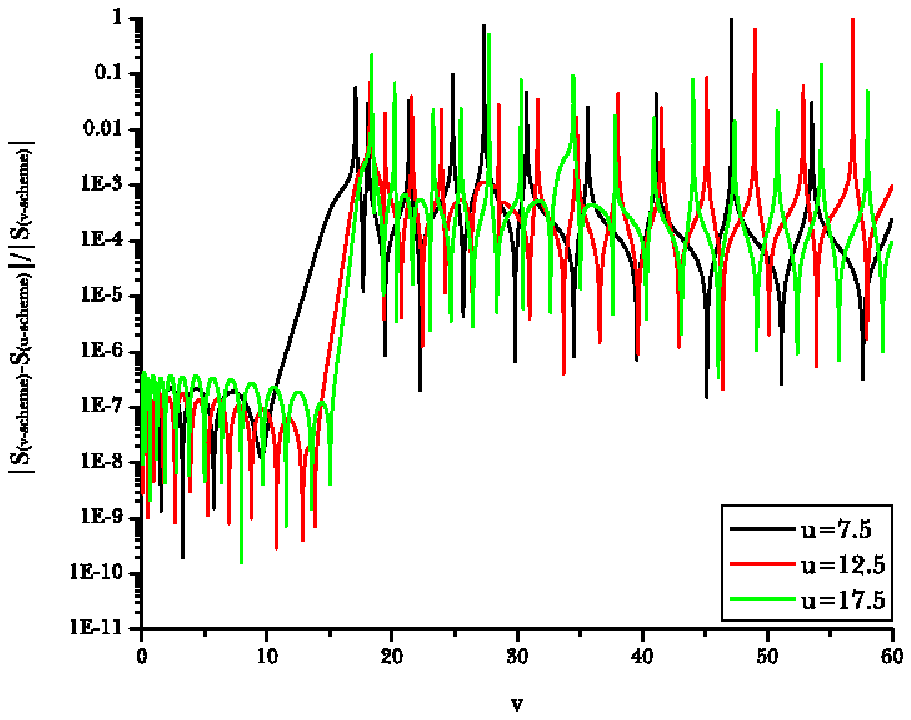,scale=0.75} {\label{fig:convergence_S_type1}Convergence tests of $S$ for the condition $(\Delta u=0.1, S_{0}=0.1, \Lambda=0.001)$ with potential $V_{\mathrm{poly}}(S)$. This shows the second order convergence.}{\label{fig:consistency_S_type1}Consistency tests of $S$ by comparing $S_{v-\mathrm{scheme}}$ and $S_{u-\mathrm{scheme}}$.}

Then, the next question is that, does the nature exclude such combination of exotic fields? One of physically reasonable processes of an exotic matter or the violation of energy conditions come from Hawking radiation \cite{Hawkingradiation}\cite{Birrell:1982ix}. For a black hole case, Hawking radiation of negative energy tends to inside of the black hole; however, here we need out-going negative energy flux. Therefore, it is unclear whether Hawking radiation may be helpful on this issue. However, if we can control the negative energy flux of Hawking radiation to the out-going direction, it will definitely make our first assumption reasonable. These problems should be discussed later.

\acknowledgments{The authors would like to thank Ewan Stewart, Gungwon Kang, Heeseung Zoe, and Sungwook Hong for discussions and encouragements.
DY and DH were supported by BK21 and the Korea Research Foundation Grant funded by the Korean government (MOEHRD; KRF-313-2007-C00164, KRF-341-2007-C00010). JH was supported in part by the JSPS Postdoctoral Fellowship For Foreign Researchers and the Grant-in-Aid for Scientific Research Fund of the JSPS (19-07795).}

\appendix

\section{\label{sec:convergence}Convergence and consistency tests}

\DOUBLEFIGURE[t]{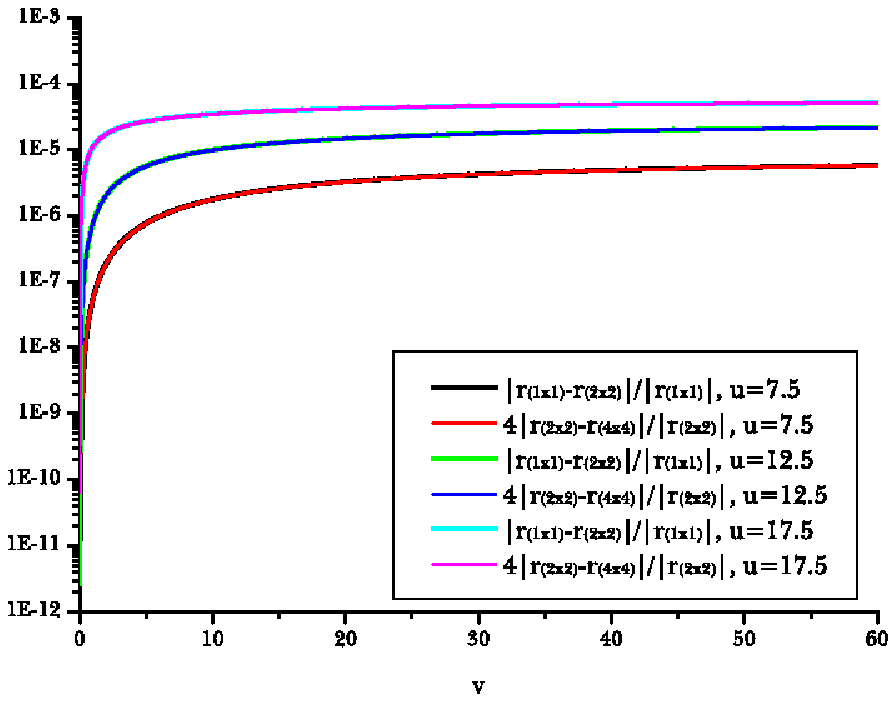,scale=0.75}{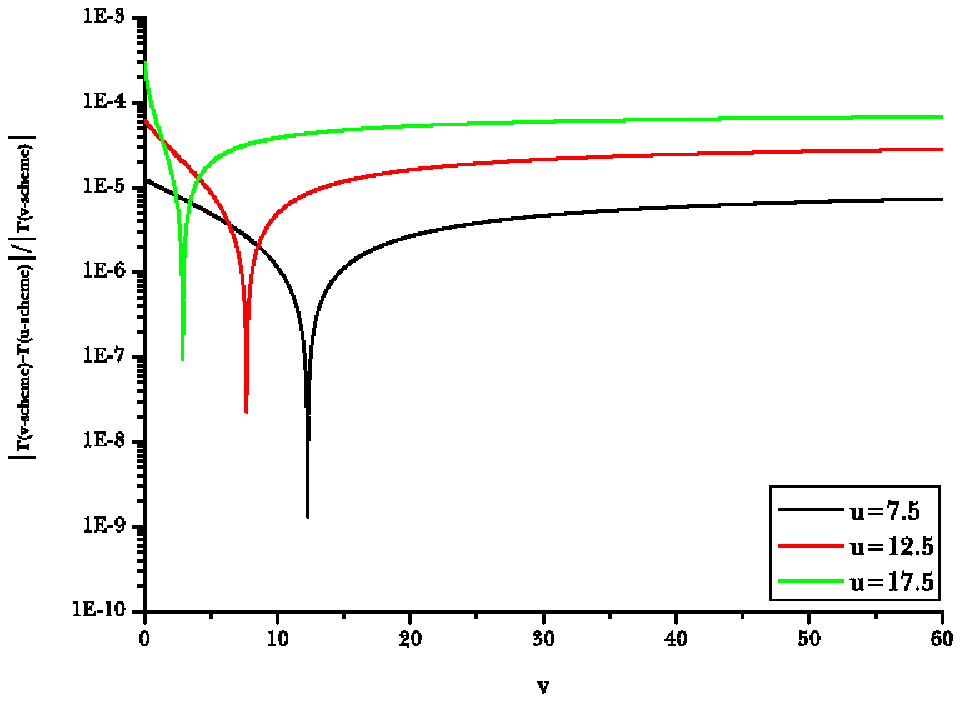,scale=0.75} {\label{fig:convergence_R_type2}Convergence tests of $r$ for the condition $(\Delta u=0.1, S_{0}=0.3, \Lambda=0.001)$ with potential $V_{\mathrm{poly}}(S)$. This shows the second order convergence.}{\label{fig:consistency_R_type2}Consistency tests of $r$ by comparing $r_{v-\mathrm{scheme}}$ and $r_{u-\mathrm{scheme}}$.}
\DOUBLEFIGURE[t]{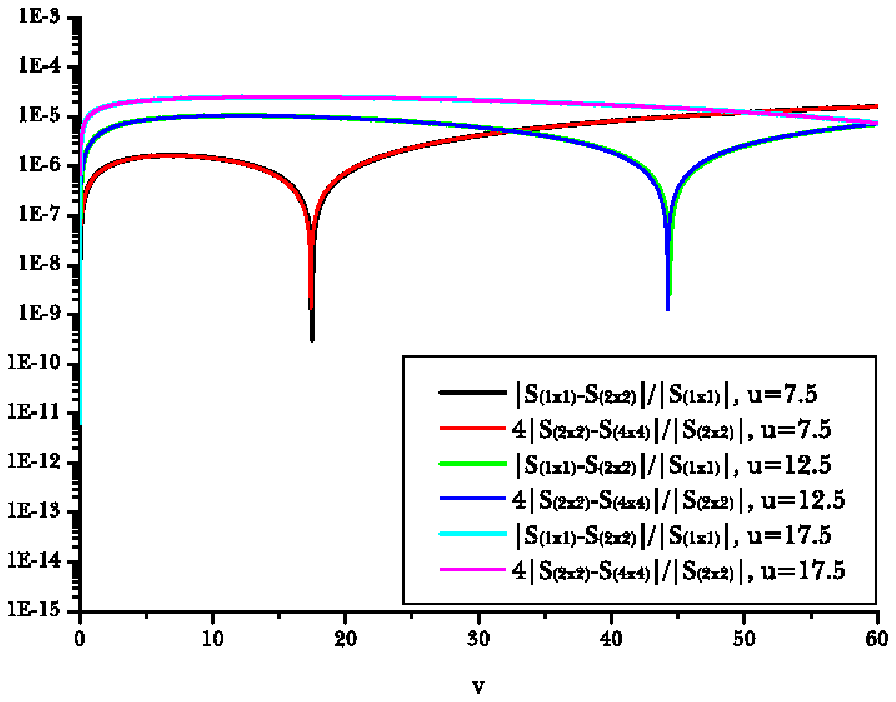,scale=0.75}{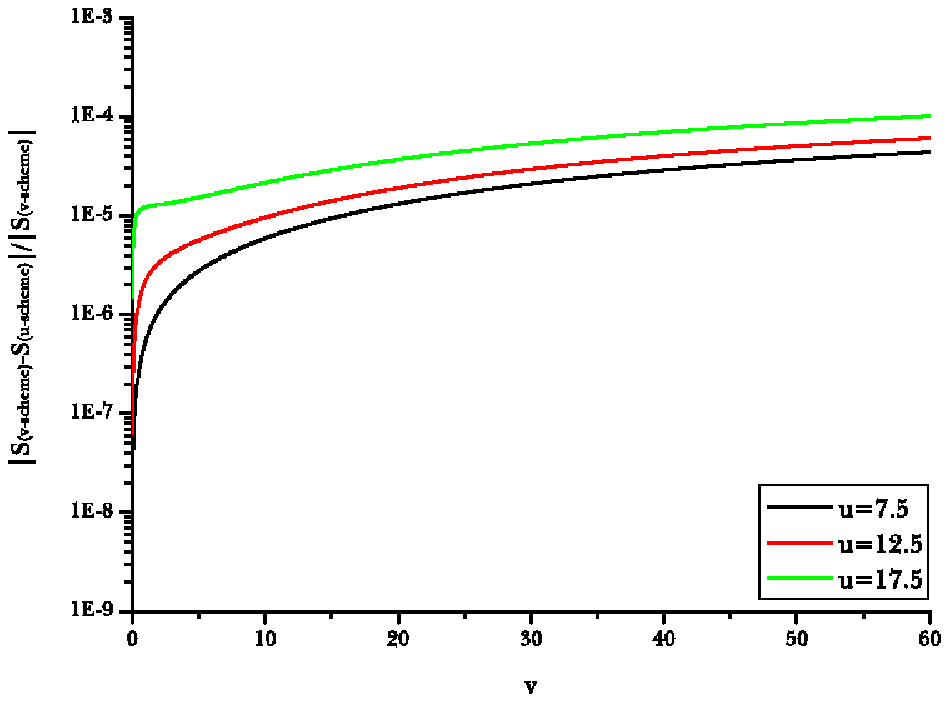,scale=0.75} {\label{fig:convergence_S_type2}Convergence tests of $S$ for the condition $(\Delta u=0.1, S_{0}=0.3, \Lambda=0.001)$ with potential $V_{\mathrm{poly}}(S)$. This shows the second order convergence.}{\label{fig:consistency_S_type2}Consistency tests of $S$ by comparing $S_{v-\mathrm{scheme}}$ and $S_{u-\mathrm{scheme}}$.}

In this appendix, we discuss convergence and consistency tests for our simulations of each type. Also, we briefly comment on the instability for fast-rolling fields. Here, we check the consistency by comparing the $v$-scheme and the $u$-scheme for certain $u=\mathrm{constant}$ surfaces. We check the convergence by comparing $1 \times 1$, $2 \times 2$ finer, and $4 \times 4$ finer simulations for certain $u=\mathrm{constant}$ surfaces. We observed $u=7.5, 12.5, 17.5$ slices.

We compared two independent evolutions by using equations for $r_{,uu}$ and $r_{,vv}$. Then the equation for $r_{,uv}$ remains a constraint equation. Therefore, we checked the following quantity to check the constraint equation:
\begin{eqnarray}
\frac{|-r_{,u}r_{,v}/r - \alpha^{2}/4r + 2\pi\alpha^2 r V(S) - r_{,uv}|}{|-r_{,u}r_{,v}/r| + |- \alpha^{2}/4r| + |2\pi\alpha^2 r V(S)| + |- r_{,uv}|}.
\end{eqnarray}
We checked the constraint equation for $v=10, 20, 40$ slices.

Note that simulations in this paper are based on a numerical code of \cite{HHSY}. And these results could be reproduced consistently from an independent code of \cite{Hansen:2005am}.

\DOUBLEFIGURE[t]{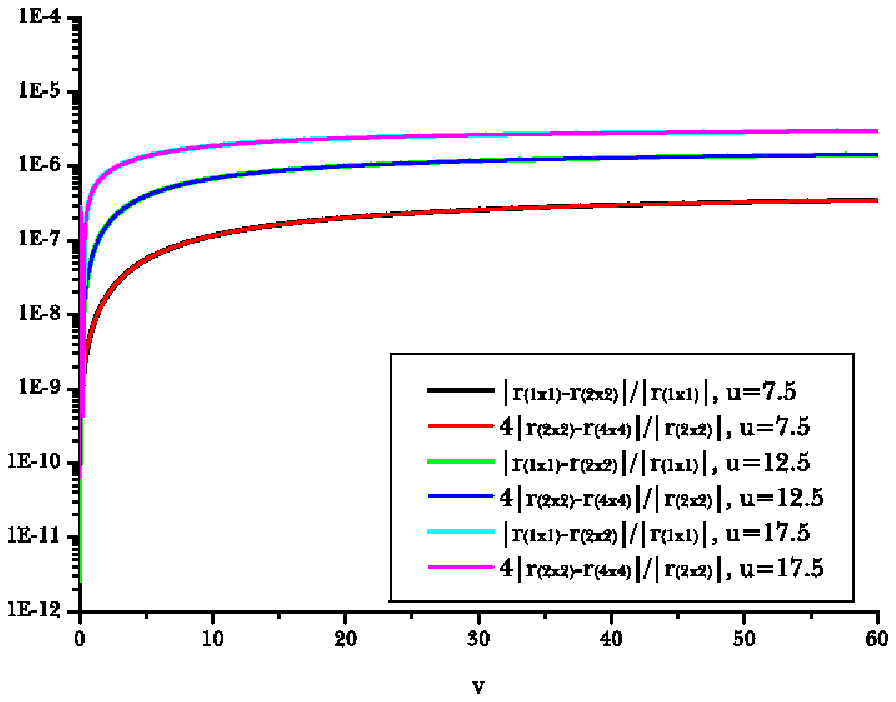,scale=0.75}{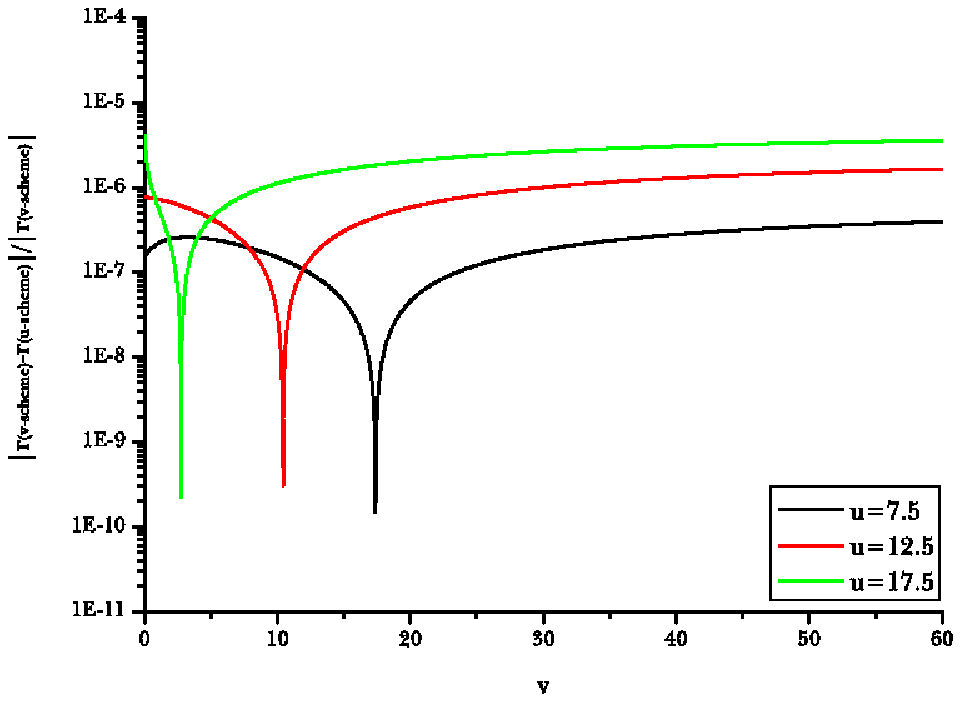,scale=0.75} {\label{fig:convergence_R_type3}Convergence tests of $r$ for the condition $(\Delta u=0.3, S_{0}=0.1, \Lambda=0.01, N=10^{8})$ with potential $V_{\mathrm{phi4}}(S)$. This shows the second order convergence.}{\label{fig:consistency_R_type3}Consistency tests of $r$ by comparing $r_{v-\mathrm{scheme}}$ and $r_{u-\mathrm{scheme}}$.}
\DOUBLEFIGURE[t]{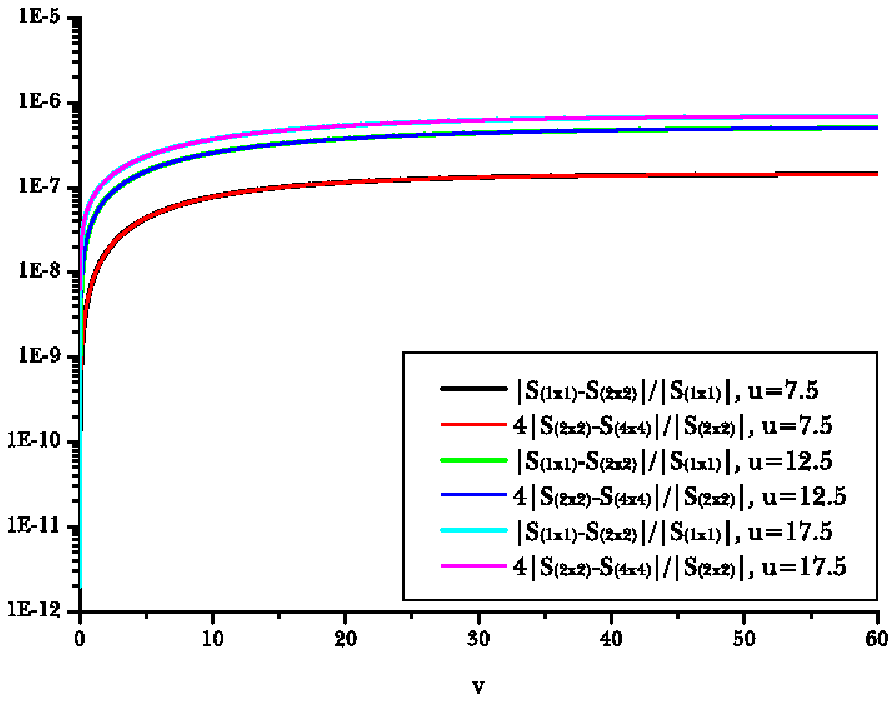,scale=0.75}{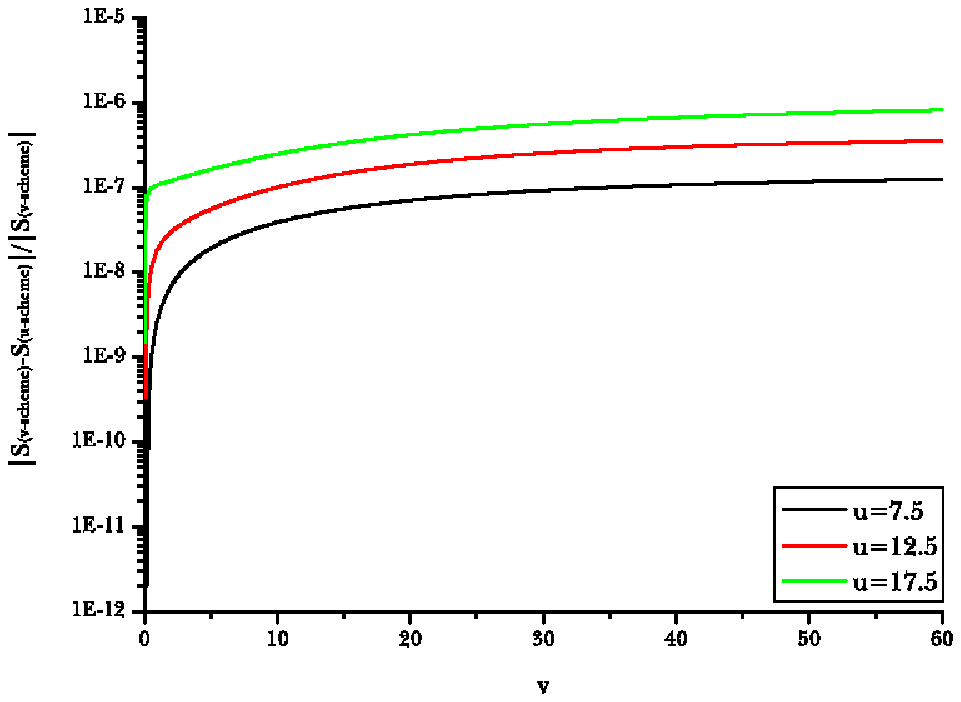,scale=0.75} {\label{fig:convergence_S_type3}Convergence tests of $S$ for the condition $(\Delta u=0.3, S_{0}=0.1, \Lambda=0.01, N=10^{8})$ with potential $V_{\mathrm{phi4}}(S)$. This shows the second order convergence.}{\label{fig:consistency_S_type3}Consistency tests of $S$ by comparing $S_{v-\mathrm{scheme}}$ and $S_{u-\mathrm{scheme}}$.}

\DOUBLEFIGURE[t]{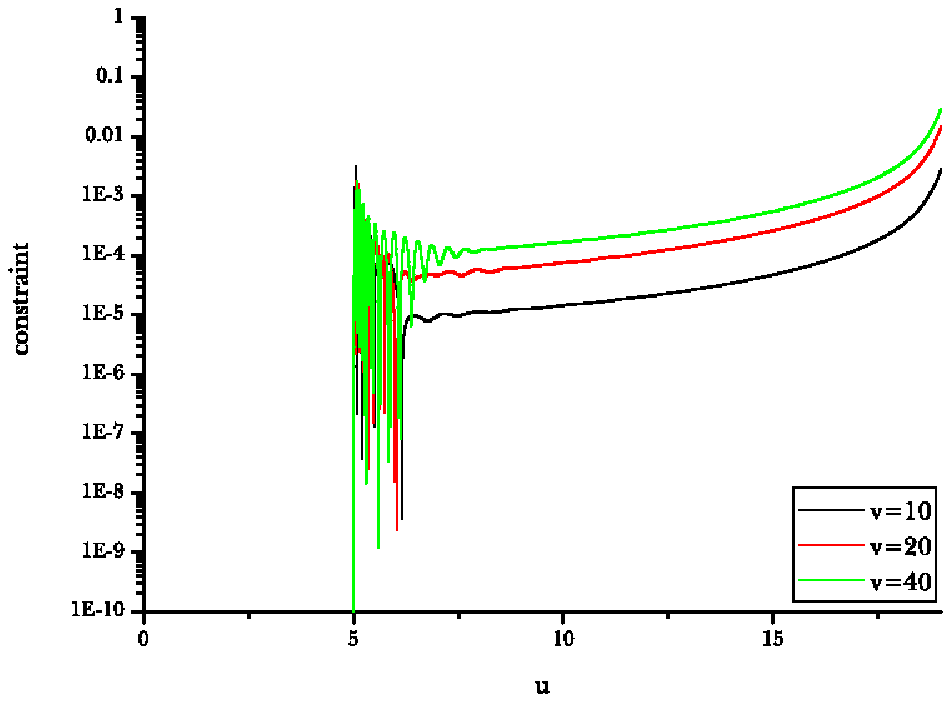,scale=0.75}{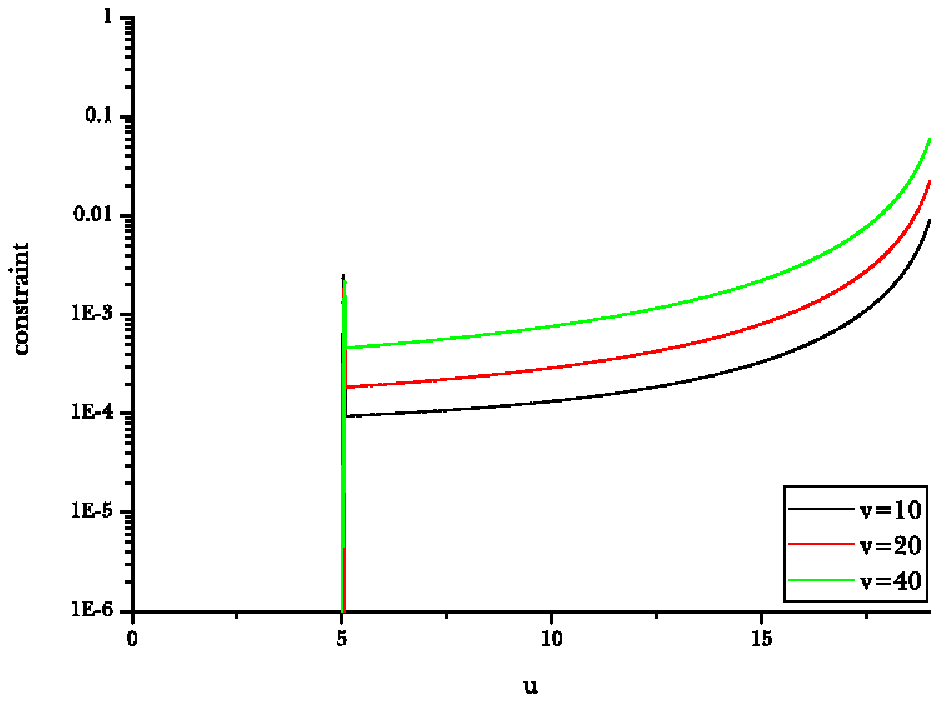,scale=0.75} {\label{fig:constraint_type1}The constraint equation for the condition $(\Delta u=0.1, S_{0}=0.1, \Lambda=0.001)$ with potential $V_{\mathrm{poly}}(S)$.}{\label{fig:constraint_type2}The constraint equation for the condition $(\Delta u=0.1, S_{0}=0.3, \Lambda=0.001)$ with potential $V_{\mathrm{poly}}(S)$.}
\FIGURE{\epsfig{file=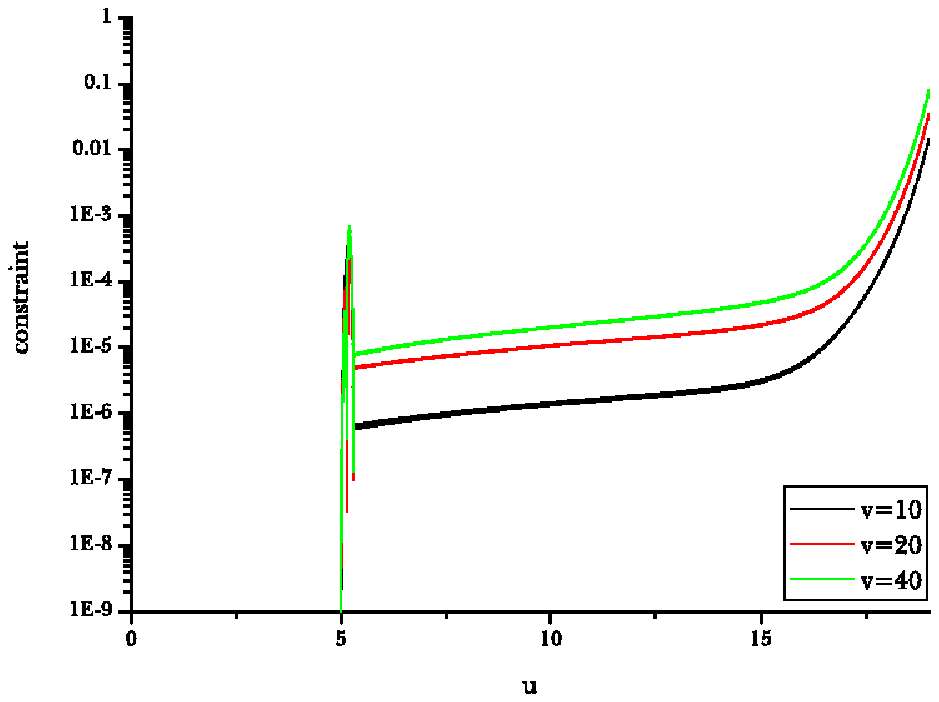,scale=0.75} \caption{The constraint equation for the condition $(\Delta u=0.3, S_{0}=0.1, \Lambda=0.01, N=10^{8})$ with potential $V_{\mathrm{phi4}}(S)$.} \label{fig:constraint_type3}}

\subsection{\label{sec:app_type1}Type~1}

We checked consistency and convergence for the condition $(\Delta u=0.1, S_{0}=0.1, \Lambda=0.001)$ with potential $V_{\mathrm{poly}}(S)$. Figure~\ref{fig:convergence_R_type1} and Figure~\ref{fig:consistency_R_type1} show that errors for $r$ is less than $10^{-2}\%$. Figure~\ref{fig:convergence_S_type1} and Figure~\ref{fig:consistency_S_type1} show that errors for $S$ is almost less than $10^{-2}\%$, but there are some peaks which are order of few percents.

This phenomenon comes from the fast-rolling of the field around the true vacuum. Near the true vacuum, as the field rolls around $0$ point, a small deviation from $0$ may look like a big error since we need to divide by $0$. This is the instability of fast-rolling fields. If its back-reaction to $r$ is sufficiently small, the instability is negligible. Type~1 holds this case. However, if there exists inflation via fast-rolling fields, one cannot ignore its back-reaction, and to maintain sufficiently small error for $S$, we need more and more finer simulations. This is a practical reason why we consider slow-rolling inflation only.

Figure~\ref{fig:convergence_R_type1} and Figure~\ref{fig:convergence_S_type1} show $|(1 \times 1) - (2 \times 2)| / |(1 \times 1)|$ and $4 |(2 \times 2) - (4 \times 4)| / |(2 \times 2)|$ for functions $r$ and $S$, where $(n \times n)$ means an $n \times n$ finer simulation. In these figures, for each $u$ slice, two curves are quite close and seem to be degenerated. These show that our simulations converge on the second order.

Figure~\ref{fig:constraint_type1} shows the constraint equation for some slices. The violation of the constraint equation is less than $1\%$ for almost all integrated domains ($0 \lesssim u \lesssim 18$).

\subsection{\label{sec:app_type2}Type~2}

We checked consistency and convergence for the condition $(\Delta u=0.1, S_{0}=0.3, \Lambda=0.001)$ with potential $V_{\mathrm{poly}}(S)$. Figure~\ref{fig:convergence_R_type2} and Figure~\ref{fig:consistency_R_type2} show that errors for $r$ is less than $10^{-2}\%$. Figure~\ref{fig:convergence_S_type2} and Figure~\ref{fig:consistency_S_type2} show that errors for $S$ is less than $10^{-2}\%$. Also, Type~2 solution converges to second order as well. In this case, there was no instability of fields, since the field values of the inside false vacuum slowly moves.

Figure~\ref{fig:constraint_type2} shows the constraint equation for some slices. The violation of the constraint equation is less than $1\%$ for almost all integrated domains ($0 \lesssim u \lesssim 18$).

\subsection{\label{sec:app_type34}Type~3 and Type~4}

We checked consistency and convergence for the condition $(\Delta u=0.3, S_{0}=0.1, \Lambda=0.01, N=10^{8})$ with potential $V_{\mathrm{phi4}}(S)$. Figure~\ref{fig:convergence_R_type3} and Figure~\ref{fig:consistency_R_type3} show that errors for $r$ is less than $10^{-3}\%$. Figure~\ref{fig:convergence_S_type3} and Figure~\ref{fig:consistency_S_type3} show that errors for $S$ is less than $10^{-4}\%$. Also, Type~3 solution converges to second order as well.

Finally, Figure~\ref{fig:constraint_type3} shows the constraint equation for some slices. The violation of the constraint equation is less than $0.1\%$ for almost all integrated domains ($0 \lesssim u \lesssim 18$).


\begin{thebibliography}{999}

%\cite{Guth:1980zm}
\bibitem{Guth:1980zm}
  A.~H.~Guth, {\it The inflationary universe: a possible solution to the horizon and flatness problems,} \prd{23}{1981}{347}.

%\cite{Linde:1981mu}
\bibitem{Linde:1981mu}
  A.~D.~Linde, {\it A new inflationary universe scenario: a possible solution of the horizon, flatness, homogeneity, isotropy and primordial monopole problems,} \plb{108}{1982}{389};\\
  A.~J.~Albrecht and P.~J.~Steinhardt, {\it Cosmology for grand unified theories with radiatively induced symmetry breaking,} \prl{48}{1982}{1220}.

%\cite{Susskind:2003kw}
\bibitem{Susskind:2003kw}
  L.~Susskind, {\it The anthropic landscape of string theory,} [\hepth{0302219}].

%\cite{Coleman:1980aw}
\bibitem{Coleman:1980aw}
  S.~R.~Coleman, {\it The fate of the false vacuum. 1. semiclassical theory,} \prd{15}{1977}{2929} [Erratum-ibid. \prd{16}{1977}{1248}];\\
  C.~G.~Callan and S.~R.~Coleman, {\it The fate of the false vacuum. 2. first quantum corrections,} \prd{16}{1977}{1762}; \\
  S.~R.~Coleman and F.~De Luccia, {\it Gravitational effects on and of vacuum decay,} \prd{21}{1980}{3305}.

%\cite{Lee:1987qc}
\bibitem{Lee:1987qc}
  K.~M.~Lee and E.~J.~Weinberg, {\it Decay of the true vacuum in curved space-time,} \prd{36}{1987}{1088}.

%\cite{Ansoldi:2007qu}
\bibitem{Ansoldi:2007qu}
  S.~Ansoldi and L.~Sindoni, {\it Shell-mediated tunnelling between (anti-)de Sitter vacua,} \prd{76}{2007}{064020} [\grqc{0704.1073}].

%\cite{Blau:1986cw}
\bibitem{Blau:1986cw}
  S.~K.~Blau, E.~I.~Guendelman and A.~H.~Guth, {\it The dynamics of false vacuum bubbles,} \prd{35}{1987}{1747}.

%\cite{Aguirre:2005xs}
\bibitem{Aguirre:2005xs}
  A.~Aguirre and M.~C.~Johnson, {\it Dynamics and instability of false vacuum bubbles,} \prd{72}{2005}{103525} [\grqc{0508093}]; \\
  A.~Aguirre and M.~C.~Johnson, {\it Two tunnels to inflation,} \prd{73}{2006}{123529} [\grqc{0512034}].

%\cite{Alberghi:1999kd}
\bibitem{Alberghi:1999kd}
  G.~L.~Alberghi, D.~A.~Lowe and M.~Trodden, {\it Charged false vacuum bubbles and the AdS/CFT correspondence,} \jhep{9907}{1999}{020} [\hepth{9906047}].

%\cite{Freivogel:2005qh}
\bibitem{Freivogel:2005qh}
  B.~Freivogel, V.~E.~Hubeny, A.~Maloney, R.~C.~Myers, M.~Rangamani and S.~Shenker, {\it Inflation in AdS/CFT,} \jhep{0603}{2006}{007} [\hepth{0510046}].

%\cite{Sato:1981bf}
\bibitem{Sato:1981bf}
  K.~Sato, M.~Sasaki, H.~Kodama and K.~i.~Maeda, {\it Creation of wormholes by first order phase transition of a vacuum in the early universe,} \ptp{65}{1981}{1443}; \\
  H.~Kodama, M.~Sasaki, K.~Sato and K.~i.~Maeda, {\it Fate of wormholes created by first order phase transition in the early universe,} \ptp{66}{1981}{2052}; \\
  K.~i.~Maeda, K.~Sato, M.~Sasaki and H.~Kodama, {\it Creation of de Sitter-Schwarzschild wormholes by a cosmological first order phase transition,} \plb{108}{1982}{98}; \\
  K.~Sato, H.~Kodama, M.~Sasaki and K.~i.~Maeda, {\it Multiproduction of universes by first order phase transition of a vacuum,} \plb{108}{1982}{103}.

%\cite{Lee:2006vka}
\bibitem{Lee:2006vka}
  W.~Lee, B.~H.~Lee, C.~H.~Lee and C.~Park, {\it The false vacuum bubble nucleation due to a nonminimally coupled scalar field,} \prd{74}{2006}{123520} [\hepth{0604064}]; \\
  B.~H.~Lee, C.~H.~Lee, W.~Lee, S.~Nam and C.~Park, {\it Dynamics of false vacuum bubbles with the negative tension due to nonminimal coupling,} \prd{77}{2008}{063502} [\hepth{0710.4599}].

%\cite{Farhi:1986ty}
\bibitem{Farhi:1986ty}
  E.~Farhi and A.~H.~Guth, {\it An obstacle to creating a universe in the laboratory,} \plb{183}{1987}{149}.

%\cite{Farhi:1989yr}
\bibitem{Farhi:1989yr}
  E.~Farhi, A.~H.~Guth and J.~Guven, {\it Is it possible to create a universe in the laboratory by quantum tunneling?,} \npb{339}{1990}{417}; \\
  W.~Fischler, D.~Morgan and J.~Polchinski, {\it Quamtum necleation of false vacuum bubbles,} \prd{41}{1990}{2638}; \\
  W.~Fischler, D.~Morgan and J.~Polchinski, {\it Quantization of false vacuum bubbles: a Hamiltonian treatment of gravitational tunneling,} \prd{42}{1990}{4042}.

%\cite{Banks:2002nm}
\bibitem{Banks:2002nm}
  T.~Banks, {\it Heretics of the false vacuum: Gravitational effects on and of vacuum decay. II,} [\hepth{0211160}].

%\cite{Maldacena:1997re}
\bibitem{Maldacena:1997re}
  J.~M.~Maldacena, {\it The large N limit of superconformal field theories and supergravity,} \atmp{2}{1998}{231} [\ijtp{38}{1999}{1113}] [\hepth{9711200}].

%\cite{Piran}
\bibitem{Piran}
  T.~Piran and A.~Strominger, {\it Numerical analysis of black hole evaporation,} \prd{48}{1993}{4729} [\hepth{9304148}]; \\
  R.~Parentani and T.~Piran, {\it The internal geometry of an evaporating black hole,} \prl{73}{1994}{2805} [\hepth{9405007}]; \\
  S.~Ayal and T.~Piran, {\it Spherical collapse of a massless scalar field with semiclassical corrections,} \prd{56}{1997}{4768} [\grqc{9704027}]; \\
  S.~Hod and T.~Piran, {\it Mass inflation in dynamical gravitational collapse of a charged scalar field,} \prl{81}{1998}{1554} [\grqc{9803004}]; \\
  S.~Hod and T.~Piran, {\it The inner structure of black holes,} \grg{30}{1998}{1555} [\grqc{9902008}]; \\
  E.~Sorkin and T.~Piran, {\it The effects of pair creation on charged gravitational collapse,} \prd{63}{2001}{084006} [\grqc{0009095}]; \\
  E.~Sorkin and T.~Piran, {\it Formation and evaporation of charged black holes,}\prd{63}{2001}{124024} [\grqc{0103090}]; \\
  Y.~Oren and T.~Piran, {\it On the collapse of charged scalar fields,} \prd{68}{2003}{044013} [\grqc{0306078}].

%\cite{HHSY}
\bibitem{HHSY}
  S.~E.~Hong, D.~Hwang, E.~D.~Stewart and D.~Yeom, {\it The causal structure of dynamical charged black holes,} [\grqc{0808.1709}].

%\cite{Hansen:2005am}
\bibitem{Hansen:2005am}
  J.~Hansen, A.~Khokhlov and I.~Novikov, {\it Physics of the interior of a spherical, charged black hole with a scalar field,} \prd{71}{2005}{064013} [\grqc{0501015}]; \\
  A.~Doroshkevich, J.~Hansen, I.~Novikov and A.~Shatskiy, {\it Passage of radiation through wormholes,} [\grqc{0812.0702}].

%\cite{Israel:1966rt}
\bibitem{Israel:1966rt}
  W.~Israel, {\it Singular hypersurfaces and thin shells in general relativity,} {\it Nuovo Cim.} {\bf B 44S10} (1966) 1 [Erratum-ibid. {\bf B 48} (1967) 463][{\it Nuovo Cim.} {\bf B 44} (1966) 1].

%\cite{Hawking:1973uf}
\bibitem{Hawking:1973uf}
  S.~W.~Hawking and G.~F.~R.~Ellis, {\it ``The large scale structure of space-time,''} Cambridge, Cambridge University Press (1973).

%\cite{Wald:1984rg}
\bibitem{Wald:1984rg}
  R.~M.~Wald, {\it ``General relativity,''} Chicago, Chicago University Press (1984).

%\cite{Hamade:1995ce}
\bibitem{Hamade:1995ce}
  R.~S.~Hamade and J.~M.~Stewart, {\it The spherically symmetric collapse of a massless scalar field,} \cqg{13}{1996}{497} [\grqc{9506044}].

%\cite{Waugh:1986jh}
\bibitem{Waugh:1986jh}
  B.~Waugh and K.~Lake, {\it Double null coordinates for the Vaidya metric,} \prd{34}{1986}{2978}.

%\cite{nr}
\bibitem{nr}
  W.~H.~Press, S.~A.~Teukolsky, W.~T.~Vetterling and B.~P.~Flannery, {\it ``Numerical Recipes: The Art of Scientific Computing,'' 3rd ed.,}
  Cambridge, Cambridge University Press (2007).

%\cite{Sen:2002nu}
%\bibitem{Sen:2002nu}
%  A.~Sen, {\it Rolling tachyon,} \jhep{0204}{2002}{048} [\hepth{0203211}]; \\
%  A.~Sen, {\it Tachyon matter,} \jhep{0207}{2002}{065} [\hepth{0203265}].

%\cite{Fewster:2005vh}
\bibitem{Fewster:2005vh}
  M.~S.~Morris and K.~S.~Thorne, {\it Wormholes in space-time and their use for interstellar travel: A tool for teaching general relativity,} {\it Am.\ J.\ Phys.\ } {\bf 56} (1988) 395; \\
  M.~S.~Morris, K.~S.~Thorne and U.~Yurtsever, {\it Wormholes, time machines, and the weak energy condition,} \prl{61}{1988}{1446}; \\
  S.~W.~Hawking, {\it The chronology protection conjecture,} \prd{46}{1992}{603}; \\
  M.~Alcubierre, {\it The warp drive: hyper-fast travel within general relativity,} \cqg{11}{1994}{L73} [\grqc{0009013}]; \\
  S.~V.~Krasnikov, {\it Hyper-fast interstellar travel in general relativity,} \prd{57}{1998}{4760} [\grqc{9511068}]; \\
  C.~J.~Fewster and T.~A.~Roman, {\it Problems with wormholes which involve arbitrarily small amounts of exotic matter,} [\grqc{0510079}]; \\
  K.~A.~Bronnikov and J.~C.~Fabris, {\it Regular phantom black holes,} \prl{96}{2006}{251101} [\grqc{0511109}]; \\
  K.~A.~Bronnikov, V.~N.~Melnikov and H.~Dehnen, {\it Regular black holes and black universes,} \grg{39}{2007}{973} [\grqc{0611022}]; \\
  M.~Cataldo, S.~del Campo, P.~Minning and P.~Salgado, {\it Evolving Lorentzian wormholes supported by phantom matter and cosmological constant,} \prd{79}{2009}{024005} [\grqc{0812.4436}].

%\cite{Caldwell:1999ew}
\bibitem{Caldwell:1999ew}
  R.~R.~Caldwell, {\it A phantom menace?,} \plb{545}{2002}{23} [\astroph{9908168}]; \\
  G.~W.~Gibbons, {\it Phantom matter and the cosmological constant,} [\hepth{0302199}].

%\cite{local_horizon}
%\bibitem{local_horizon}
%  A.~Ashtekar and B.~Krishnan, {\it Isolated and dynamical horizons and their applications,} {\it Living Rev.\ Rel.\ } {\bf 7} (2004) 10 [\grqc{0407042}]; \\
%  S.~A.~Hayward, {\it Marginal surfaces and apparent horizons,} [\grqc{9303006}]; \\
%  S.~A.~Hayward, {\it Energy and entropy conservation for dynamical black holes,} \prd{70}{2004}{104027} [\grqc{0408008}].

%\cite{Yeom:2008qw}
\bibitem{Yeom:2008qw}
  D.~Yeom and H.~Zoe, {\it Constructing a counterexample to the black hole complementarity,} \prd{78}{2008}{104008} [\grqc{0802.1625}]; \\
  S.~E.~Hong, D.~Hwang, D.~Yeom and H.~Zoe, {\it Black hole complementarity with local horizons and Horowitz-Maldacena's proposal,} \jhep{0812}{2008}{080} [\grqc{0809.1480}]; \\
  D.~Yeom and H.~Zoe, {\it Black hole complementarity gets troubled by charged black holes,} [\grqc{0811.1637}].

%\cite{inforpara}
\bibitem{inforpara}
  S.~W.~Hawking, {\it Breakdown of predictability in gravitational collapse,} \prd{14}{1976}{2460}.

%\cite{Yeom:2009zp}
\bibitem{Yeom:2009zp}
  D.~Yeom and H.~Zoe, {\it Semi-classical black holes with large N re-scaling and information loss problem,} [\hepth{0907.0677}].

%\cite{Hawkingradiation}
\bibitem{Hawkingradiation}
  S.~W.~Hawking, {\it Particle creation by black holes,} \cmp{43}{1975}{199} [Erratum-ibid.\ {\bf 46} (1976) 206].

%\cite{Birrell:1982ix}
\bibitem{Birrell:1982ix}
  N.~D.~Birrell and P.~C.~W.~Davies,
  {\it ``Quantum fields in curved space,''}
  Cambridge, Cambridge University Press (1982).

\end{thebibliography}
\end{document}